\documentclass[12pt]{article}

\usepackage{xcolor}
\definecolor{nosaka}{rgb}{0.7, 0.3, 0.0}
\usepackage{cite}
\usepackage{amsmath}
\usepackage{amssymb}
\usepackage{bm}
\usepackage{color}
\usepackage{graphicx}
\numberwithin{equation}{section}

\allowdisplaybreaks

\DeclareMathOperator{\Tr}{Tr}

\DeclareMathOperator{\Det}{Det}

\makeatletter
\newsavebox{\@brx}
\newcommand{\llangle}[1][]{\savebox{\@brx}{\(\m@th{#1\langle}\)}%
  \mathopen{\copy\@brx\kern-0.5\wd\@brx\usebox{\@brx}}}
\newcommand{\rrangle}[1][]{\savebox{\@brx}{\(\m@th{#1\rangle}\)}%
  \mathclose{\copy\@brx\kern-0.5\wd\@brx\usebox{\@brx}}}
\makeatother

\renewcommand{\thefootnote}{\fnsymbol{footnote}}

\newcounter{aff}

\newcommand{\strokedint}{\hspace{0.15cm}-\hspace{-0.45cm}\int}
\newcommand{\strokedintnotdisplay}{\hspace{0.06cm}-\hspace{-0.38cm}\int\hspace{0.07cm}}

\begin{document}
\begin{titlepage}
\begin{flushright}
{\footnotesize NITEP 172, OCU-PHYS 581}
\end{flushright}
\begin{center}
{\Large\bf
40 Bilinear Relations of $q$-Painlev\'e VI\\[6pt]
from ${\cal N}=4$ Super Chern-Simons Theory
}\\
\bigskip\bigskip
{\large
Sanefumi Moriyama\footnote{\tt moriyama@omu.ac.jp}
and
Tomoki Nosaka\footnote{\tt nosaka@yukawa.kyoto-u.ac.jp}
}\\
\bigskip
${}^\dagger$\,{\it Kavli Institute for Theoretical Sciences,}\\
{\it University of Chinese Academy of Sciences, Beijing, China 100190}\\[3pt]
${}^*$\,{\it Department of Physics, Graduate School of Science,}\\
{\it Osaka Metropolitan University, Osaka, Japan 558-8585}\\[3pt]
${}^*$\,{\it Nambu Yoichiro Institute of Theoretical and Experimental Physics (NITEP),}\\
{\it Osaka Metropolitan University, Osaka, Japan 558-8585}\\[3pt]
${}^*$\,{\it Osaka Central Advanced Mathematical Institute (OCAMI),}\\
{\it Osaka Metropolitan University, Osaka, Japan 558-8585}
\end{center}

\begin{abstract}
We investigate partition functions of the circular-quiver supersymmetric Chern-Simons theory which corresponds to the $q$-deformed Painlev\'e VI equation.
From the partition functions with the lowest rank vanishing, where the circular quiver reduces to a linear one, we find 40 bilinear relations.
The bilinear relations extend naturally to higher ranks if we regard these partition functions as those in the lowest order of the grand canonical partition functions in the fugacity.
Furthermore, we show that these bilinear relations are a powerful tool to determine some unknown partition functions.
We also elaborate the relation with some previous works on $q$-Painlev\'e equations.
\end{abstract}

\end{titlepage}

\setcounter{footnote}{0}
\renewcommand{\thefootnote}{$\dagger$\arabic{footnote}}

\tableofcontents

\section{Introduction\label{intro}}

Gauge theories with extended supersymmetries play a central role in understanding the mathematical structure of string theory.
Due to their enormous symmetries, these theories are often integrable \cite{DW}.
These integrable structures allow us to solve the theories and sometimes the solvability further helps us to understand the integrabilities more deeply.
In this paper we focus on the integrability of a three-dimensional ${\cal N}=4$ supersymmetric Chern-Simons theory, which is a generalization of the Aharony-Bergman-Jafferis-Maldacena (ABJM) theory \cite{ABJM}.
We find a large set of non-trivial relations to constrain the partition function, which are associated to Painlev\'e equations \cite{KNY}.
In the introduction, after explaining the theory, we shall summarize our main results and sketch how we have tracked them down.

The ABJM theory \cite{ABJM} is the three-dimensional ${\cal N}=6$ super Chern-Simons theory with gauge group $\text{U}(N)_k\times\text{U}(N)_{-k}$ (subscripts $(k,-k)$ denoting the Chern-Simons levels) and two pairs of bifundamental matters.
It was proposed that this theory describes the worldvolume theory of $N$ M2-branes on the background ${\mathbb C}^4/{\mathbb Z}_k$.
The description originates from the IIB brane configuration of D3-branes on a circle with an NS5-brane and a $(1,k)$5-brane placed perpendicularly at different positions and tilted relatively by an angle to preserve the supersymmetries.

The theory we focus on here is a three-dimensional ${\cal N}=4$ super Chern-Simons theory of circular-quiver type with gauge group $\text{U}(N)_0\times\text{U}(N)_k\times\text{U}(N)_0\times\text{U}(N)_{-k}$ and bifundamental matters connecting adjacent gauge group factors \cite{GW}.
This theory describes the worldvolume theory of $N$ M2-branes on the background $[{\mathbb C}^2/{\mathbb Z}_2\times{\mathbb C}^2/{\mathbb Z}_2]/{\mathbb Z}_k$ \cite{IK}.
The brane configuration consists of D3-branes with two NS5-branes and two $(1,k)$5-branes.

From the gravity side \cite{KT}, it was expected that the degrees of freedom of $N$ M2-branes behaves as $N^{\frac{3}{2}}$ in the large $N$ limit.
After reducing the infinite-dimensional path integral of the partition function into a finite-dimensional matrix model \cite{KWY} and confirming the large $N$ behavior from the free energy of the ABJM theory \cite{DMP1,DMP2}, the perturbative corrections to the ABJM partition function $\exp(N^{\frac{3}{2}})$ was found to sum up to the Airy function \cite{FHM}.
The integral representation of the Airy function, then, naturally leads us to the grand canonical ensemble, where the grand partition function is expressed as the Fredholm determinant $\Det(1+\kappa\widehat H^{-1})$ of free fermions with a spectral operator $\widehat H$ and a fugacity $\kappa$ \cite{MP}.
One of the main applications of the Fermi gas formalism is the full non-perturbative expression of the grand partition function including two types of instanton effects, the worldsheet instantons and the membrane instantons \cite{HMO2,CM,HMO3,HMMO}, based on the exact values obtained from the Fredholm determinant \cite{HMO1,PY}.
The results are summarized by the free energy of refined topological strings \cite{HMMO} on a local Calabi-Yau threefold whose mirror curve is given by the classical limit of ${\widehat H}$.

In the Fermi gas formalism, the spectral operator $\widehat H$ of the ABJM theory is constructed by translating the NS5-brane into $\widehat{\cal X}=\widehat X^{\frac{1}{2}}+\widehat X^{-\frac{1}{2}}$ and the $(1,k)$5-brane into $\widehat{\cal P}=\widehat P^{\frac{1}{2}}+\widehat P^{-\frac{1}{2}}$ respectively with the canonical operators $\widehat X=e^{\widehat x}$ and $\widehat P=e^{\widehat p}$ satisfying the canonical commutation relation $[\widehat x,\widehat p]=i\hbar$ and the Planck constant $\hbar$ identified with the Chern-Simons level by $\hbar=2\pi k$.
The Fermi gas formalism is generalized\footnote{
The Fermi gas formalism is also generalized to other supersymmetric gauge theories such as \cite{BDF}.
}
to various super Chern-Simons theories \cite{MP,MN1}, where the spectral operators are given by the same rule.

Geometrically, the spectral operator for the ABJM theory is classically the algebraic curve ${\mathbb P}^1\times{\mathbb P}^1$ with symmetries of the $A_1$ Weyl group, whose quantum counterpart \cite{MiMo,ACDKV} further governs physical quantities.
For the brane configuration of our theory with two NS5-branes and two $(1,k)$5-branes \cite{MN3,MNN}, hence two $\widehat{\cal X}$ and two $\widehat{\cal P}$, the spectral operator $\widehat H$ consists of terms $\widehat X^m\widehat P^n$ with $m,n=1,0,-1$.
Classically, this is the famous algebraic curve with symmetries of the $D_5$ Weyl group and hence the grand partition function is summarized by the $D_5$ character \cite{MNY,KMN} as expected.

The super Chern-Simons theories of circular-quiver type can be generalized to those with rank differences \cite{HLLLP2,ABJ}, which are interpreted as the worldvolume theories of $\min(N_i)$ M2-branes with various fractional M2-branes on the same orbifold backgrounds as those for the uniform ranks.
In studying partition functions with fractional branes, two formalisms associated respectively to open strings \cite{MM,HO} and closed strings \cite{H,KMZ,MS2,MN5,closed} were proposed for the grand partition functions.
For the open string formalism we correct the Fredholm determinant by multiplying with quantities similar to loop operators, while for the closed string formalism we modify the spectral operator itself.
Practically, to obtain the spectral operator in the closed string formalism, it is easier to extrapolate it from those without fractional branes \cite{KM,FMN} by using the Hanany-Witten transitions \cite{HW}.
As pointed out in \cite{KM}, in the extrapolation we need to fix an interval where the Hanany-Witten transitions are forbidden in brane configurations and avoid uncritical uses of similarity transformations for spectral operators.

After rephrasing matrix models into spectral operators, the correspondence between spectral theories and topological strings was proposed \cite{GHM1} as a generalization of \cite{HMMO}.
As long as the (normalized)\footnote{
Note that the grand partition function with rank differences needs to be normalized by the partition function with the lowest rank vanishing which is not necessarily equipped with the group-theoretical structure.
This fact is important in later studies of bilinear relations for the grand partition function.
}
grand partition function is given by the Fredholm determinant of a spectral operator in a Laurent polynomial of ${\widehat X}$ and ${\widehat P}$, the perturbative behavior of the Airy function or $N^{\frac{3}{2}}$ is always guaranteed.
For this reason, it is natural to consider that M2-branes are described by spectral operators instead of matrix models. 
From this viewpoint the relation between the partition function of M2-branes and quantum curves with symmetries of exceptional Weyl groups were clarified \cite{M,MY}.
As we see below, these properties of the partition function of M2-branes suggest a relation to $q$-Painlev\'e equations.

Painlev\'e equations appear in many interesting physical systems such as \cite{MW}.
Aside from their applications, Painlev\'e equations have a fascinating history by themselves.
They originate from classification of differential equations with the condition of no movable branch points.
One can also construct $q$-deformed Painlev\'e equations by uplifting the time derivatives to $q$-differences \cite{Grammaticos:1991zz,Ramani:1991zz,JS} and generalize the notion of Painlev\'e equations.
In the modern viewpoint, more interesting aspects of Painlev\'e equations are added.
Namely, in relations to integrabilities, tau functions are introduced also for Painlev\'e equations and Painlev\'e equations are reformulated in terms of bilinear equations for tau functions.
Furthermore, it was found that Painlev\'e equations are classified systematically by algebraic curves \cite{S,KNY}, where the $q$-deformations are also incorporated naturally.

The classification of Painlev\'e equations by algebraic curves suggests their appearance in physical systems characterized by the same algebraic curves.
Indeed, a particularly interesting aspect is the connection between $q$-Painlev\'e equations and the five-dimensional ${\cal N}=1$ $\text{SU}(2)$ super Yang-Mills theories \cite{BS3,JNS,Matsuhira:2018qtx}, where the tau functions are given by the discrete Fourier transform of the Nekrasov partition functions (Nekrasov-Okounkov partition functions) of the gauge theories.
These five-dimensional theories are characterized by the genus-one curves with exceptional Weyl groups \cite{Seiberg:1996bd}, each of which precisely coincides with the algebraic curve associated with the corresponding $q$-Painlev\'e equation.
For $N_\text{f}\le 4$, this is the $q$-uplift of the connection between Painlev\'e equations and the four-dimensional ${\cal N}=2$ $\text{SU}(2)$ super Yang-Mills theories \cite{GIL,Iorgov:2014vla}.\footnote{
A similar connection with integrable systems is also known for the four (or five) dimensional pure super Yang-Mills theories with a general gauge group $G$, where the corresponding integrable system is the non-autonomous ($q$-)Toda chain of affine $G$ type \cite{Bonelli:2017ptp,Bonelli:2021rrg,Bonelli:2022iob,Gavrylenko:2023ewx}.
}


Since the partition functions of M2-branes are characterized by quantum algebraic curves as mentioned above, it is natural to expect that there is also a connection between these partition functions and ($q$-)Painlev\'e equations.
Indeed, it was found that the grand partition function of the ABJM theory, whose corresponding curve is $\mathbb{P}^1\times \mathbb{P}^1$, solves the Hirota bilinear equation for the tau function of the third $q$-Painlev\'e equation $q\text{P}_{\text{III}_3}$ which is also associated with $\mathbb{P}^1\times \mathbb{P}^1$ in the Sakai classification \cite{Zamolodchikov:1994uw,GHM2,BGT3}.
Here we turn to our super Chern-Simons theory associated to the curve with symmetries of the $D_5$ Weyl group.
This time the corresponding Painlev\'e equation is the sixth $q$-Painlev\'e equation $q\text{P}_\text{VI}$.
Compared with the simpler curve ${\mathbb P}^1\times{\mathbb P}^1$ where many interesting studies of the bilinear equations for the grand partition functions are initiated, our curve allows us to detect the full group-theoretical structure of the bilinear equations.
For example, instead of values for partition functions with discrete rank differences in \cite{BGT3}, our partition functions are obtained also as functions of the continuous Fayet-Iliopoulos (FI) parameters, where the bilinear relations are subject to more stringent checks.
Note also that, after taking care of the $D_5$ Weyl group, non-trivial coefficients not fixed by the Weyl group appear for our bilinear relations of $q\text{P}_{\text{VI}}$.

In this paper we discuss bilinear relations for the above-mentioned super Chern-Simons theory.
As listed later in tables \ref{bilinear1} and \ref{bilinear2} (and in section \ref{higher} for higher orders in the fugacity $\kappa$)
we find 40 bilinear relations.
We may want to summarize the bilinear relations for the grand partition functions symbolically as
\begin{align}
&e^{\frac{\pi i}{2k}(\sigma_1 c+\sigma_2 d+\sigma_3 e)}S\Xi^\text{(bilinear)}_{a_\pm b_\pm c_\varnothing d_\varnothing e_\varnothing}(\kappa)
+e^{-\frac{\pi i}{2k}(\sigma_1 c+\sigma_2 d+\sigma_3 e)}S\Xi^\text{(bilinear)}_{a_\pm b_\mp c_\varnothing d_\varnothing e_\varnothing}(-\kappa)\nonumber\\
&\quad\quad\quad\quad+S\Xi^\text{(bilinear)}_{a_\varnothing b_\varnothing c_{\pm \sigma_1} d_{\pm \sigma_2} e_{\pm \sigma_3}}(\mp i\kappa)=0,
\label{bischematical}
\end{align}
where $\Xi^\text{(bilinear)}$ are bilinear forms of the grand partition functions with various shifts of the parameters appearing in subscripts and $S$ are $\kappa$-independent prefactors.
The details of these bilinear relations are clarified further around the tables.

The discovery of the 40 bilinear relations is phenomenological.
Our analysis relies heavily on the computation of the grand partition function in \cite{BGKNT} combining the two formalisms associated to open strings and closed strings.
The nice expression allows us to compute the lowest order of the grand partition function in the fugacity $\kappa$ for any value of $k$, where the circular quiver reduces to a linear one.
With the help of a computer we can evaluate all terms of the lowest order for various rank differences.
Firstly, it is interesting to observe that the expression is only non-vanishing in the fundamental domain proposed in \cite{FMMN,FMS}, where duality cascades in three dimensions are shown to terminate uniquely.
In particular for $k=1,2$, the exact expressions of the grand partition function at the lowest order are simple enough, which allows us to discover the bilinear relations \eqref{bischematical} heuristically.
Namely, we proceed as follows.
We assume the structure of bilinear relations by expressing bilinear partition functions with two orthogonal shifts in two variables in terms of those with shifts in the remaining variables \eqref{bischematical} as already known in \cite{Ohta,JNS}.
Under the assumption, we can compute additions with various factors $S$ for the first two terms, where the result is only simplified when we have correct factors of $S$.
We can then divide by the third term to find out the expression for the bilinear relations.
We also generalize our discussions to higher ranks, where the domain of non-vanishing partition functions extends from the original fundamental domain, and find that the fugacity $\kappa$ needs to be corrected with simple coefficients as in \eqref{bischematical}.

Note that a similar analysis was worked out in \cite{BGKNT} by consulting the 8 bilinear relations in \cite{JNS}.
We point out that there are, however, some room for improvements.
\begin{itemize}
\item The Fredholm determinant was identified as the tau function with a non-trivial overall factor in \cite{BGKNT}.
Here we can replace the factor by the partition function with the lowest rank vanishing, which has a physically sound origin.
\item Moreover, although a non-trivial phase factor was introduced for higher ranks as coefficients of $\kappa$ in \cite{BGKNT}, since the grand partition function is expected to be given by the Fredholm determinant of the spectral operator invariant under the Weyl group once being normalized by the partition function with the lowest rank vanishing, the phase should also be invariant under the Weyl group.
This helps us to simplify the coefficients for $\kappa$ as in \eqref{bischematical}.
\item Although the structure of bilinear relations are fixed directly from the group-theoretical viewpoint \cite{Ohta}, the constraint does not apply to the coefficients.
For this reason, it is desirable to fix the coefficients also for the remaining 32 bilinear relations.
\item Besides, we find that the coefficients found from the current three-dimensional super Chern-Simons theories look somewhat simpler than those for the five-dimensional super Yang-Mills theories in \cite{JNS}.
We hope that the structures found here are helpful in understanding the relations to the five-dimensional theories as well.
\end{itemize}

The content of this paper is given as follows.
After recapitulating the three-dimensional ${\cal N}=4$ super Chern-Simons theory in the next section, we turn to the study of bilinear relations.
We first identify the bilinear relations for the partition function with the lowest rank vanishing in section \ref{bilinear} and move to the full grand partition function in section \ref{higher}.
We then discuss the covariance under the Weyl group in section \ref{sec_Weylcovariance} and exhibit some applications in section \ref{app}.
After discussing comparison with some previous works in section \ref{comparison}, finally we conclude by listing some future directions in section \ref{conclusion}.
Appendix \ref{computation} is devoted to technical details of the evaluations of partition functions and various results on partition functions are given in appendices \ref{lowestpf}, \ref{higherrank1} and \ref{checkedlist}.

\section{Super Chern-Simons theory}

In this section, we recapitulate the three-dimensional ${\cal N}=4$ super Chern-Simons theory we consider in this paper.
We first explain the theory with its brane configuration and discuss its symmetries of the $D_5$ Weyl group.
We then proceed to computations of the partition function.

\subsection{Brane configuration and Weyl group}

\begin{table}[ht!]
\begin{center}
\begin{tabular}{c|ccccc}
&$0\quad 1\quad 2$&$6$&$3\quad 7$&$4\quad 8$&$5\quad 9$\\\hline
D3-branes&$-\;-\;-$&$-$&&\\
NS5-branes&$-\;-\;-$&&$-\;\phantom{--}$&$-\;\phantom{--}$&$-\;\phantom{--}$\\
$(1,k)$5-branes&$-\;-\;-$&&$[3,7]_k$&$[4,8]_k$&$[5,9]_k$\\
\end{tabular}
\end{center}
\caption{Directions which various branes extend to.}
\label{branedirection}
\end{table}

\begin{figure}[!t]
\centering\includegraphics[width=10cm]{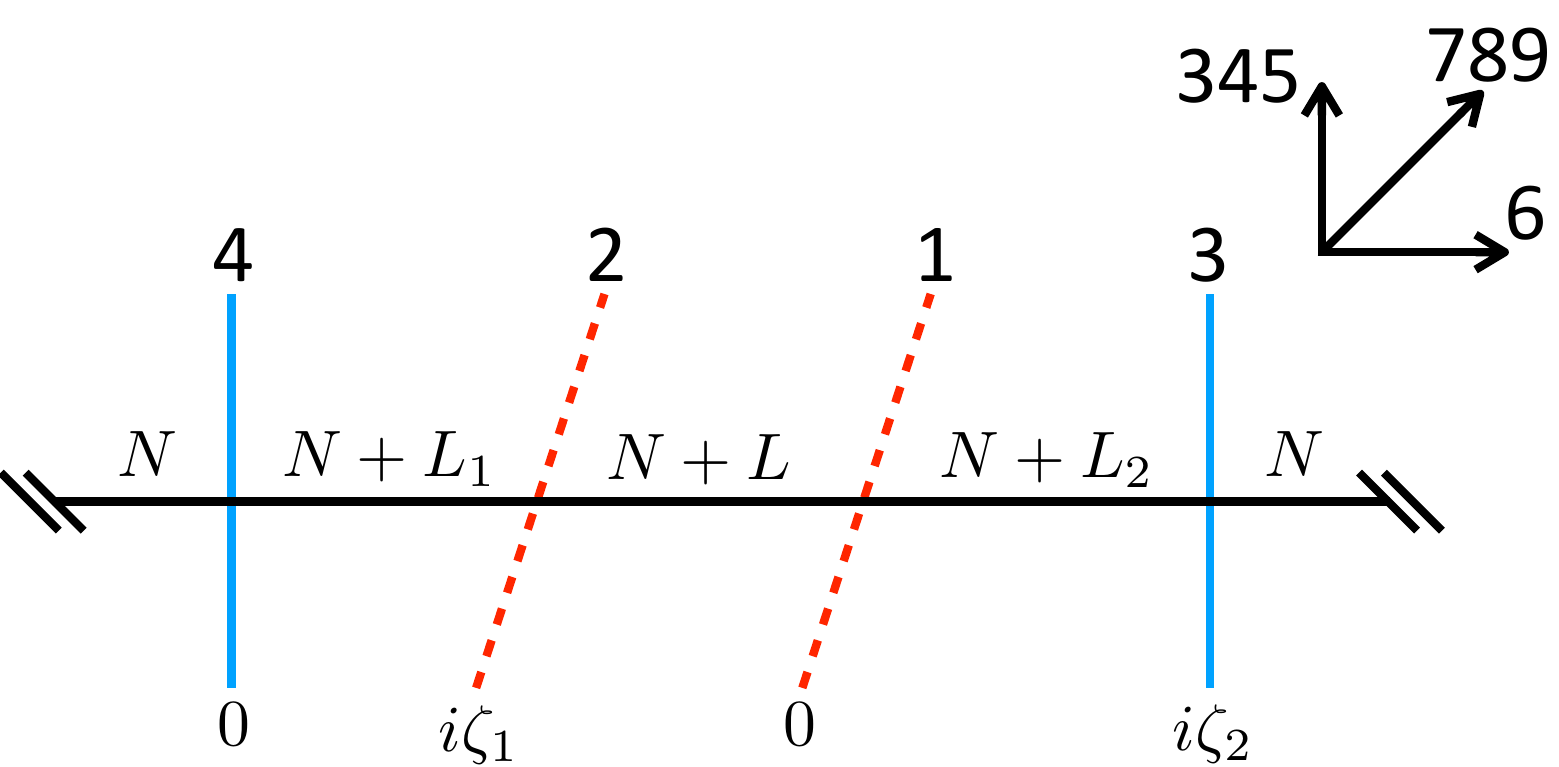}
\caption{Cartoon of the brane configuration considered in this paper.
The solid vertical blue lines are NS5-branes and the dashed tilted red lines are $(1,k)$5-branes.
}
\label{cartoon}
\end{figure}

Let us review the brane configuration we consider in this paper.
We compactify IIB string theory on a circle and place two NS5-branes and two $(1,k)$5-branes on the circle at different positions with the two types of 5-branes tilted relatively by an angle $\arctan k$.
We further add D3-branes connecting various pairs of 5-branes along the compactified direction.
See table \ref{branedirection} for directions which various branes extend to.
In discussing the brane configuration, we cut off the circle as in \cite{KM}\footnote{
As explained in \cite{KM} it is inevitable to cut the circle into a segment to avoid ambiguity in charge conservations when we discuss the Hanany-Witten transition.
}
by indicating the cut with the endpoints of brackets and abbreviate NS5-branes and $(1,k)$5-branes as $\circ$ and $\bullet$ respectively with labels.
Namely, the brane configuration we consider is given by
\begin{align}
&{\langle}N_1\mathop{\circ}^4_0
N_2\mathop{\bullet}^2_{i\zeta_1}
N_3\mathop{\bullet}^1_0
N_4\mathop{\circ}^3_{i\zeta_2}{\rangle}
={\langle}N\mathop{\circ}^4_0
N+L_1\mathop{\bullet}^2_{i\zeta_1}
N+L\mathop{\bullet}^1_0
N+L_2\mathop{\circ}^3_{i\zeta_2}{\rangle}.
\label{braneconfigurationofBGKNT}
\end{align}
Here we denote the number of D3-branes in each interval with\footnote{To fix the arguments we often restrict ourselves to the range of $L\le L_1$ and $L\le L_2$.} $N$, $L_1$, $L_2$ and $L$ being non-negative.
Namely, the first rank $N_1=N$ is assumed to be the lowest and the overall rank $N$ is separated from the remaining relative ranks $(L_1,L_2,L)$.
Also, we label the 5-branes $\bullet$ and $\circ$ by Arabic numerals above them and indicate the orthogonal positions $i\zeta_1,i\zeta_2$ below them. 
See figure \ref{cartoon} for a cartoon of the brane configuration.
It is known that by dualizing to IIA string theory and lifting to M-theory, the system describes M2-branes with fractional M2-branes on the background $[{\mathbb C}^2/{\mathbb Z}_2\times{\mathbb C}^2/{\mathbb Z}_2]/{\mathbb Z}_k$ \cite{ABJM,IK}.
Here, however, we stick to the picture of IIB string theory.

The corresponding effective theory on D3-branes (and hence, M2-branes, after applying various dualities) is the three-dimensional ${\cal N}=4$ super Chern-Simons theory with the gauge group \cite{Kitao:1998mf,Bergman:1999na}
\begin{align}
&\text{U}(N_1)_{0,-i\zeta_2}\times
\text{U}(N_2)_{k,i\zeta_1}\times
\text{U}(N_3)_{0,-i\zeta_1}\times
\text{U}(N_4)_{-k,i\zeta_2}\nonumber\\
&=\text{U}(N)_{0,-i\zeta_2}\times
\text{U}(N+L_1)_{k,i\zeta_1}\times
\text{U}(N+L)_{0,-i\zeta_1}\times
\text{U}(N+L_2)_{-k,i\zeta_2}.
\label{fournode}
\end{align}
Here the first subscripts denote the levels for the gauge group, while the second ones denote the FI parameters.
Let us express the partition function of the gauge theory by $Z'_{k,{\bm M}}(N)$ where relative ranks $(L_1,L_2,L)$ and FI parameters $(\zeta_1,\zeta_2)$ are denoted collectively by ${\bm M}$.
By applying the localization technique, the partition function defined by the infinite-dimensional path integral reduces to a finite-dimensional matrix model \cite{KWY,Jafferis:2010un,Hama:2010av}
\begin{align}
Z'_{k,\bm{M}}(N)&=\frac{i^{-\frac{N_2^2}{2}+\frac{N_4^2}{2}}}{N_1!N_2!N_3!N_4!}\int
\prod_{i=1}^{N_1}\frac{d\lambda_i^{(1)}}{2\pi}
\prod_{i=1}^{N_2}\frac{d\lambda_i^{(2)}}{2\pi}
\prod_{i=1}^{N_3}\frac{d\lambda_i^{(3)}}{2\pi}
\prod_{i=1}^{N_4}\frac{d\lambda_i^{(4)}}{2\pi}\nonumber\\
&\quad\times e^{\frac{ik}{4\pi}\sum_{i=1}^{N_2}(\lambda_i^{(2)})^2
-\frac{ik}{4\pi}\sum_{i=1}^{N_4}(\lambda_i^{(4)})^2}
e^{-i\zeta_1(\sum_{i=1}^{N_2}\lambda_i^{(2)}-\sum_{i=1}^{N_3}\lambda_i^{(3)})-i\zeta_2(\sum_{i=1}^{N_4}\lambda_i^{(4)}-\sum_{i=1}^{N_1}\lambda_i^{(1)})}\nonumber\\
&\quad\times\prod_{a=1}^4\frac{\prod_{i<j}^{N_a}(2\sinh\frac{\lambda_i^{(a)}-\lambda_j^{(a)}}{2})^2}
{\prod_{i=1}^{N_a}\prod_{j=1}^{N_{a+1}}2\cosh\frac{\lambda_i^{(a)}-\lambda_j^{(a+1)}}{2}},
\label{Zprime1}
\end{align}
with $N_5=N_1$ and $\lambda_j^{(5)}=\lambda_j^{(1)}$.
If we consider the grand canonical ensemble of the partition function, aside from the partition function at $N=0$, the whole tower of the summation in $N$ is given by the Fredholm determinant \cite{MP,MNN}
\begin{align}
\sum_{N=0}^\infty Z_{k,{\bm M}}'(N)\kappa^N
=Z_{k,{\bm M}}'(0)\Det(1+\kappa {\widehat H}_{k,{\bm M}}^{-1}),
\label{XiHinvold}
\end{align}
with the spectral operator $\widehat H_{k,{\bm M}}$ being a Laurent polynomial of $\widehat X$ and $\widehat P$,
\begin{align}
\widehat H_{k,{\bm M}}=\sum_{m,n=1,0,-1}c_{m,n}e^{-\frac{i\hbar}{2}mn}
\widehat X^m\widehat P^n,
\label{H}
\end{align}
where ${\widehat X}=e^{\widehat x}$ and ${\widehat P}=e^{\widehat p}$ are exponentiated canonical operators, satisfying $[\widehat x,\widehat p]=i\hbar$ with $\hbar=2\pi k$.
The coefficients $c_{m,n}$ are some functions of $(L_1,L_2,L,\zeta_1,\zeta_2)$.
Since the Newton polygon of the spectral operator $\widehat H_{k,{\bm M}}$ in the Fredholm determinant is identical to that of the $D_5$ curve, as in previous works \cite{MNY,KMN}, we conjecture that the spectral operator $\widehat H_{k,{\bm M}}$ is invariant under the $D_5$ Weyl group (up to similarity transformations).
For this reason, we introduce new variables $(M_0,M_1,M_3,Z_1,Z_3)$ suitable for the $D_5$ Weyl group to parameterize the brane configuration
\begin{align}
{\langle}N
\mathop{\bullet}^1_{Z_1}N+M_0+M_1+k
\mathop{\bullet}^2_{0}N+2M_0+2k
\mathop{\circ}^3_{Z_3}N+M_0+M_3+k
\mathop{\circ}^4_{0}{\rangle},
\label{braneconfig}
\end{align}
and identify them with $(L_1,L_2,L,\zeta_1,\zeta_2)$ afterwards.

\begin{figure}[!t]
\centering\includegraphics[scale=0.6,angle=-90]{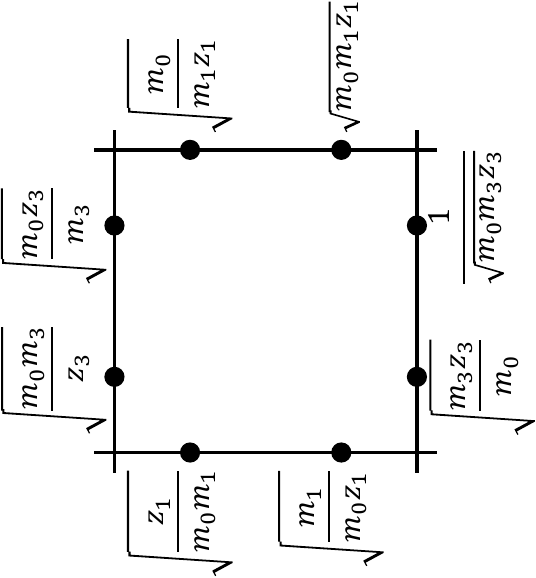}
\caption{Point configuration of asymptotic values for the quantum curve with symmetries of the $D_5$ Weyl group.}
\label{config}
\end{figure}

It is known that the curve is characterized by the eight asymptotic values at $X,P\rightarrow 0,\infty$, which are subject to a constraint from Vieta's formula and two redundancies of scalings coming from trivial similarity transformations.
The curve corresponding to the brane configuration \eqref{braneconfig} with various relative ranks and FI parameters was studied by extrapolating the asymptotic values linearly from those for the brane configurations without rank differences, which were found to be \cite{KM}
\begin{align}
&\big(\infty,-\sqrt{m_0m_1z_1}\big),\quad\Big(\infty,-\sqrt{\frac{m_0}{m_1z_1}}\Big),\quad
\Big(-\sqrt{\frac{m_0z_3}{m_3}},\infty\Big),\quad\Big(-\sqrt{\frac{m_0m_3}{z_3}},\infty\Big),
\nonumber\\
&\Big(0,-\sqrt{\frac{z_1}{m_0m_1}}\Big),\quad\Big(0,-\sqrt{\frac{m_1}{m_0z_1}}\Big),\quad
\Big(-\sqrt{\frac{m_3z_3}{m_0}},0\Big),\quad\Big(-\frac{1}{\sqrt{m_0m_3z_3}},0\Big),
\label{asymptoticvalues}
\end{align}
(symbolically given in\footnote{
Here we have changed notations from some of our previous works \cite{MNN,FMMN} by
$m_0=(m_1)^\text{\cite{FMMN}}$,
$m_1=\big(m_2/m_3\big)^\text{\cite{FMMN}}$,
$m_3=\big(1/(m_2m_3)\big)^\text{\cite{FMMN}}$.
} figure \ref{config}) with
\begin{align}
m_0=e^{2\pi iM_0},\quad
m_1=e^{2\pi iM_1},\quad
m_3=e^{2\pi iM_3},\quad
z_1=e^{2\pi iZ_1},\quad
z_3=e^{2\pi iZ_3}.
\label{additive}
\end{align}
Note that if the square roots look too cumbersome, we can apply the similarity transformations to shift the asymptotic values along each direction to avoid them.
Symmetries of the $D_5$ Weyl group for the curve are generated by the reflections on the asymptotic values
\begin{align}
s_1&:(M_0,M_1,M_3,Z_1,Z_3)\mapsto(M_0,M_1,Z_3,Z_1,M_3),\nonumber\\
s_2&:(M_0,M_1,M_3,Z_1,Z_3)\mapsto(M_0,M_1,-Z_3,Z_1,-M_3),\nonumber\\
s_3&:(M_0,M_1,M_3,Z_1,Z_3)\mapsto(M_3,M_1,M_0,Z_1,Z_3),\nonumber\\
s_4&:(M_0,M_1,M_3,Z_1,Z_3)\mapsto(-M_1,-M_0,M_3,Z_1,Z_3),\nonumber\\
s_5&:(M_0,M_1,M_3,Z_1,Z_3)\mapsto(M_0,Z_1,M_3,M_1,Z_3),\nonumber\\
s_0&:(M_0,M_1,M_3,Z_1,Z_3)\mapsto(M_0,-Z_1,M_3,-M_1,Z_3).
\label{WD5}
\end{align}

Now let us relate the new brane configuration \eqref{braneconfig} with the original configuration \eqref{braneconfigurationofBGKNT} via the Hanany-Witten transition as\footnote{
To obtain the second line we have used, besides the Hanany-Witten transitions, the symmetry of the partition function
\begin{align}
{\langle}
N_1\mathop{\bullet}_{Z_1}
N_2\mathop{\bullet}_{Z_2}
N_3\mathop{\circ}_{Z_3}
N_4\mathop{\circ}_{Z_4}
{\rangle}
={\langle}
N_1\mathop{\bullet}_{Z_1+a}
N_2\mathop{\bullet}_{Z_2+a}
N_3\mathop{\circ}_{Z_3+b}
N_4\mathop{\circ}_{Z_4+b}
{\rangle},
\end{align}
for arbitrary values of $N_1,N_2,N_3,N_4,Z_1,Z_2,Z_3,Z_4$ and $a,b$.
Namely, the overall shift $a+b$ does not matter as the FI parameters come from the differences of $Z_i$, while the relative shift $a-b$ is irrelevant since it can be absorbed by the uniform shift of integration variables in \eqref{Zprime1}.
}
\begin{align}
&{\langle}
N\mathop{\circ}^4_0
N+L_1\mathop{\bullet}^2_{i\zeta_1}
N+L\mathop{\bullet}^1_0
N+L_2\mathop{\circ}^3_{i\zeta_2}
{\rangle}\nonumber\\
&\qquad={\langle}
N\mathop{\bullet}^1_{-i\zeta_1}
N+L_2-L+k\mathop{\bullet}^2_0
N+L_2-L_1+2k\mathop{\circ}^3_{i\zeta_2}
N-L_1+2k\mathop{\circ}^4_0
{\rangle}.
\label{identify}
\end{align}
By comparing the two brane configurations, we can identify the two sets of parameters as
\begin{align}
&L_1-k=-M_0-M_3,\quad
L_2-k=M_0-M_3,\quad
L-k=-M_1-M_3,\quad
\zeta_1=iZ_1,\quad
\zeta_2=-iZ_3,\nonumber\\
&M_0=-\frac{L_1}{2}+\frac{L_2}{2},\quad
M_1=-L+\frac{L_1}{2}+\frac{L_2}{2},\quad
M_3=k-\frac{L_1}{2}-\frac{L_2}{2}.
\label{LM}
\end{align}
Note that while the variables $(M_0,M_1,M_3,Z_1,Z_3)$ are useful when we argue the properties of the partition functions associated with the $D_5$ Weyl group, for actual calculations of the exact values the variables $(L_1,L_2,L,\zeta_1,\zeta_2)$ are more useful.
Therefore in the remaining part of this paper we use both of the two conventions interchangeably depending on the context.
\begin{figure}[!t]
\centering\includegraphics[scale=0.6,angle=-90]{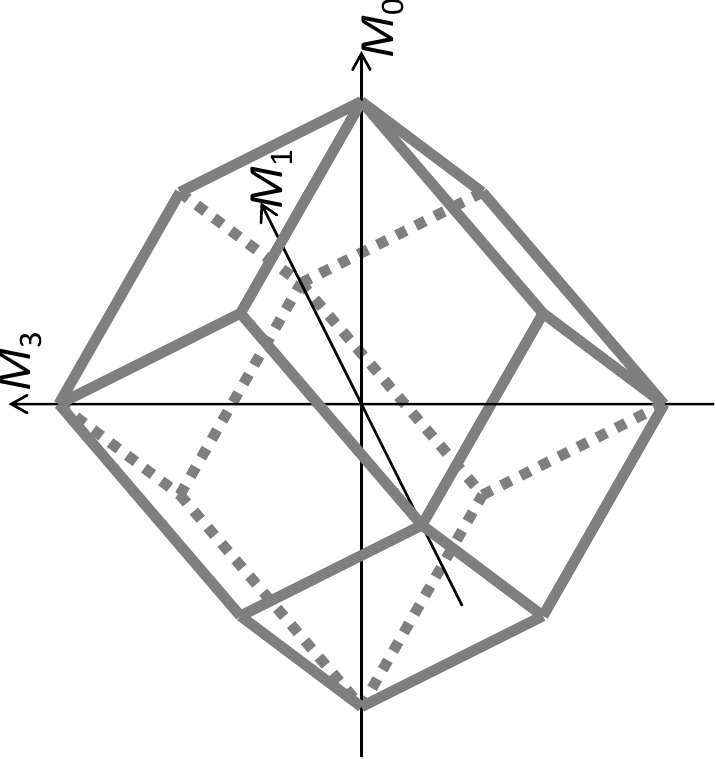}
\caption{The fundamental domain \eqref{funddomain} of supersymmetric brane configurations for the super Chern-Simons theory considered here is the rhombic dodecahedron.}
\label{fd}
\end{figure}

Note that not all values of ranks $(M_0,M_1,M_3)$ are allowed when the overall rank $N$ is vanishing.
Indeed, in \cite{FMMN,FMS} it was found that the brane configurations are subject to breakdown of supersymmetries due to negative ranks appearing in the Hanany-Witten transitions unless
\begin{align}
\pm M_0\pm M_1\le k,\quad
\pm M_0\pm M_3\le k,\quad
\pm M_1\pm M_3\le k.
\label{funddomain}
\end{align}
On the other hand, when the overall rank $N$ is large enough, we reduce to the domain \eqref{funddomain} after duality cascades.
Duality cascades for three-dimensional super Chern-Simons theories were discussed in \cite{Evslin:2009pk,HK}.
In \cite{FMMN,FMS} duality cascades were defined as the following process.
\begin{itemize}
\item
Set one of the lowest ranks as the reference.
\item
Apply the Hanany-Witten transitions arbitrarily without crossing the reference.
\item
Change references, if lower ranks compared with the reference appear.
\item
Increase the overall rank uniformly when we encounter negative ranks (since we are considering the grand partition function with sources of the overall rank \eqref{XiHinvold}).
\item
Repeat the last three steps until no more lower ranks appear.
\end{itemize}
As discussed there, duality cascades always end uniquely regardless of initial brane configurations for all the super Chern-Simons theories of circular-quiver type.
For this reason, the parameter domain of final destinations of duality cascades (where duality cascades occur no more) is called the fundamental domain of super Chern-Simons theories in duality cascades.
See figure \ref{fd} for the fundamental domain \eqref{funddomain} for our case.
Hereafter we mainly restrict ourselves in the fundamental domain when discussing the partition functions, unless we mention explicitly. 
Besides, from our parameterization of the ranks in \eqref{braneconfig}, $(M_0,M_1,M_3)$ have to be all integers or all half-integers.

\subsection{Exact results on partition function}

The partition function \eqref{Zprime1} was studied in \cite{BGKNT} by combining the viewpoints from open strings and closed strings and a simple result was found.
The result is summarized as
\begin{align}
Z'_{k,\bm{M}}(N)=e^{i(\Theta'_{k,{\bm M}}-\Theta_{k,{\bm M}})}Z_{k,\bm{M}}(N),
\label{Zprime2}
\end{align}
with
\begin{align}
&Z_{k,\bm{M}}(N)
=e^{i\Theta_{k,{\bm M}}}\frac{Z_k^\text{CS}(L_1)Z_k^\text{CS}(L_2)}{N!}
\int\prod_{n=1}^N\frac{d\mu_n}{2\pi}\nonumber\\
&\qquad\times\det\begin{pmatrix}
[\langle\mu_m|\widehat D_1\widehat D_2|\mu_n\rangle]_{(m,n)}&
[\langle\mu_m|\widehat D_1\widehat d_2|{-t_{L,s}}\rrangle]_{(m,s)}\\
[\llangle t_{L,r}|\widehat d_1\widehat D_2|\mu_n\rangle]_{(r,n)}&
[\llangle t_{L,r}|\widehat d_1\widehat d_2|{-t_{L,s}}\rrangle]_{(r,s)}
\end{pmatrix}.
\label{Z0kM}
\end{align}
Here for the overall factors, we have introduced
\begin{align}
&\Theta'_{k,{\bm M}}=\theta_k(L_1,0)+\theta_k(L_1-L,\zeta_1)-\theta_k(L_2,\zeta_2)-\theta_k(L_2-L,0),\nonumber\\
&\Theta_{k,{\bm M}}=\frac{\pi}{k}(M_0-M_1-M_3)Z_1Z_3-\pi Z_1Z_3-2\pi M_1(Z_1+Z_3),\nonumber\\
&\theta_k(n,\zeta)=\frac{\pi}{k}\Bigl(\frac{n^3-n}{12}-n\zeta^2\Bigr),\quad
Z_k^{\text{CS}}(n)=\frac{1}{k^{\frac{n}{2}}}\prod_{j<j'}^n2\sin\frac{\pi(j'-j)}{k}.
\label{Thetanew}
\end{align}
Note that two phases $\Theta'_{k,{\bm M}}$ and $\Theta_{k,{\bm M}}$ are given in different sets of variables due to their different origins:
$\Theta'_{k,{\bm M}}$ comes directly from the matrix model, while $\Theta_{k,{\bm M}}$ is introduced to simplify the coefficients of the bilinear relations obeyed by $Z_{k,{\bm M}}(0)$, as listed later in tables \ref{bilinear1} and \ref{bilinear2}.
The matrix elements of the $(N+L)\times (N+L)$ matrix in the determinant in \eqref{Z0kM} are written in the notation of one-dimensional quantum mechanics, where we introduce the canonical operators of position $\widehat x$ and momentum $\widehat p$ obeying $[\widehat x,\widehat p]=2\pi ik$ as well as their eigenstates normalized as
\begin{align}
\langle x|x'\rangle=2\pi\delta(x-x'),\quad
\langle\!\langle p|p'\rangle\!\rangle=2\pi\delta(p-p'),\quad
\langle x|p\rangle\!\rangle=\frac{1}{\sqrt{k}}e^{\frac{ixp}{2\pi k}}.
\end{align}
Then, the operators ${\widehat D}_1,{\widehat D}_2,{\widehat d}_1,{\widehat d}_2$ are written as
\begin{align}
&\widehat{D}_{1}=e^{-\frac{i\zeta_1}{k}\widehat{x}}S_{L_1}\left({\widehat x}\right)\frac{1}{2\cosh\frac{\widehat{p}-\pi iL}{2}}e^{\frac{i\zeta_1}{k}\widehat{x}}C_{L_1}\left({\widehat x}\right),\quad
\widehat{d}_{1}=e^{\frac{i\zeta_1}{k}\widehat{x}}C_{L_1}\left({\widehat x}\right),\nonumber \\
&\widehat{D}_{2}=C_{L_2}\left({\widehat x}+2\pi \zeta_2\right)\frac{1}{2\cosh\frac{\widehat{p}+\pi iL}{2}}S_{L_2}\left({\widehat x}+2\pi \zeta_2\right),\quad
\widehat{d}_{2}=C_{L_2}\left({\widehat x}+2\pi \zeta_2\right),
\label{DVIanddVI}
\end{align}
with
\begin{align}
S_{n}\left(x\right)  =i^{n}\frac{\prod_{r=1}^{n}2\sinh\frac{x-t_{n,r}}{2k}}{2\cosh\frac{x+\pi in}{2}},\quad
C_{n}\left(x\right)  =\frac{1}{\prod_{r=1}^{n}2\cosh\frac{x-t_{n,r}}{2k}},
\label{SandC}
\end{align}
satisfying $S_n(x)=C_{k-n}(x)$ for integral levels $k$.
We have also introduced $\langle\!\langle t_{n,r}|$, $|{-t_{n,r}}\rangle\!\rangle$ with $t_{n,r}$ defined as
\begin{align}
t_{n,r}=2\pi i\Bigl(\frac{n+1}{2}-r\Bigr),\quad (r=1,2,\cdots,n).
\end{align}

Taking into account the phase change in \eqref{Zprime2}, we redefine the grand partition function in \eqref{XiHinvold} accordingly and remove the primes,
\begin{align}
\Xi_{k,{\bm M}}(\kappa)=
\sum_{N=0}^\infty Z_{k,{\bm M}}(N)\kappa^N
=Z_{k,{\bm M}}(0)\Det(1+\kappa {\widehat H}_{k,{\bm M}}^{-1}).
\label{XiHinv}
\end{align}
Note that when $L=0$, ${\widehat H}_{k,{\bm M}}^{-1}$ in \eqref{XiHinv} is given by ${\widehat D}_1{\widehat D}_2$, since the determinant in \eqref{Z0kM} reduces only to the upper-left block.
We can then obtain the parameters of the curve directly by calculating the inverse ${\widehat D}_2^{-1}{\widehat D}_1^{-1}$ \cite{KMZ,KM,BGKNT}.
This gives a consistent result as the asymptotic values obtained by the extrapolation \eqref{asymptoticvalues}.
Furthermore, besides the asymptotic values, it was found \cite{BGKNT}
that the coefficient $c_{0,0}$ and the overall factor in ${\widehat H}_{k,{\bm M}}$ are also compatible with the $D_5$ Weyl group in the subspace $L=0$.
The partition function with the lowest rank vanishing $Z_{k,{\bm M}}(0)$, however, does not have to be invariant.

Empirically the partition functions serve as the tau functions in integrable models \cite{W,KMMOZ}.
In our three-dimensional theories, the tau function is expected to be the partition function in the grand canonical ensemble.
Then, the grand partition function should satisfy bilinear equations of integrable models.
Indeed as found in \cite{BGT3} this is the case for the ABJM theory.
However, since the symmetry is the $A_1$ Weyl group and there is only one direction for the weight space, it is not straightforward to generalize bilinear relations to our case.
By consulting the results in \cite{JNS}, we expect that the bilinear terms with shifts in two variables in two orthogonal directions can be expressed by those with shifts in the remaining variables.
In the next section, we search for bilinear relations from this expectation.

\section{Bilinear relations at lowest order in fugacity}\label{bilinear}

After reviewing the partition function and the $D_5$ Weyl group in the previous section, let us utilize the expression of the partition function to explore bilinear relations.
At first we only need the grand partition function of the lowest order, namely, the partition function with the lowest rank vanishing $N=0$, which is given as
\begin{align}
Z_{k,\bm{M}}(0)
=e^{i\Theta_{k,{\bm M}}}Z^\text{CS}_k(L_1)Z^\text{CS}_k(L_2)
\det\begin{pmatrix}
[\llangle t_{L,r}|\widehat d_1^\text{VI}\widehat d_2^\text{VI}|{-t_{L,s}}\rrangle]_{(r,s)}
\end{pmatrix},
\label{ZN0}
\end{align}
for $L\le L_1$ and $L\le L_2$, as can be read off from \eqref{Z0kM}.

\begin{figure}[!t]
\centering\includegraphics[scale=0.6,angle=-90]{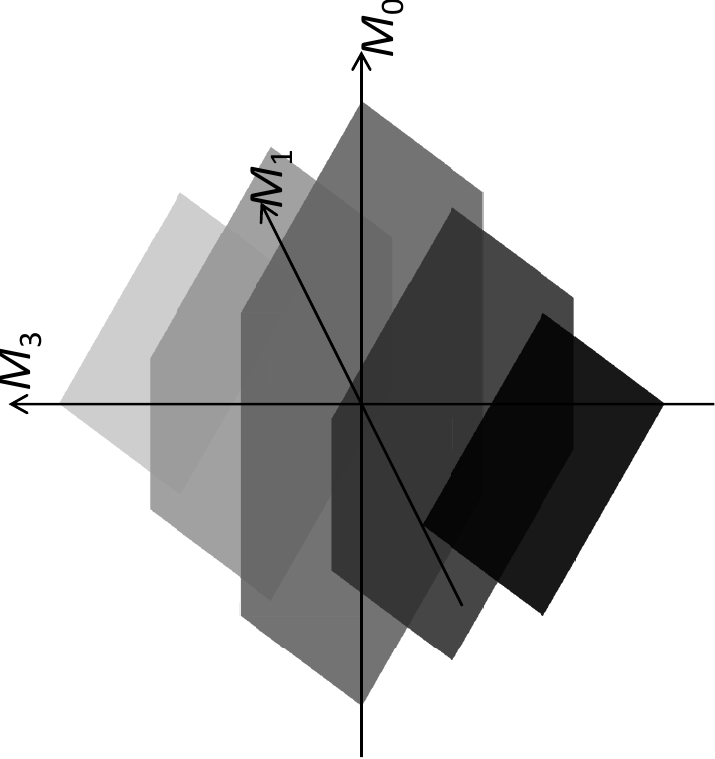}\includegraphics[scale=0.6,angle=-90]{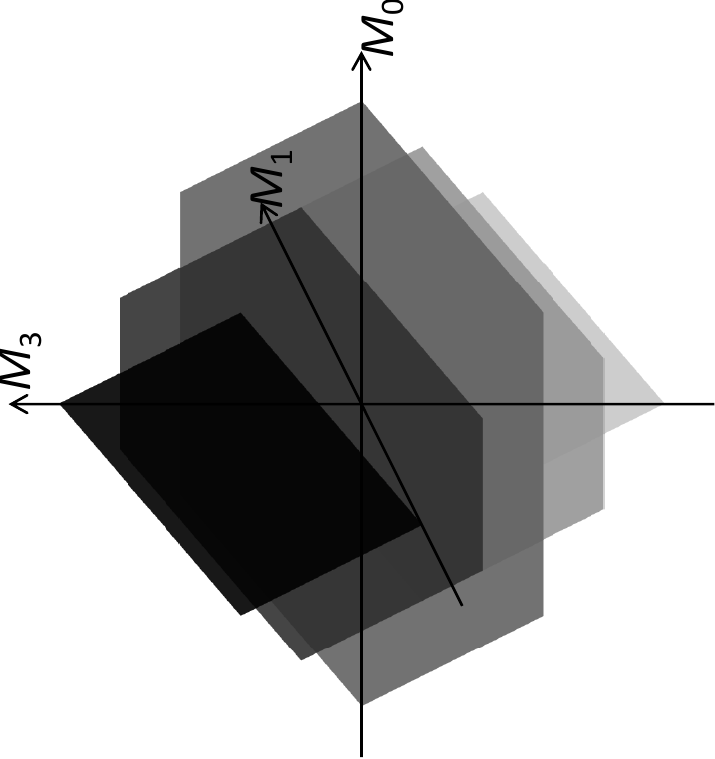}
\caption{
The matrix size of the determinants increases as we move away from the boundary.}
\label{layerfigure}
\end{figure}

Note that, among the three relative ranks, the rank $L$ plays an important role in controlling the complexity of the expression.
Namely, if we concentrate on partition functions at $N=0$ \eqref{ZN0}, $L$ serves as the matrix size appearing in the determinant expression.
For $L=0$, the linear quiver splits into two, leaving only the pure Chern-Simons partition functions $Z_k^\text{CS}(L_1)Z_k^\text{CS}(L_2)$.
Physically, in the variables $(L_1,L_2,L)$ \eqref{LM}, we expand partition functions from $(M_0,M_1,M_3)=(0,0,k)$ and the matrix size $L$ increases as we move away from the boundary $-M_1-M_3+k=0$ into internal layers.
Note that there are several identifications applicable.
For example, we can exchange the labels for NS5-branes or those for $(1,k)$5-branes.
These identifications are related by the Weyl reflections exchanging the signs of $(M_1,Z_1)$ or those of $(M_3,Z_3)$ simultaneously.
Also, we can switch the identification between NS5-branes and $(1,k)$5-branes.
This is an outer-automorphism of the Weyl group.
Depending on the identifications, we expand partition functions either from $(M_0,M_1,M_3)=(0,\pm k,0)$ or $(0,0,\pm k)$ and the matrix size $L$ increases differently.
See figure \ref{layerfigure}.
These figures already imply a deep connection to Painlev\'e bilinear equations, where the solution is given by determinants with the matrix sizes different for each layer (see for example \cite{KMNOY}).

The partition function at $N=0$ \eqref{ZN0} can be evaluated explicitly, by inserting the identity for a complete set of basis $1=\int_{-\infty}^\infty \frac{dx}{2\pi}|x\rangle\langle x|$ into each matrix element, as
\begin{align}
&Z_{k,\bm{M}}(0)
=e^{i\Theta_{k,{\bm M}}}Z_k^\text{CS}(L_1)Z_k^\text{CS}(L_2)\nonumber\\
&\quad\times\det\begin{pmatrix}
[I_{k,L_1+L_2}(L+1-r-s+i\zeta_1,\{t_{L_1,r'}\}_{r'=1}^{L_1}\cup\{-2\pi \zeta_2+t_{L_2,r'}\}_{r'=1}^{L_2})]_{(r,s)}
\end{pmatrix},
\label{Z0kM0formula}
\end{align}
where $I_{k,n}(\alpha,\{\beta_a\})$ denotes the integration\footnote{
Note that the integration $I_{k,n}(\alpha,\{\beta_a\})$ \eqref{Inalphabeta} is convergent only for $|\text{Re}[\frac{\alpha}{k}]|<\frac{n}{2k}$, which is
\begin{align}
L-\frac{L_1}{2}-\frac{L_2}{2}+|\text{Im}[\zeta_1]|<1,
\label{Z(0)conditiongeneral}
\end{align}
in terms of $(L_1,L_2,L,\zeta_1,\zeta_2)$.
}
\begin{align}
I_{k,n}(\alpha,\{\beta_a\})=\int_{-\infty}^\infty \frac{dx}{2\pi k}\frac{e^{\frac{\alpha x}{k}}}{\prod_{a=1}^n2\cosh\frac{x-\beta_a}{2k}}
=\frac{1}{e^{-\pi i\alpha}-(-1)^ne^{\pi i\alpha}}\sum_{a=1}^n\frac{e^{\frac{\alpha \beta_a}{k}}}{\prod_{a'(\neq a)}2i\sinh\frac{\beta_a-\beta_{a'}}{2k}},
\label{Inalphabeta}
\end{align}
and the corresponding parameters $\alpha$ and $\beta_a$ of the integration $I_{k,n}(\alpha,\{\beta_a\})$ in the determinant are given by
\begin{align}
&\alpha=L+1-r-s+i\zeta_1,\nonumber\\
&\beta_a=
\begin{cases}
2\pi i\bigl(\frac{L_1+1}{2}-a\bigr),&\quad (a=1,2,\cdots,L_1),\\
-2\pi iZ_3+2\pi i\bigl(\frac{L_2+1}{2}-(a-L_1)\bigr),&\quad (a=L_1+1,L_1+2,\cdots,L_1+L_2).
\end{cases}
\end{align}

Especially, from this expression we can study the partition function at $N=0$ for small integral $k$, such as $k=1$ and $k=2$.
The results\footnote{
We have to confess that some exact expressions for $k=2$, $L=4$ are difficult to obtain and are in fact determined from the 40 bilinear relations later.
Nevertheless, we can check the validity of these expressions at generic values of $k$.
}
are listed in appendix \ref{lowestpf}.
Note that sometimes the expression for the partition function is indefinite for integral levels $k$.
For these cases we need to shift the level away from integers and take the limit
\begin{align}
Z_{k,(M_0(k),M_1(k),M_3(k),Z_1,Z_3)}(0)=\lim_{k'\to k}Z_{k',(M_0(k),M_1(k),M_3(k),Z_1,Z_3)}(0),
\label{limit}
\end{align}
afterwards.
Namely, though the level $k$ also appears in changing from the variables $(L_1,L_2,L)$ to $(M_0,M_1,M_3)$ in \eqref{LM}, we keep it intact in shifting away from integers.
Also, note that, strictly speaking, we have obtained the expression \eqref{Z0kM0formula} in a range of parameters $L\le L_1$ and $L\le L_2$ and the integration converges only for \eqref{Z(0)conditiongeneral}.
Nevertheless, we apply the expression to find the results in appendix \ref{lowestpf} sometimes outside the range of validity.
This is because the condition $L\le L_1$ and $L\le L_2$ is only imposed for simplicity and we can often compute in different ranges to find the identical results.
The convergence condition may also be overcome by the principal-value prescription or other regularizations.
As a result, we find that the expression vanishes outside of the fundamental domain.
This is a non-trivial sign that we are on the correct path of evaluations despite the assumption of validity we have made.

It is also surprising that partition functions at different points ${\bm M}$ related by the $D_5$ Weyl group (obtained by applying \eqref{Z0kM0formula}) take different values.
Namely, although the grand partition function normalized by $Z_{k,{\bm M}}(0)$ is rewritten into the Fredholm determinant with the spectral operator conjectured to be invariant under the Weyl group, the symmetry does not apply to $Z_{k,{\bm M}}(0)$.
We will come back to this point in section \ref{sec_dualitycascade}.

Interestingly, note that the results in appendices \ref{lowestpf} and \ref{higherrank1} are given cleanly by trigonometric functions only when these results are written in terms of $Z_1$ and $Z_3$.
For example, $Z_{k,{\bm M}}(0)$ with $k=1$ and $M_0=M_1=M_3=0$ (which is denoted as $Z^{(0)}_{0,0,0}$ in \eqref{Z0000}) would contain both trigonometric functions and hyperbolic functions if we express them with $\zeta_1$ and $\zeta_2$.

\begin{table}[t!]
\begin{tabular}{c|r}
$(M_1,M_3)$
&$e^{-\frac{\pi i}{2k}(M_0+Z_1+Z_3)}\widetilde Z_{\varnothing\pm\pm\varnothing\varnothing}
+e^{\frac{\pi i}{2k}(M_0+Z_1+Z_3)}\widetilde Z_{\varnothing\pm\mp\varnothing\varnothing}
+\widetilde Z_{\pm\varnothing\varnothing\pm\pm}=0$\\
&$e^{-\frac{\pi i}{2k}(-M_0-Z_1+Z_3)}\widetilde Z_{\varnothing\pm\pm\varnothing\varnothing}
+e^{\frac{\pi i}{2k}(-M_0-Z_1+Z_3)}\widetilde Z_{\varnothing\pm\mp\varnothing\varnothing}
+\widetilde Z_{\mp\varnothing\varnothing\mp\pm}=0$\\
&$e^{-\frac{\pi i}{2k}(-M_0+Z_1-Z_3)}\widetilde Z_{\varnothing\pm\pm\varnothing\varnothing}
+e^{\frac{\pi i}{2k}(-M_0+Z_1-Z_3)}\widetilde Z_{\varnothing\pm\mp\varnothing\varnothing}
+\widetilde Z_{\mp\varnothing\varnothing\pm\mp}=0$\\
&$e^{-\frac{\pi i}{2k}(M_0-Z_1-Z_3)}\widetilde Z_{\varnothing\pm\pm\varnothing\varnothing}
+e^{\frac{\pi i}{2k}(M_0-Z_1-Z_3)}\widetilde Z_{\varnothing\pm\mp\varnothing\varnothing}
+\widetilde Z_{\pm\varnothing\varnothing\mp\mp}=0$\\\hline
$(M_1,Z_1)$
&$e^{-\frac{\pi i}{2k}(M_0+M_3+Z_3)}S_1^+\widetilde Z_{\varnothing\pm\varnothing\pm\varnothing}
+e^{\frac{\pi i}{2k}(M_0+M_3+Z_3)}S_1^-\widetilde Z_{\varnothing\pm\varnothing\mp\varnothing}
+S_3^+\widetilde Z_{\pm\varnothing\pm\varnothing\pm}=0$\\
&$e^{-\frac{\pi i}{2k}(-M_0-M_3+Z_3)}S_1^+\widetilde Z_{\varnothing\pm\varnothing\pm\varnothing}
+e^{\frac{\pi i}{2k}(-M_0-M_3+Z_3)}S_1^-\widetilde Z_{\varnothing\pm\varnothing\mp\varnothing}
+S_3^-\widetilde Z_{\mp\varnothing\mp\varnothing\pm}=0$\\
&$e^{-\frac{\pi i}{2k}(-M_0+M_3-Z_3)}S_1^+\widetilde Z_{\varnothing\pm\varnothing\pm\varnothing}
+e^{\frac{\pi i}{2k}(-M_0+M_3-Z_3)}S_1^-\widetilde Z_{\varnothing\pm\varnothing\mp\varnothing}
+S_3^-\widetilde Z_{\mp\varnothing\pm\varnothing\mp}=0$\\
&$e^{-\frac{\pi i}{2k}(M_0-M_3-Z_3)}S_1^+\widetilde Z_{\varnothing\pm\varnothing\pm\varnothing}
+e^{\frac{\pi i}{2k}(M_0-M_3-Z_3)}S_1^-\widetilde Z_{\varnothing\pm\varnothing\mp\varnothing}
+S_3^+\widetilde Z_{\pm\varnothing\mp\varnothing\mp}=0$\\\hline
$(M_1,Z_3)$
&$e^{-\frac{\pi i}{2k}(M_0+M_3+Z_1)}\widetilde Z_{\varnothing\pm\varnothing\varnothing\pm}
+e^{\frac{\pi i}{2k}(M_0+M_3+Z_1)}\widetilde Z_{\varnothing\pm\varnothing\varnothing\mp}
+\widetilde Z_{\pm\varnothing\pm\pm\varnothing}=0$\\
&$e^{-\frac{\pi i}{2k}(-M_0-M_3+Z_1)}\widetilde Z_{\varnothing\pm\varnothing\varnothing\pm}
+e^{\frac{\pi i}{2k}(-M_0-M_3+Z_1)}\widetilde Z_{\varnothing\pm\varnothing\varnothing\mp}
+\widetilde Z_{\mp\varnothing\mp\pm\varnothing}=0$\\
&$e^{-\frac{\pi i}{2k}(-M_0+M_3-Z_1)}\widetilde Z_{\varnothing\pm\varnothing\varnothing\pm}
+e^{\frac{\pi i}{2k}(-M_0+M_3-Z_1)}\widetilde Z_{\varnothing\pm\varnothing\varnothing\mp}
+\widetilde Z_{\mp\varnothing\pm\mp\varnothing}=0$\\
&$e^{-\frac{\pi i}{2k}(M_0-M_3-Z_1)}\widetilde Z_{\varnothing\pm\varnothing\varnothing\pm}
+e^{\frac{\pi i}{2k}(M_0-M_3-Z_1)}\widetilde Z_{\varnothing\pm\varnothing\varnothing\mp}
+\widetilde Z_{\pm\varnothing\mp\mp\varnothing}=0$\\\hline
$(M_3,Z_1)$
&$e^{-\frac{\pi i}{2k}(M_0+M_1+Z_3)}\widetilde Z_{\varnothing\varnothing\pm\pm\varnothing}
+e^{\frac{\pi i}{2k}(M_0+M_1+Z_3)}\widetilde Z_{\varnothing\varnothing\pm\mp\varnothing}
+\widetilde Z_{\pm\pm\varnothing\varnothing\pm}=0$\\
&$e^{-\frac{\pi i}{2k}(-M_0-M_1+Z_3)}\widetilde Z_{\varnothing\varnothing\pm\pm\varnothing}
+e^{\frac{\pi i}{2k}(-M_0-M_1+Z_3)}\widetilde Z_{\varnothing\varnothing\pm\mp\varnothing}
+\widetilde Z_{\mp\mp\varnothing\varnothing\pm}=0$\\
&$e^{-\frac{\pi i}{2k}(-M_0+M_1-Z_3)}\widetilde Z_{\varnothing\varnothing\pm\pm\varnothing}
+e^{\frac{\pi i}{2k}(-M_0+M_1-Z_3)}\widetilde Z_{\varnothing\varnothing\pm\mp\varnothing}
+\widetilde Z_{\mp\pm\varnothing\varnothing\mp}=0$\\
&$e^{-\frac{\pi i}{2k}(M_0-M_1-Z_3)}\widetilde Z_{\varnothing\varnothing\pm\pm\varnothing}
+e^{\frac{\pi i}{2k}(M_0-M_1-Z_3)}\widetilde Z_{\varnothing\varnothing\pm\mp\varnothing}
+\widetilde Z_{\pm\mp\varnothing\varnothing\mp}=0$\\\hline
$(M_3,Z_3)$
&$e^{-\frac{\pi i}{2k}(M_0+M_1+Z_1)}S_3^+\widetilde Z_{\varnothing\varnothing\pm\varnothing\pm}
+e^{\frac{\pi i}{2k}(M_0+M_1+Z_1)}S_3^-\widetilde Z_{\varnothing\varnothing\pm\varnothing\mp}
+S_1^+\widetilde Z_{\pm\pm\varnothing\pm\varnothing}=0$\\
&$e^{-\frac{\pi i}{2k}(-M_0-M_1+Z_1)}S_3^+\widetilde Z_{\varnothing\varnothing\pm\varnothing\pm}
+e^{\frac{\pi i}{2k}(-M_0-M_1+Z_1)}S_3^-\widetilde Z_{\varnothing\varnothing\pm\varnothing\mp}
+S_1^-\widetilde Z_{\mp\mp\varnothing\pm\varnothing}=0$\\
&$e^{-\frac{\pi i}{2k}(-M_0+M_1-Z_1)}S_3^+\widetilde Z_{\varnothing\varnothing\pm\varnothing\pm}
+e^{\frac{\pi i}{2k}(-M_0+M_1-Z_1)}S_3^-\widetilde Z_{\varnothing\varnothing\pm\varnothing\mp}
+S_1^-\widetilde Z_{\mp\pm\varnothing\mp\varnothing}=0$\\
&$e^{-\frac{\pi i}{2k}(M_0-M_1-Z_1)}S_3^+\widetilde Z_{\varnothing\varnothing\pm\varnothing\pm}
+e^{\frac{\pi i}{2k}(M_0-M_1-Z_1)}S_3^-\widetilde Z_{\varnothing\varnothing\pm\varnothing\mp}
+S_1^+\widetilde Z_{\pm\mp\varnothing\mp\varnothing}=0$\\\hline
$(Z_1,Z_3)$
&$e^{-\frac{\pi i}{2k}(M_0+M_1+M_3)}\widetilde Z_{\varnothing\varnothing\varnothing\pm\pm}
+e^{\frac{\pi i}{2k}(M_0+M_1+M_3)}\widetilde Z_{\varnothing\varnothing\varnothing\pm\mp}
+\widetilde Z_{\pm\pm\pm\varnothing\varnothing}=0$\\
&$e^{-\frac{\pi i}{2k}(-M_0-M_1+M_3)}\widetilde Z_{\varnothing\varnothing\varnothing\pm\pm}
+e^{\frac{\pi i}{2k}(-M_0-M_1+M_3)}\widetilde Z_{\varnothing\varnothing\varnothing\pm\mp}
+\widetilde Z_{\mp\mp\pm\varnothing\varnothing}=0$\\
&$e^{-\frac{\pi i}{2k}(-M_0+M_1-M_3)}\widetilde Z_{\varnothing\varnothing\varnothing\pm\pm}
+e^{\frac{\pi i}{2k}(-M_0+M_1-M_3)}\widetilde Z_{\varnothing\varnothing\varnothing\pm\mp}
+\widetilde Z_{\mp\pm\mp\varnothing\varnothing}=0$\\
&$e^{-\frac{\pi i}{2k}(M_0-M_1-M_3)}\widetilde Z_{\varnothing\varnothing\varnothing\pm\pm}
+e^{\frac{\pi i}{2k}(M_0-M_1-M_3)}\widetilde Z_{\varnothing\varnothing\varnothing\pm\mp}
+\widetilde Z_{\pm\mp\mp\varnothing\varnothing}=0$
\end{tabular}
\caption{40 bilinear relations for $Z_{k,{\bm M}}(0)$ in the three-dimensional ${\cal N}=4$ super Chern-Simons theory of a circular quiver with four nodes.
}
\label{bilinear1}
\end{table}

\begin{table}[t!]
\begin{tabular}{c|r}
$(M_0,M_1)$
&$e^{-\frac{\pi i}{2k}(M_3+Z_1+Z_3)}\widetilde Z_{\pm\pm\varnothing\varnothing\varnothing}
+e^{\frac{\pi i}{2k}(M_3+Z_1+Z_3)}\widetilde Z_{\pm\mp\varnothing\varnothing\varnothing}
+S_3^+\widetilde Z_{\varnothing\varnothing\pm\pm\pm}=0$\\
&$e^{-\frac{\pi i}{2k}(-M_3-Z_1+Z_3)}\widetilde Z_{\pm\pm\varnothing\varnothing\varnothing}
+e^{\frac{\pi i}{2k}(-M_3-Z_1+Z_3)}\widetilde Z_{\pm\mp\varnothing\varnothing\varnothing}
+S_3^-\widetilde Z_{\varnothing\varnothing\mp\mp\pm}=0$\\
&$e^{-\frac{\pi i}{2k}(-M_3+Z_1-Z_3)}\widetilde Z_{\pm\pm\varnothing\varnothing\varnothing}
+e^{\frac{\pi i}{2k}(-M_3+Z_1-Z_3)}\widetilde Z_{\pm\mp\varnothing\varnothing\varnothing}
+S_3^+\widetilde Z_{\varnothing\varnothing\mp\pm\mp}=0$\\
&$e^{-\frac{\pi i}{2k}(M_3-Z_1-Z_3)}\widetilde Z_{\pm\pm\varnothing\varnothing\varnothing}
+e^{\frac{\pi i}{2k}(M_3-Z_1-Z_3)}\widetilde Z_{\pm\mp\varnothing\varnothing\varnothing}
+S_3^-\widetilde Z_{\varnothing\varnothing\pm\mp\mp}=0$\\\hline
$(M_0,M_3)$
&$e^{-\frac{\pi i}{2k}(M_1+Z_1+Z_3)}\widetilde Z_{\pm\varnothing\pm\varnothing\varnothing}
+e^{\frac{\pi i}{2k}(M_1+Z_1+Z_3)}\widetilde Z_{\pm\varnothing\mp\varnothing\varnothing}
+S_1^+\widetilde Z_{\varnothing\pm\varnothing\pm\pm}=0$\\
&$e^{-\frac{\pi i}{2k}(-M_1-Z_1+Z_3)}\widetilde Z_{\pm\varnothing\pm\varnothing\varnothing}
+e^{\frac{\pi i}{2k}(-M_1-Z_1+Z_3)}\widetilde Z_{\pm\varnothing\mp\varnothing\varnothing}
+S_1^+\widetilde Z_{\varnothing\mp\varnothing\mp\pm}=0$\\
&$e^{-\frac{\pi i}{2k}(-M_1+Z_1-Z_3)}\widetilde Z_{\pm\varnothing\pm\varnothing\varnothing}
+e^{\frac{\pi i}{2k}(-M_1+Z_1-Z_3)}\widetilde Z_{\pm\varnothing\mp\varnothing\varnothing}
+S_1^-\widetilde Z_{\varnothing\mp\varnothing\pm\mp}=0$\\
&$e^{-\frac{\pi i}{2k}(M_1-Z_1-Z_3)}\widetilde Z_{\pm\varnothing\pm\varnothing\varnothing}
+e^{\frac{\pi i}{2k}(M_1-Z_1-Z_3)}\widetilde Z_{\pm\varnothing\mp\varnothing\varnothing}
+S_1^-\widetilde Z_{\varnothing\pm\varnothing\mp\mp}=0$\\\hline
$(M_0,Z_1)$
&$e^{-\frac{\pi i}{2k}(M_1+M_3+Z_3)}\widetilde Z_{\pm\varnothing\varnothing\pm\varnothing}
+e^{\frac{\pi i}{2k}(M_1+M_3+Z_3)}\widetilde Z_{\pm\varnothing\varnothing\mp\varnothing}
+S_3^+\widetilde Z_{\varnothing\pm\pm\varnothing\pm}=0$\\
&$e^{-\frac{\pi i}{2k}(-M_1-M_3+Z_3)}\widetilde Z_{\pm\varnothing\varnothing\pm\varnothing}
+e^{\frac{\pi i}{2k}(-M_1-M_3+Z_3)}\widetilde Z_{\pm\varnothing\varnothing\mp\varnothing}
+S_3^-\widetilde Z_{\varnothing\mp\mp\varnothing\pm}=0$\\
&$e^{-\frac{\pi i}{2k}(-M_1+M_3-Z_3)}\widetilde Z_{\pm\varnothing\varnothing\pm\varnothing}
+e^{\frac{\pi i}{2k}(-M_1+M_3-Z_3)}\widetilde Z_{\pm\varnothing\varnothing\mp\varnothing}
+S_3^-\widetilde Z_{\varnothing\mp\pm\varnothing\mp}=0$\\
&$e^{-\frac{\pi i}{2k}(M_1-M_3-Z_3)}\widetilde Z_{\pm\varnothing\varnothing\pm\varnothing}
+e^{\frac{\pi i}{2k}(M_1-M_3-Z_3)}\widetilde Z_{\pm\varnothing\varnothing\mp\varnothing}
+S_3^+\widetilde Z_{\varnothing\pm\mp\varnothing\mp}=0$\\\hline
$(M_0,Z_3)$
&$e^{-\frac{\pi i}{2k}(M_1+M_3+Z_1)}\widetilde Z_{\pm\varnothing\varnothing\varnothing\pm}
+e^{\frac{\pi i}{2k}(M_1+M_3+Z_1)}\widetilde Z_{\pm\varnothing\varnothing\varnothing\mp}
+S_1^+\widetilde Z_{\varnothing\pm\pm\pm\varnothing}=0$\\
&$e^{-\frac{\pi i}{2k}(-M_1-M_3+Z_1)}\widetilde Z_{\pm\varnothing\varnothing\varnothing\pm}
+e^{\frac{\pi i}{2k}(-M_1-M_3+Z_1)}\widetilde Z_{\pm\varnothing\varnothing\varnothing\mp}
+S_1^-\widetilde Z_{\varnothing\mp\mp\pm\varnothing}=0$\\
&$e^{-\frac{\pi i}{2k}(-M_1+M_3-Z_1)}\widetilde Z_{\pm\varnothing\varnothing\varnothing\pm}
+e^{\frac{\pi i}{2k}(-M_1+M_3-Z_1)}\widetilde Z_{\pm\varnothing\varnothing\varnothing\mp}
+S_1^+\widetilde Z_{\varnothing\mp\pm\mp\varnothing}=0$\\
&$e^{-\frac{\pi i}{2k}(M_1-M_3-Z_1)}\widetilde Z_{\pm\varnothing\varnothing\varnothing\pm}
+e^{\frac{\pi i}{2k}(M_1-M_3-Z_1)}\widetilde Z_{\pm\varnothing\varnothing\varnothing\mp}
+S_1^-\widetilde Z_{\varnothing\pm\mp\mp\varnothing}=0$
\end{tabular}
\caption{40 bilinear relations for $Z_{k,{\bm M}}(0)$ in the three-dimensional ${\cal N}=4$ super Chern-Simons theory of a circular quiver with four nodes (continued).}
\label{bilinear2}
\end{table}

Using the expressions for $k=1$ and $k=2$, we can search for bilinear relations as explained in section \ref{intro}.
We find that the partition function satisfies 40 bilinear relations listed in tables \ref{bilinear1} and \ref{bilinear2}.
Typically, bilinear relations relate three bilinear terms with the arguments shifted to the closest lattice points in the opposite directions.
In the current situation, the shifts in the three terms are arranged as follows.
We use the variables ${\bm M}=(M_0,M_1,M_3,Z_1,Z_3)$ and consider the partition function $Z_{M_0,M_1,M_3,Z_1,Z_3}=Z_{k,{\bm M}}(N=0)$, where we abbreviate the level $k$ and the overall rank $N$ since the level $k$ is common and the overall rank $N$ always refers to the lowest one $N=0$ in the bilinear relations in this section.
Then, the sum of partition functions obtained by shifting two arguments in the two orthogonal directions in the first two terms is expressed by the third term with the rest arguments shifted.
When an argument of $Z_{M_0,M_1,M_3,Z_1,Z_3}$ in the bilinear relations is shifted by $\pm 1/2$ we denote it by $\pm$, otherwise by $\varnothing$.
Namely, for example, if two variables $(M_1,M_3)$ in the bilinear relations are shifted we denote them by
\begin{align}
\widetilde Z_{\varnothing\pm\pm\varnothing\varnothing}
=Z_{M_0,M_1+\frac{1}{2},M_3+\frac{1}{2},Z_1,Z_3}
Z_{M_0,M_1-\frac{1}{2},M_3-\frac{1}{2},Z_1,Z_3},\nonumber\\
\widetilde Z_{\varnothing\pm\mp\varnothing\varnothing}
=Z_{M_0,M_1+\frac{1}{2},M_3-\frac{1}{2},Z_1,Z_3}
Z_{M_0,M_1-\frac{1}{2},M_3+\frac{1}{2},Z_1,Z_3},
\label{ZZtilde}
\end{align}
while if the rest $(M_0,Z_1,Z_3)$ are shifted we denote by
\begin{align}
\widetilde Z_{\pm\varnothing\varnothing\pm\pm}
=Z_{M_0+\frac{1}{2},M_1,M_3,Z_1+\frac{1}{2},Z_3+\frac{1}{2}}
Z_{M_0-\frac{1}{2},M_1,M_3,Z_1-\frac{1}{2},Z_3-\frac{1}{2}},\nonumber\\
\widetilde Z_{\mp\varnothing\varnothing\mp\pm}
=Z_{M_0-\frac{1}{2},M_1,M_3,Z_1-\frac{1}{2},Z_3+\frac{1}{2}}
Z_{M_0+\frac{1}{2},M_1,M_3,Z_1+\frac{1}{2},Z_3-\frac{1}{2}},\nonumber\\
\widetilde Z_{\mp\varnothing\varnothing\pm\mp}
=Z_{M_0-\frac{1}{2},M_1,M_3,Z_1+\frac{1}{2},Z_3-\frac{1}{2}}
Z_{M_0+\frac{1}{2},M_1,M_3,Z_1-\frac{1}{2},Z_3+\frac{1}{2}},\nonumber\\
\widetilde Z_{\pm\varnothing\varnothing\mp\mp}
=Z_{M_0+\frac{1}{2},M_1,M_3,Z_1-\frac{1}{2},Z_3-\frac{1}{2}}
Z_{M_0-\frac{1}{2},M_1,M_3,Z_1+\frac{1}{2},Z_3+\frac{1}{2}}.
\end{align}
The structure of the coefficients for these bilinear relations are quite clear.
As in the case of $(M_1,M_3)$, $(M_1,Z_3)$, $(M_3,Z_3)$ and $(Z_1,Z_3)$, when we shift two variables in the first two terms and express the results by shifting the rest variables, the first two terms need to be multiplied by an exponential phase factor depending on the shifts of the remaining variables.
Note that the two exponential factors in the first two terms are chosen so that the parity of the equation is conserved.
Namely, if we denote the two bilinear partition functions $\widetilde Z_{\varnothing\pm\pm\varnothing\varnothing}$ and $\widetilde Z_{\varnothing\pm\mp\varnothing\varnothing}$ in \eqref{ZZtilde} as even and odd respectively, even and odd terms are multiplied by the phase factor with odd and even numbers of negative phases such as $e^{-\frac{\pi i}{2k}M_0}$ respectively.
Though this also applies to the remaining cases, we need to introduce
\begin{align}
S_1^\pm=2\sin\frac{\pi(M_1\pm Z_1)}{k},\quad
S_3^\pm=2\sin\frac{\pi(M_3\pm Z_3)}{k},
\label{S13}
\end{align}
in addition when both $M_1$ and $Z_1$ or both $M_3$ and $Z_3$ are shifted simultaneously.
The rule for introducing these factors can be traced back to the point configurations of asymptotic values for the spectral operator in figure \ref{config}.
In the figure, the two variables $(M_1,Z_1)$ (and $(M_3,Z_3)$) are responsible for the vertical (and horizontal) asymptotic values, while $M_0$ connects these two sets of variables.

From the structure observed above from tables \ref{bilinear1} and \ref{bilinear2}, we can summarize the bilinear relations by changing notations slightly as
\begin{align}
&e^{-\frac{\pi i}{2k}(\sigma_1 c+\sigma_2 d+\sigma_3 e)}S^{(1)}_{{\bm M}}\widetilde Z_{a_\pm b_\pm c_\varnothing d_\varnothing e_\varnothing}
+e^{\frac{\pi i}{2k}(\sigma_1 c+\sigma_2 d+\sigma_3 e)}S^{(2)}_{{\bm M}}\widetilde Z_{a_\pm b_\mp c_\varnothing d_\varnothing e_\varnothing}\nonumber\\
&\qquad+S^{(3)}_{{\bm M}}\widetilde Z_{a_\varnothing b_\varnothing c_{\pm\sigma_1} d_{\pm\sigma_2} e_{\pm\sigma_3}}=0,
\label{biliearZtilde}
\end{align}
Here $(a,b,c,d,e)$ is\footnote{
This $e$ should not be confused with the Euler's constant which is also denoted as $e$ in this paper.
}
a permutation of $(M_0,M_1,M_3,Z_1,Z_3)$ and $(\sigma_1,\sigma_2,\sigma_3)$ is either of $(+,+,+)$, $(+,-,-)$, $(-,+,-)$, $(-,-,+)$.
The subscripts $(a,b,c,d,e)$ with signs ($\pm$ or $\varnothing$) of $\widetilde Z$ denote the directions of the shifts of the parameters ${\bm M}=(M_0,M_1,M_3,Z_1,Z_3)$
\begin{align}
&\widetilde Z_{a_\pm b_\pm c_\varnothing d_\varnothing e_\varnothing}
=\prod_\pm Z_{k,{\bm M}_\alpha\pm\frac{1}{2}(\delta^a_\alpha+\delta^b_\alpha)},\quad
\widetilde Z_{a_\pm b_\mp c_\varnothing d_\varnothing e_\varnothing}
=\prod_\pm Z_{k,{\bm M}_\alpha\pm\frac{1}{2}(\delta^a_\alpha-\delta^b_\alpha)},\nonumber\\
&\widetilde Z_{a_\varnothing b_\varnothing c_{\pm\sigma_1} d_{\pm\sigma_2} e_{\pm\sigma_3}}
=\prod_\pm Z_{k,{\bm M}_\alpha\pm\frac{1}{2}(\sigma_1\delta^c_\alpha+\sigma_2\delta^d_\alpha+\sigma_3\delta^e_\alpha)}.
\label{ZZZ}
\end{align}
Thus, the 40 bilinear relations are labelled by the first two components of the permutation and the three signs $(a,b;\sigma_1,\sigma_3,\sigma_3)$.
Aside from the phase factors, we also need to introduce the coefficients $S_{{\bm M}}$ as in \eqref{S13} depending on the shifts in subscripts of ${\widetilde Z}$.
Note that by labelling $\sigma_1,\sigma_2,\sigma_3$ with the name of the associated component in $(M_0,M_1,M_3,Z_1,Z_3)$ such as $\sigma_{M_0}$, we can write the coefficients in the 40 bilinear relations as
\begin{align}
(a,b)=(M_1,Z_1)&:\quad S^{(1)}_{\bm M}=S_1^+,\quad S^{(2)}_{\bm M}=S_1^-,\quad
S^{(3)}_{\bm M}=S_3^{\sigma_{M_0}},\nonumber\\
(a,b)=(M_3,Z_3)&:\quad S^{(1)}_{\bm M}=S_3^+,\quad S^{(2)}_{\bm M}=S_3^-,\quad
S^{(3)}_{\bm M}=S_1^{\sigma_{M_0}},\nonumber\\
(a,b)=(M_0,M_1)&:\quad S^{(1)}_{\bm M}=S^{(2)}_{\bm M}=1,\quad S^{(3)}_{\bm M}=S_3^{\sigma_{Z_1}},\nonumber\\
(a,b)=(M_0,M_3)&:\quad S^{(1)}_{\bm M}=S^{(2)}_{\bm M}=1,\quad S^{(3)}_{\bm M}=S_1^{\sigma_{Z_3}},\nonumber\\
(a,b)=(M_0,Z_1)&:\quad S^{(1)}_{\bm M}=S^{(2)}_{\bm M}=1,\quad S^{(3)}_{\bm M}=S_3^{\sigma_{M_1}},\nonumber\\
(a,b)=(M_0,Z_3)&:\quad S^{(1)}_{\bm M}=S^{(2)}_{\bm M}=1,\quad S^{(3)}_{\bm M}=S_1^{\sigma_{M_3}},
\label{S(a)compact}
\end{align}
and $S^{(1)}_{\bm M}=S^{(2)}_{\bm M}=S^{(3)}_{\bm M}=1$ for the other $(a,b)$.

Note that the bilinear relations found originally from $Z'_{k,{\bm M}}(0)$ \eqref{Zprime2} have exponential phase factors which seem irregular.
To simplify the exponential phase factors as those listed in tables \ref{bilinear1} and \ref{bilinear2}, we need to multiply another phase factor $e^{i\Theta_{k,{\bm M}}}$ \eqref{Thetanew} different from $e^{i\Theta'_{k,{\bm M}}}$ and use $Z_{k,{\bm M}}(0)$ instead of $Z_{k,{\bm M}}'(0)$.

After the bilinear relations are identified we can check them for various values of levels $k$ and ranks $(M_0,M_1,M_3)$ numerically.
We have performed this check even for generic non-rational values of $k$, which provide highly non-trivial checks for the precise expressions of the coefficients.
In the next section, we shall show that the bilinear relations \eqref{ZZZ} extend naturally to the grand partition functions, where the above partition functions serve as the lowest ranks.

\section{Bilinear relations for higher orders in fugacity}
\label{higher}

In this section we study how the bilinear relations \eqref{biliearZtilde} for the partition functions with the lowest rank vanishing, $N=0$, extend to higher ranks, as the relations for the grand partition functions $\Xi_{k,{\bm M}}(\kappa)$ \eqref{XiHinv}.

First, let us guess the uplifted bilinear relations.
We assume that the bilinear relations enjoy the same structure as those for the lowest order \eqref{biliearZtilde}
by replacing the bilinears of partition functions $\widetilde Z$ \eqref{ZZZ} with the bilinears of full grand partition functions $\widetilde\Xi(\gamma_{{\bm M}},\kappa)$ as
\begin{align}
&e^{-\frac{\pi i}{2k}(\sigma_1 c+\sigma_2 d+\sigma_3 e)}S^{(1)}_{{\bm M}}
\widetilde\Xi_{a_\pm b_\pm c_\varnothing d_\varnothing e_\varnothing}(\gamma_{{\bm M}}^{(1)},\kappa)
+e^{\frac{\pi i}{2k}(\sigma_1 c+\sigma_2 d+\sigma_3 e)}S^{(2)}_{{\bm M}}
\widetilde\Xi_{a_\pm b_\mp c_\varnothing d_\varnothing e_\varnothing}(\gamma_{{\bm M}}^{(2)},\kappa)\nonumber\\
&\qquad+S^{(3)}_{{\bm M}}
\widetilde\Xi_{a_\varnothing b_\varnothing c_{\pm\sigma_1} d_{\pm \sigma_2} e_{\pm\sigma_3}}(\gamma_{{\bm M}}^{(3)},\kappa)=0,
\label{bilinearwithkappa}
\end{align}
with
\begin{align}
&\widetilde\Xi_{a_\pm b_\pm c_\varnothing d_\varnothing e_\varnothing}(\gamma_{{\bm M}}^{(1)},\kappa)
=\prod_\pm\Xi_{k,{\bm M}_\alpha\pm\frac{1}{2}(\delta^a_\alpha+\delta^b_\alpha)}
((\gamma^{(1)}_{{\bm M}})^{\pm 1}\kappa),\nonumber\\
&\widetilde\Xi_{a_\pm b_\mp c_\varnothing d_\varnothing e_\varnothing}(\gamma_{{\bm M}}^{(2)},\kappa)
=\prod_\pm\Xi_{k,{\bm M}_\alpha\pm\frac{1}{2}(\delta^a_\alpha-\delta^b_\alpha)}
((\gamma^{(2)}_{{\bm M}})^{\pm 1}\kappa),\nonumber\\
&\widetilde\Xi_{a_\varnothing b_\varnothing c_{\pm \sigma_1} d_{\pm \sigma_2} e_{\pm \sigma_3}}(\gamma_{{\bm M}}^{(3)},\kappa)
=\prod_\pm\Xi_{k,{\bm M}_\alpha\pm\frac{1}{2}(\sigma_1\delta^c_\alpha+\sigma_2\delta^d_\alpha+\sigma_3\delta^e_\alpha)}((\gamma^{(3)}_{{\bm M}})^{\pm 1}\kappa).
\label{Xiwithgamma}
\end{align}
Here the coefficients $(\gamma^{(1)}_{{\bm M}},\gamma^{(2)}_{{\bm M}},\gamma^{(3)}_{{\bm M}})$ do not affect the observations in the previous section at $\kappa=0$, and are to be determined by requiring the bilinear relations at higher orders in $\kappa$.
In principle, $(\gamma^{(1)}_{{\bm M}},\gamma^{(2)}_{{\bm M}},\gamma^{(3)}_{{\bm M}})$ are not necessarily the same for each of 40 equations.
However, considering the (conjectured) invariance of the spectral operator under the $D_5$ Weyl group, they should not be completely irregular.

Let us first determine $(\gamma_{\bm M}^{(1)},\gamma_{\bm M}^{(2)},\gamma_{\bm M}^{(3)})$ from the partition functions $Z_{k,{\bm M}}(N)$ for $k=1$ and $N=1$ and then see that the choice applies to other cases.
As explained in appendix \ref{computation}, we can calculate the partition functions $Z_{k,{\bm M}}(N)$, the expansion coefficients of $\Xi_{k,{\bm M}}(\kappa)$ in $\kappa$, explicitly at each order and for each values of $k,L_1,L_2,L\in\mathbb{Z}$ with two methods.
By using method II in section \ref{sec_Okuyama} we can compute the partition functions explicitly.
Although the calculation is cumbersome in general even for $k=1$, from the invariance of the spectral operator ${\widehat H}_{k,{\bm M}}$ under the $D_5$ Weyl group, it is natural to expect that all values of the partition function for $N=1$ (or more generally for any $N$) in the fundamental domain \eqref{funddomain} reduce to four types
\begin{align}
&{\cal X}_0
=\frac{Z^{(1)}_{-1,0,0}}{Z^{(0)}_{-1,0,0}}
=\frac{Z^{(1)}_{0,-1,0}}{Z^{(0)}_{0,-1,0}}
=\frac{Z^{(1)}_{0,0,-1}}{Z^{(0)}_{0,0,-1}}
=\frac{Z^{(1)}_{0,0,1}}{Z^{(0)}_{0,0,1}}
=\frac{Z^{(1)}_{0,1,0}}{Z^{(0)}_{0,1,0}}
=\frac{Z^{(1)}_{1,0,0}}{Z^{(0)}_{1,0,0}},\quad
{\cal Z}_0
=\frac{Z^{(1)}_{0,0,0}}{Z^{(0)}_{0,0,0}},\nonumber\\
&{\cal Y}_1
=\frac{Z^{(1)}_{-\frac{1}{2},-\frac{1}{2},-\frac{1}{2}}}{Z^{(0)}_{-\frac{1}{2},-\frac{1}{2},-\frac{1}{2}}}
=\frac{Z^{(1)}_{-\frac{1}{2},\frac{1}{2},\frac{1}{2}}}{Z^{(0)}_{-\frac{1}{2},\frac{1}{2},\frac{1}{2}}}
=\frac{Z^{(1)}_{\frac{1}{2},-\frac{1}{2},\frac{1}{2}}}{Z^{(0)}_{\frac{1}{2},-\frac{1}{2},\frac{1}{2}}}
=\frac{Z^{(1)}_{\frac{1}{2},\frac{1}{2},-\frac{1}{2}}}{Z^{(0)}_{\frac{1}{2},\frac{1}{2},-\frac{1}{2}}},\nonumber\\
&{\cal Y}_0
=\frac{Z^{(1)}_{-\frac{1}{2},-\frac{1}{2},\frac{1}{2}}}{Z^{(0)}_{-\frac{1}{2},-\frac{1}{2},\frac{1}{2}}}
=\frac{Z^{(1)}_{-\frac{1}{2},\frac{1}{2},-\frac{1}{2}}}{Z^{(0)}_{-\frac{1}{2},\frac{1}{2},-\frac{1}{2}}}
=\frac{Z^{(1)}_{\frac{1}{2},-\frac{1}{2},-\frac{1}{2}}}{Z^{(0)}_{\frac{1}{2},-\frac{1}{2},-\frac{1}{2}}}
=\frac{Z^{(1)}_{\frac{1}{2},\frac{1}{2},\frac{1}{2}}}{Z^{(0)}_{\frac{1}{2},\frac{1}{2},\frac{1}{2}}}.
\end{align}
Here on the right-hand side (and also in appendices \ref{lowestpf} and \ref{higherrank1}) we adopt the abbreviated notation
\begin{align}
Z^{(N)}_{M_0,M_1,M_3}=e^{-i\Theta_{k,{\bm M}}}Z_{k,{\bm M}}(N)=e^{-i\Theta'_{k,{\bm M}}}Z'_{k,{\bm M}}(N).
\label{abb}
\end{align}
Namely, despite the definition of \eqref{Zprime2} by removing the phase $e^{i\Theta'_{k,{\bm M}}}$ coming from partition functions and introducing another $e^{-i\Theta_{k,{\bm M}}}$ \eqref{Thetanew} suitably to simplify the bilinear relations, we remove both of them.
Also, we drop indices of the level $k=1$ and the FI parameters $Z_1,Z_3$ and denote $N$ only by superscripts $(N)$ for simplicity.
Note that ${\cal Y}_0$ is easily obtained from ${\cal Y}_1$ by changing the signs of $Z_1$, ${\cal Y}_0={\cal Y}_1|_{Z_1\rightarrow -Z_1}$, due to the invariance under the reflection $(M_3,Z_1)\rightarrow (-M_3,-Z_1)$ in the $D_5$ Weyl group.
In appendix \ref{sec_Okuyama} we compute the values for $(L_1,L_2,L)=(0,0,0)$, $(1,0,0)$, $(1,1,1)$, which correspond to $(M_0,M_1,M_3)=(0,0,1)$, $(-1/2,1/2,1/2)$, $(0,0,0)$ respectively and find
\begin{align}
{\cal X}_0&=\frac{1}{4}Z_1Z_3\csc(\pi Z_1)\csc(\pi Z_3),\nonumber\\
{\cal Y}_1
&=\frac{-1}{16\pi}\sec(\pi Z_1)\sec(\pi Z_3)\Big(2-4\pi iZ_1Z_3
+2\pi\big(Z_1\tan(\pi Z_1)+Z_3\tan(\pi Z_3)\big)\nonumber\\
&\quad+\pi i\tan(\pi Z_1)\tan(\pi Z_3)-e^{2\pi iZ_1Z_3}\pi\sec(\pi Z_1)\sec(\pi Z_3)\Big),\nonumber\\
{\cal Y}_0
&=\frac{-1}{16\pi}\sec(\pi Z_1)\sec(\pi Z_3)\Big(2+4\pi iZ_1Z_3
+2\pi\big(Z_1\tan(\pi Z_1)+Z_3\tan(\pi Z_3)\big)\nonumber\\
&\quad-\pi i\tan(\pi Z_1)\tan(\pi Z_3)-e^{-2\pi iZ_1Z_3}\pi\sec(\pi Z_1)\sec(\pi Z_3)\Big),\nonumber\\
{\cal Z}_0
&=\frac{-1}{4\pi}\csc(\pi Z_1)\csc(\pi Z_3)\Big(\pi Z_1Z_3-\pi\cot(\pi Z_1)\cot(\pi Z_3)-\cot(\pi Z_1Z_3)\nonumber\\
&\quad+\pi\big(Z_1\cot(\pi Z_1)+Z_3\cot(\pi Z_3)\big)\cot(\pi Z_1Z_3)\Big).
\label{XYZ}
\end{align}
By using these exact values we find that all of the 40 bilinear relations are satisfied also at higher orders\footnote{
In fact, we have only checked for the second-lowest order $N=1$.
} in $\kappa$ if we choose $(\gamma_{{\bm M}}^{(1)}, \gamma_{{\bm M}}^{(2)}, \gamma_{{\bm M}}^{(3)})=(1,-1,-i)$.
With this choice of $(\gamma_{\bm M}^{(1)},\gamma_{\bm M}^{(2)},\gamma_{\bm M}^{(3)})$ we have checked the bilinear relations also for $k>1$ at several points $(L_1,L_2,L,\zeta_1)$ with $2\pi i\zeta_1\in\mathbb{Z}$ fixed by using method I explained in appendix \ref{method1}, as listed in appendix \ref{checkedlist}.

In summary our proposal for the bilinear relations of the grand partition functions is
\begin{align}
&e^{-\frac{\pi i}{2k}(\sigma_1 c+\sigma_2 d+\sigma_3 e)}S^{(1)}_{{\bm M}}
\widetilde\Xi_{a_\pm b_\pm c_\varnothing d_\varnothing e_\varnothing}(1,\kappa)
+e^{\frac{\pi i}{2k}(\sigma_1 c+\sigma_2 d+\sigma_3 e)}S^{(2)}_{{\bm M}}
\widetilde\Xi_{a_\pm b_\mp c_\varnothing d_\varnothing e_\varnothing}(-1,\kappa)\nonumber\\
&\qquad+S^{(3)}_{{\bm M}}
\widetilde\Xi_{a_\varnothing b_\varnothing c_{\pm\sigma_1} d_{\pm \sigma_2} e_{\pm\sigma_3}}(-i,\kappa)=0.
\label{bilinearwithkappa2}
\end{align}
Note that the coefficients $(\gamma_{{\bm M}}^{(1)}, \gamma_{{\bm M}}^{(2)}, \gamma_{{\bm M}}^{(3)})=(1,-1,-i)$ found here are reminiscent of those in \cite{BGT3}, where similar bilinear relations were found for the ABJM theory.
Since there is only one relative rank for the ABJM theory, the two factors in the first two bilinear terms in \eqref{bilinearwithkappa2} are degenerate to identical factors.
Here for the current $D_5$ case we can lift the degeneracies with more relative ranks and detect the full expression in \eqref{bilinearwithkappa2}.

\section{Covariance under Weyl group}
\label{sec_Weylcovariance}

In the previous two sections we have found that the grand partition functions satisfy the 40 bilinear relations.
Due to the extra factors $(S^{(1)}_{{\bm M}},S^{(2)}_{{\bm M}},S^{(3)}_{{\bm M}})$ \eqref{S(a)compact} these bilinear relations do not enjoy the covariance under the Weyl group at first sight.
Nevertheless, since the grand partition function normalized by the lowest order is the Fredholm determinant of the spectral operator $\Det(1+\kappa{\widehat H}_{k,{\bm M}}^{-1})$, which is (conjectured to be) invariant under the Weyl group \eqref{XiHinv}, it is natural to expect the covariance of the bilinear relations.
In this section we shall explain how the covariance is realized.

Let us regard the bilinear relations \eqref{bilinearwithkappa} for the grand partition function as those for the Fredholm determinant $\text{Det}(1+\kappa{\widehat H}_{k,\bm M}^{-1})$, and rewrite them schematically as
\begin{align}
&f_{1,{\bm M}}^{(a,b;\sigma_1,\sigma_2,\sigma_3)}\prod_\pm \text{Det}(1+\kappa{\widehat H}_{k,{\bm M}_\alpha\pm\frac{1}{2}(\delta^a_\alpha+\delta^b_\alpha)}^{-1})
+f_{2,{\bm M}}^{(a,b;\sigma_1,\sigma_2,\sigma_3)}\prod_\pm \text{Det}(1-\kappa{\widehat H}_{k,{\bm M}_\alpha\pm\frac{1}{2}(\delta^a_\alpha-\delta^b_\alpha)}^{-1})\nonumber \\
&\quad +f_{3,{\bm M}}^{(a,b;\sigma_1,\sigma_2,\sigma_3)}\prod_\pm \text{Det}(1\mp i\kappa{\widehat H}_{k,{\bm M}_\alpha\pm\frac{1}{2}(\sigma_1\delta^c_\alpha+\sigma_2\delta^d_\alpha+\sigma_3\delta^e_\alpha)}^{-1})
=0.
\label{bilineareqforDet}
\end{align}
Here $f_{1,{\bm M}}^{(a,b;\sigma_1,\sigma_2,\sigma_3)}$, $f_{2,{\bm M}}^{(a,b;\sigma_1,\sigma_2,\sigma_3)}$, $f_{3,{\bm M}}^{(a,b;\sigma_1,\sigma_2,\sigma_3)}$ are the bilinears of $Z_{k,{\bm M}'}(0)$
\begin{align}
&f^{(a,b;\sigma_1,\sigma_2,\sigma_3)}_{1,{\bm M}}=e^{-\frac{\pi i}{2k}(\sigma_1c+\sigma_2d+\sigma_3e)}S_{\bm M}^{(1)}\prod_\pm Z_{k,{\bm M}_\alpha\pm\frac{1}{2}(\delta^a_\alpha+\delta^b_\alpha)}(0),\nonumber \\
&f^{(a,b;\sigma_1,\sigma_2,\sigma_3)}_{2,{\bm M}}=e^{\frac{\pi i}{2k}(\sigma_1c+\sigma_2d+\sigma_3e)}S_{\bm M}^{(2)}\prod_\pm Z_{k,{\bm M}_\alpha\pm\frac{1}{2}(\delta^a_\alpha-\delta^b_\alpha)}(0),\nonumber \\
&f^{(a,b;\sigma_1,\sigma_2,\sigma_3)}_{3,{\bm M}}=S_{\bm M}^{(3)}\prod_\pm Z_{k,{\bm M}_\alpha\pm\frac{1}{2}(\sigma_1\delta^c_\alpha+\sigma_2\delta^d_\alpha+\sigma_3\delta^e_\alpha)}(0).
\label{f=ZZ}
\end{align}
Note that the argument ${\bm M}'$ of $Z_{k,{\bm M}'}(0)$ are shifted from ${\bm M}$ according to $(a,b;\sigma_1,\sigma_2,\sigma_3)$.
As we see below, the Weyl covariance is realized by the non-trivial quartic identities of $Z_{k,{\bm M}}(0)$.

In the viewpoint of the relations for the Fredholm determinant \eqref{bilineareqforDet}, besides the original relations \eqref{bilineareqforDet} labelled by $(a,b;\sigma_1,\sigma_2,\sigma_3)$ we can generate another set of 40 relations with each element of the $D_5$ Weyl group.
For example, let us look at the relation $(a,b;\sigma_1,\sigma_2,\sigma_3)=(M_0,M_1;+++)$ in \eqref{bilineareqforDet}, evaluated at ${\bm M}=(M_0,M_3,M_1,Z_1,Z_3)$:
\begin{align}
&f_{1,(M_0,M_3,M_1,Z_1,Z_3)}^{(M_0,M_1;+++)}\prod_\pm \text{Det}(1+\kappa{\widehat H}_{k,(M_0\pm\frac{1}{2},M_3\pm\frac{1}{2},M_1,Z_1,Z_3)}^{-1})\nonumber \\
&+f_{2,(M_0,M_3,M_1,Z_1,Z_3)}^{(M_0,M_1;+++)}\prod_\pm \text{Det}(1-\kappa{\widehat H}_{k,(M_0\pm\frac{1}{2},M_3\mp\frac{1}{2},M_1,Z_1,Z_3)}^{-1})\nonumber \\
&+f_{3,(M_0,M_3,M_1,Z_1,Z_3)}^{(M_0,M_1;+++)}\prod_\pm \text{Det}(1\mp i\kappa{\widehat H}_{k,(M_0,M_3,M_1\pm\frac{1}{2},Z_1\pm\frac{1}{2},Z_3\pm\frac{1}{2})}^{-1})
=0.
\end{align}
By using the invariance of the Fredholm determinant under $w:(M_0,M_1,M_3,Z_1,Z_3)\rightarrow (M_0,M_3,M_1,Z_1,Z_3)$, we can rewrite this relation as
\begin{align}
&f_{1,(M_0,M_3,M_1,Z_1,Z_3)}^{(M_0,M_1;+++)}\prod_\pm \text{Det}(1+\kappa{\widehat H}_{k,(M_0\pm\frac{1}{2},M_1,M_3\pm\frac{1}{2},Z_1,Z_3)}^{-1})\nonumber \\
&+f_{2,(M_0,M_3,M_1,Z_1,Z_3)}^{(M_0,M_1;+++)}\prod_\pm \text{Det}(1-\kappa{\widehat H}_{k,(M_0\pm\frac{1}{2},M_1,M_3\mp\frac{1}{2},Z_1,Z_3)}^{-1})\nonumber \\
&+f_{3,(M_0,M_3,M_1,Z_1,Z_3)}^{(M_0,M_1;+++)}\prod_\pm \text{Det}(1\mp i\kappa{\widehat H}_{k,(M_0,M_1\pm\frac{1}{2},M_3,Z_1\pm\frac{1}{2},Z_3\pm\frac{1}{2})}^{-1})
=0.
\end{align}
As the relation for the Fredholm determinant, this is of the same form as the relation with the shift $(a,b;\sigma_1,\sigma_2,\sigma_3)=(M_0,M_3;+++)$ in \eqref{bilineareqforDet} except for the coefficients of the three terms.
Since both of the two relations are satisfied for all order in $\kappa$, it would be natural to consider the two relations as identical up to an overall factor, which implies
\begin{align}
\frac{
f_{1,(M_0,M_1,M_3,Z_1,Z_3)}^{(M_0,M_3;+++)}
}{
f_{1,(M_0,M_3,M_1,Z_1,Z_3)}^{(M_0,M_1;+++)}
}
=
\frac{
f_{2,(M_0,M_1,M_3,Z_1,Z_3)}^{(M_0,M_3;+++)}
}{
f_{2,(M_0,M_3,M_1,Z_1,Z_3)}^{(M_0,M_1;+++)}
}
=
\frac{
f_{3,(M_0,M_1,M_3,Z_1,Z_3)}^{(M_0,M_3;+++)}
}{
f_{3,(M_0,M_3,M_1,Z_1,Z_3)}^{(M_0,M_1;+++)}
}.
\end{align}
Since $f^{(a,b;\sigma_1,\sigma_2,\sigma_3)}_{1,{\bm M}}$, $f^{(a,b;\sigma_1,\sigma_2,\sigma_3)}_{2,{\bm M}}$, $f^{(a,b;\sigma_1,\sigma_2,\sigma_3)}_{3,{\bm M}}$ are bilinears of $Z_{k,{\bm M}'}(0)$ \eqref{f=ZZ}, these are quartic relations among $Z_{k,{\bm M}'}(0)$ evaluated at different points ${\bm M}'$.
With different choices of the initial shift $(a,b;\sigma_1,\sigma_2,\sigma_3)$ and the element $w$ of the $D_5$ Weyl group, we also obtain different quartic relations for $Z_{k,{\bm M}'}(0)$.
We have checked for several examples, by using the exact expression of $Z_{k,{\bm M}}(0)$ at various values of $(L_1,L_2,L)$ obtained by the formula \eqref{Z0kM0formula}, that these quartic relations are indeed satisfied.

Note that once we assume these quartic relations and any one of the 40 bilinear relations for $Z_{k,{\bm M}}(0)$ in tables \ref{bilinear1} and \ref{bilinear2} we can derive all the other 39 bilinear relations for $Z_{k,{\bm M}}(0)$, while the converse is not guaranteed.
In this sense we might be able to regard the quartic relations (plus one bilinear relation) more fundamental than the 40 bilinear relations \eqref{bilinearwithkappa2}.
Currently we do not have a clear explanation for how these relations follow from the definition of $Z_{k,{\bm M}}(0)$.

So far, the grand partition function is considered to be defined only in the fundamental domain.
Since the bilinear relations are valid also at the boundary of the fundamental domain, we can regard the grand partition functions to extend naturally outside the original fundamental domain.
We shall see this explicitly in the next section.

\section{Applications beyond fundamental domain}
\label{app}

We have found that the 40 bilinear relations hold for partition functions at $N=0$ in section \ref{bilinear} and continue to be correct for higher ranks in section \ref{higher}.
Here we would like to point out that the 40 bilinear relations are not only beautiful but also useful.
Although we have mainly provided partition functions at $N=0$ so far and seen that they are non-vanishing within the fundamental domain of the rhombic dodecahedron (figure \ref{fd}), partition functions of higher ranks can have non-trivial regimes of non-vanishing values.
We have computed partition functions at $N=1$ within the original fundamental domain in section \ref{higher}.
Using the bilinear relations proposed in section \ref{higher} at general points of $(L_1,L_2,L)$, we can compute partition functions outside of the original fundamental domain and study how the fundamental domain extends for partition functions at $N=1$.
Typically, the fundamental domain for $N=1$ (and higher $N$) is larger than that for $N=0$.
We shall explain this computation for the case of $k=1$ by also specifying partition functions at $N=1$ outside of the original fundamental domain.
As discussed later in section \ref{sec_dualitycascade}, the results show a picture consistent with the Painlev\'e equations and the duality cascades.
Note also that the 40 bilinear relations are in general overdetermining.
They provide not just unknown partition functions, but also non-trivial checks of the consistency.

\begin{figure}[!t]
\centering\includegraphics[scale=0.6,angle=-90]{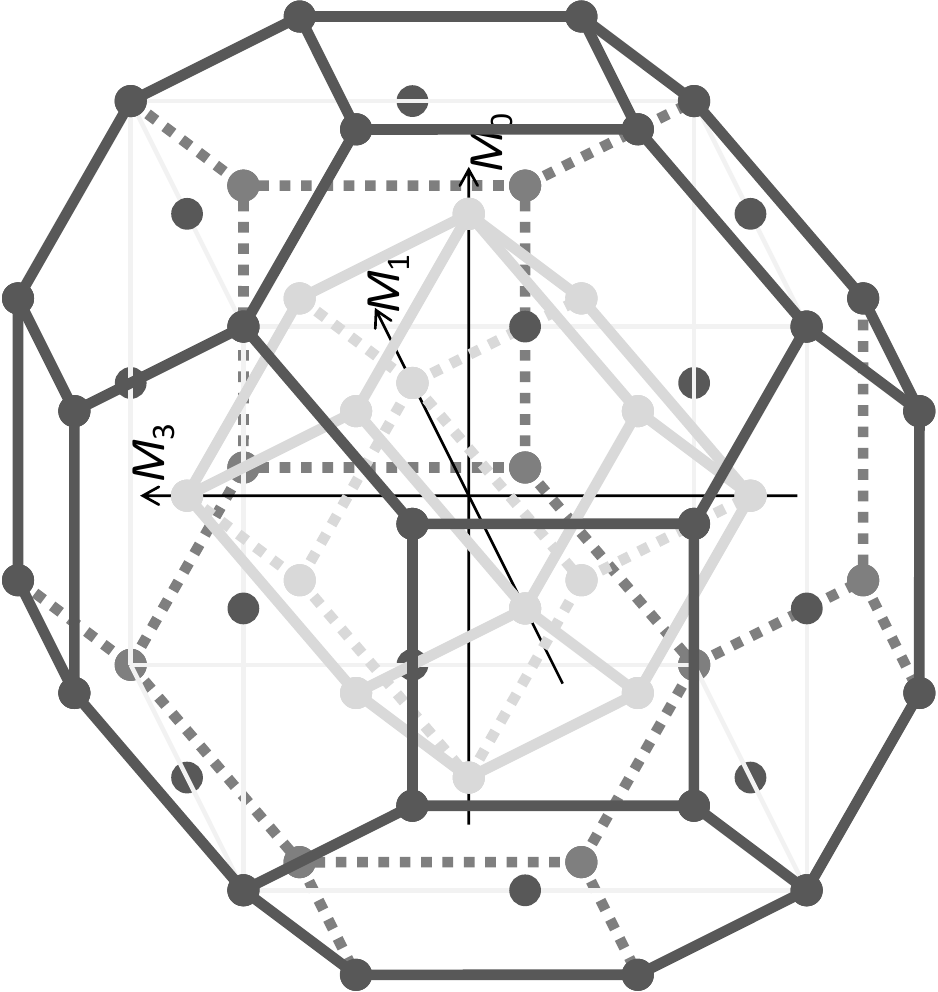}
\caption{
The domain where the partition functions of $N=1$ are non-vanishing.
Partition functions of $N\ge 1$ can be non-vanishing outside of the original fundamental domain for $N=0$ of the rhombic dodecahedron depicted in light grey.
}
\label{extendedfig}
\end{figure}

The bilinear relations evaluated at ${\bm M}=(M_0,M_1,M_3,Z_1,Z_3)$ connect partition functions at different values of ${\bm M}$ shifted to the nearest lattice points.
Hence, when evaluated at the boundaries of the original fundamental domain, the bilinear relations require non-trivial constraints for partition functions outside of it.
For example, if we consider the bilinear relation with shifts $(M_1,M_3;+++)$ at the ``base'' point $(0,1/2,1/2)$, the three terms are given by bilinears of partition functions at $(0,1,1)$, $(0,0,0)$ for the first one, those at $(0,1,0)$, $(0,0,1)$ for the second one and those at $(\pm 1/2,1/2,1/2)$ for the third one with the values of $Z_1$ and $Z_3$ also shifted.
Since only the partition function at $(0,1,1)$ among them is outside of the original fundamental domain and unknown, we can use the bilinear relation to fix the value.
Namely, let us expand the bilinear relation with $\kappa$.
Then, its zeroth order is satisfied as a result of section \ref{bilinear} with the partition function at $(0,1,1)$ vanishing, while the bilinear relation at the first order of $\kappa$ reads
\begin{align}
&e^{-\frac{\pi i}{2}(Z_1+Z_3)}Z_{0,0,0,Z_1,Z_3}(0)Z_{0,1,1,Z_1,Z_3}(1)
\nonumber \\
&-e^{\frac{\pi i}{2}(Z_1+Z_3)}Z_{0,1,0,Z_1,Z_3}(0)Z_{0,0,1,Z_1,Z_3}(0)\Bigl(\frac{Z_{0,1,0,Z_1,Z_3}(1)}{Z_{0,1,0,Z_1,Z_3}(0)}+\frac{Z_{0,0,1,Z_1,Z_3}(1)}{Z_{0,0,1,Z_1,Z_3}(0)}\Bigr)\nonumber \\
&-i\prod_\pm Z_{\pm \frac{1}{2},\frac{1}{2},\frac{1}{2},Z_1\pm \frac{1}{2}, Z_3\pm \frac{1}{2}}(0)\Bigl(\frac{Z_{\frac{1}{2},\frac{1}{2},\frac{1}{2},Z_1+\frac{1}{2},Z_3+\frac{1}{2}}(1)}{Z_{\frac{1}{2},\frac{1}{2},\frac{1}{2},Z_1+\frac{1}{2},Z_3+\frac{1}{2}}(0)}-\frac{Z_{-\frac{1}{2},\frac{1}{2},\frac{1}{2},Z_1-\frac{1}{2},Z_3-\frac{1}{2}}(1)}{Z_{-\frac{1}{2},\frac{1}{2},\frac{1}{2},Z_1-\frac{1}{2},Z_3-\frac{1}{2}}(0)}\Bigr)=0.
\end{align}
Here we have omitted $k$ from the subscripts since we are only considering $k=1$.
By solving this equation we obtain $Z_{0,1,1,Z_1,Z_3}(1)$ as \eqref{Z1M0_011}.
Note that although we have multiple relations to fix the same partition function and the relations are overdetermining, the partition functions determined in this way are all consistent.
We can also perform the same calculations to determine $Z_{k,{\bm M}}(N=1)$ at other points outside the fundamental domain.
The results are given in appendix \ref{higherrank1}.

We find that the domain for the non-vanishing second-lowest order of the grand partition function $\Xi_{k,{\bm M}}$ in $\kappa$ (with $N=1$) extends slightly from the original fundamental domain for the lowest order (with $N=0$).
See figure \ref{extendedfig}.
The results match the picture of duality cascades indicated in \cite{FMMN,FMS}, which shows that the more distant from the origin the relative ranks are, the higher overall rank $N$ a non-vanishing partition function requires.
Namely, we have found that the results obtained from bilinear relations of $q$-Painlev\'e equations are also consistent with duality cascades.

\section{Comparison with other works}
\label{comparison}

In this section, we comment on the relations between our results and some previous studies related to $q$-Painlev\'e equations and the super Chern-Simons theories.

\subsection{Comparison with \cite{Ohta}}

Note that our discovery of bilinear relations is not completely new from the viewpoint of Weyl groups in the long history of Painlev\'e equations.
In fact in \cite{Ohta} bilinear equations\footnote{We are grateful to Yasuhiko Yamada for valuable explanations and discussions.}
\begin{align}
q^{n_c+n_d+n_e}\tau^{ab}\tau_{ab}+q^{-n_c-n_d-n_e}\tau^a{}_b\tau_a{}^b-\tau^{cde}\tau_{cde}&=0,\nonumber\\
q^{-n_c-n_d+n_e}\tau^{ab}\tau_{ab}+q^{n_c+n_d-n_e}\tau^a{}_b\tau_a{}^b-
\tau^{cd}{}_e\tau_{cd}{}^e&=0,
\label{Otautau}
\end{align}
were already proposed purely from Weyl groups.
Here $\tau$ denotes the tau function of $q\text{P}_{\text{VI}}$ on five-dimensional lattice points with the indices being directions shifted positively for superscripts and negatively for subscripts.
Namely, if we construct representations of the affine $D_5$ Weyl group as in \cite{TM,MY}, the above bilinear equations follow from the fact that three terms in the bilinear equations are not linearly independent.
In this sense the structure of the bilinear equations is mostly fixed by the affine $D_5$ Weyl group.
The Weyl group, however, does not explain the coefficients of the bilinear equations since we can rescale the tau functions by an arbitrary function which changes the coefficients.

Therefore, comparatively, our discovery in this paper should be summarized by the facts that partition functions of the super Chern-Simons theory satisfy the $q$-Painlev\'e equations $q\text{P}_\text{VI}$ and that the coefficients for the equations not determined from the Weyl group are explicitly fixed.
It is interesting to find that the coefficients, though not determined from the Weyl group, are quite simple with only the $S_{{\bm M}}$ factors in \eqref{S(a)compact}.

\subsection{Comparison with \cite{BGKNT}}
\label{sec_comparewithBGKNT}

In this paper we have investigated the bilinear relations satisfied by the grand partition functions of the super Chern-Simons theory.
A similar study was initiated in \cite{BGKNT} for the Fredholm determinant $\text{Det}(1+\kappa {\widehat H}_{k,{\bm M}}^{-1})$, which is associated with the grand partition function, as reviewed in the introduction.
Our study in this paper relies heavily on their analysis.
Here we shall discuss some similarities and differences.

While in \cite{BGKNT} partition functions were computed only for $L=0$, here we have extended the computations to general $L$.
If we restrict ourselves to $N=0$, the explicit expression for the partition function is still available for general $L$.
As seen in section \ref{bilinear}, in increasing $L$, the matrix size of the determinant increases accordingly, which resembles the structure of Painlev\'e equations \cite{KMNOY}. 

Also, in \cite{BGKNT} the bilinear relations for the Fredholm determinant were searched for by consulting those for the Nekrasov-Okounkov partition function in \cite{JNS}.
It was found that, instead of the original Fredholm determinant, some factors ($F_{k,{\bm M}}$ and $\Omega_{k,{\bm M}}$) need to be introduced as in $F_{k,{\bm M}}\text{Det}(1+\Omega_{k,{\bm M}}\kappa {\widehat H}_{k,{\bm M}}^{-1})$.
The physical or group-theoretical origin of these factors is however unclear.

Our analysis here is based on the grand partition function $\Xi_{k,{\bm M}}(\kappa)$ instead of the Fredholm determinant and consults only the index structure of \cite{Ohta,JNS} without referring to the coefficients of \cite{JNS}.
We find that the factor $F_{k,{\bm M}}$ introduced in \cite{BGKNT} is replaced by the partition function appearing in the lowest order of the grand partition function, which has a sound physical origin.
Also, since it is conjectured that the spectral operator ${\widehat H}_{k,{\bm M}}$ in the Fredholm determinant is invariant under the $D_5$ Weyl group, it is perplexing to find a factor $\Omega_{k,{\bm M}}$ transforming non-trivially under it in \cite{BGKNT}.
Comparatively, our proposal with $(\gamma^{(1)}_{\bm M},\gamma^{(2)}_{\bm M},\gamma^{(3)}_{\bm M})=(1,-1,-i)$ being constant is more reasonable considering the $D_5$ Weyl group.
All these changes bring the proposal in \cite{BGKNT} back to those similar to \cite{BGT3} which are much more comfortable both physically and group-theoretically.

Furthermore, in \cite{BGKNT} only several bilinear relations were proposed.
Since the coefficients are not invariant under the Weyl group as stressed in the previous subsection, it is desirable to fix the coefficients of all the 40 bilinear relations.
We have succeeded in this direction and summarized the results in \eqref{bilinearwithkappa2} with \eqref{S(a)compact}.

\subsection{Comparison with \cite{A,HK}}
\label{sec_dualitycascade}

Lastly, we would like to make comparison with previous works \cite{A,HK} on Hanany-Witten transitions and duality cascades in this subsection.

As in \cite{FMMN,FMS}, under duality cascades the relative ranks $(M_0,M_1,M_3)$ are shifted translationally with the overall rank $N$ decreasing and finally we arrive at the fundamental domain \eqref{funddomain} with the lowest overall rank $N$.
It is, however, subtle when the relative ranks are located on the boundaries of the fundamental domain since relative ranks on one boundary are shifted to those on the opposite one and both should be treated on an equal footing.
Here, let us start with studying the relations among partition functions on the opposite boundaries.

Interestingly, using the partition functions given in appendix \ref{k2}, we find that the partition functions of $k=2$ located at the opposite boundaries and associated by duality cascades are related by
\begin{align}
&\frac{Z^{(0)}_{-M+1,M+1,M_3}}{Z^{(0)}_{-M-1,M-1,M_3}}\bigg|_{-1\le M\le 1}^{k=2}
=\{16\sin^4(\pi Z_1),8\cos^3(\pi Z_1),-4\sin^2(\pi Z_1),2\cos(\pi Z_1),1\},\nonumber\\
&\frac{Z^{(0)}_{M-1,M+1,M_3}}{Z^{(0)}_{M+1,M-1,M_3}}\bigg|_{-1\le M\le 1}^{k=2}
=e^{2\pi iZ_1Z_3}\{16\sin^4(\pi Z_1),8\cos^3(\pi Z_1),-4\sin^2(\pi Z_1),2\cos(\pi Z_1),1\},\nonumber\\
&\frac{Z^{(0)}_{-M+1,M_1,M+1}}{Z^{(0)}_{-M-1,M_1,M-1}}\bigg|_{-1\le M\le 1}^{k=2}
=\{16\sin^4(\pi Z_3),8\cos^3(\pi Z_3),-4\sin^2(\pi Z_3),2\cos(\pi Z_3),1\},\nonumber\\
&\frac{Z^{(0)}_{M-1,M_1,M+1}}{Z^{(0)}_{M+1,M_1,M-1}}\bigg|_{-1\le M\le 1}^{k=2}
=e^{2\pi iZ_1Z_3}\{16\sin^4(\pi Z_3),8\cos^3(\pi Z_3),-4\sin^2(\pi Z_3),2\cos(\pi Z_3),1\},\nonumber\\
&\frac{Z^{(0)}_{M_0,-M+1,M+1}}{Z^{(0)}_{M_0,-M-1,M-1}}\bigg|_{-1\le M\le 1}^{k=2}
=16e^{2\pi iZ_1Z_3}\nonumber\\
&\qquad\times\{\sin^4(\pi Z_3),\cos(\pi Z_1)\cos^3(\pi Z_3),\sin^2(\pi Z_1)\sin^2(\pi Z_3),\cos^3(\pi Z_1)\cos(\pi Z_3),\sin^4(\pi Z_1)\},\nonumber\\
&\frac{Z^{(0)}_{M_0,M-1,M+1}}{Z^{(0)}_{M_0,M+1,M-1}}\bigg|_{-1\le M\le 1}^{k=2}\nonumber\\
&\quad=\{\csc^4(\pi Z_1)\sin^4(\pi Z_3),\sec^3(\pi Z_1)\cos^3(\pi Z_3),\csc^2(\pi Z_1)\sin^2(\pi Z_3),\sec(\pi Z_1)\cos(\pi Z_3),1\},
\label{cascade2}
\end{align}
(with step $1/2$).
Here we have adopted the same abbreviation \eqref{abb} for the partition functions as in section \ref{app}.
Note that the result depends only on the variable in $-1\le M\le 1$ appearing in the subscript and is independent of the remaining variable.
A rather clean pattern with $\big(2\sin(\pi Z+\frac{n\pi}{2}))\big)^n$ is found here.

For the partition functions of $k=1$ with both $N=0$ and $N=1$ obtained in appendices \ref{k1} and \ref{higherrank1}, the structure is more impressive.
To explain the relation, we first define the lowest non-vanishing partition functions $Z^\text{low}_{M_0,M_1,M_3}$ even outside of the fundamental domain.
Namely, as explained in section \ref{bilinear}, partition functions at $N=0$, $Z^{(0)}_{M_0,M_1,M_3}$, are non-vanishing only in the fundamental domain.
In section \ref{app} and appendix \ref{higherrank1} we have studied partition functions at $N=1$, $Z^{(1)}_{M_0,M_1,M_3}$, and found that the fundamental domain is extended slightly.
For our current case with only $N=0$ and $N=1$, $Z^\text{low}_{M_0,M_1,M_3}$ denotes $Z^{(0)}_{M_0,M_1,M_3}$ when relative ranks ${\bm M}=(M_0,M_1,M_3)$ are in the fundamental domain, while $Z^{(1)}_{M_0,M_1,M_3}$ when ${\bm M}$ are outside the fundamental domain but located in the polygon in figure \ref{extendedfig}.
Then, we find
\begin{align}
&\frac{Z^\text{low}_{M_0+\frac{1}{2},M_1+\frac{1}{2},M_3}}
{Z^\text{low}_{M_0-\frac{1}{2},M_1-\frac{1}{2},M_3}}\bigg|^{k=1}_{-1\le M_1\le 1}
=\big\{8\cos^3(\pi Z_1),-4\sin^2(\pi Z_1),2\cos(\pi Z_1),1,\big(2\cos(\pi Z_1)\big)^{-1}\big\},\nonumber\\
&\frac{Z^\text{low}_{M_0-\frac{1}{2},M_1+\frac{1}{2},M_3}}
{Z^\text{low}_{M_0+\frac{1}{2},M_1-\frac{1}{2},M_3}}\bigg|^{k=1}_{-1\le M_1\le 1}
=e^{2\pi iZ_1Z_3}\big\{8\cos^3(\pi Z_1),-4\sin^2(\pi Z_1),2\cos(\pi Z_1),1,\big(2\cos(\pi Z_1)\big)^{-1}\big\},\nonumber\\
&\frac{Z^\text{low}_{M_0+\frac{1}{2},M_1,M_3+\frac{1}{2}}}
{Z^\text{low}_{M_0-\frac{1}{2},M_1,M_3-\frac{1}{2}}}\bigg|^{k=1}_{-1\le M_3\le 1}
=\big\{8\cos^3(\pi Z_3),-4\sin^2(\pi Z_3),2\cos(\pi Z_3),1,\big(2\cos(\pi Z_3)\big)^{-1}\big\},\nonumber\\
&\frac{Z^\text{low}_{M_0-\frac{1}{2},M_1,M_3+\frac{1}{2}}}
{Z^\text{low}_{M_0+\frac{1}{2},M_1,M_3-\frac{1}{2}}}\bigg|^{k=1}_{-1\le M_3\le 1}
=e^{2\pi iZ_1Z_3}\big\{8\cos^3(\pi Z_3),-4\sin^2(\pi Z_3),2\cos(\pi Z_3),1,\big(2\cos(\pi Z_3)\big)^{-1}\big\}.
\label{cascade1}
\end{align}
Note that, compared with the previous case of $k=2$ \eqref{cascade2} only on the boundary, here we can extend to all the three-dimensional lattice points related by duality cascades.
Namely, the relations \eqref{cascade1} hold regardless of the values of unspecified variables of $(M_0,M_1,M_3)$.

On the other hand, partition functions related by the Hanany-Witten transitions are proven to be identical in \cite{A,HK}.
Since the duality cascades can be regarded as subsequent applications of the Hanany-Witten transitions \cite{FMMN,FMS}, it may seem inconsistent if the partition functions change under the duality cascades as in \eqref{cascade2} and \eqref{cascade1}.
Here we note that the computations in \cite{A,HK} are always performed for the case of exchanging two 5-branes of different types, and not of identical types, which is considered in the current setup.
The first sign of this difference has already been detected in section \ref{bilinear}, where it is noted that partition functions related by the Weyl group are not identical.

Despite the ambiguity for identical 5-branes, our partition functions are consistent with the fundamental domain in figure \ref{fd} and its generalizations for $N=1$ in figure \ref{extendedfig} as well as the 40 bilinear relations.
In this sense, our partition functions are consistent with all the expectations and the studies so far.

\section{Conclusion}
\label{conclusion}

In this paper we have studied bilinear relations for the grand partition function of the three-dimensional ${\cal N}=4$ super Chern-Simons theory with the gauge group $\text{U}(N_1)_{0,-i\zeta_2}\times
\text{U}(N_2)_{k,i\zeta_1}\times
\text{U}(N_3)_{0,-i\zeta_1}\times
\text{U}(N_4)_{-k,i\zeta_2}$.
We have found 40 bilinear relations which are valid for any values of levels and ranks.
We list several future directions we wish to pursue in the future.

\begin{itemize}
\item To find the bilinear relations with simple coefficients we have replaced the original overall phase factor $\Theta_{k,{\bm M}}$ of the partition function by another one $\Theta_{k,{\bm M}}'$ \eqref{Thetanew}.
We would like to understand the physical or group-theoretical meaning of this change of overall phase factors.
\item In section \ref{app} we have studied the partition functions outside the fundamental domain for $N=0$ (see figure \ref{extendedfig}), and by using them we have found non-trivial relations among partition functions in section \ref{sec_dualitycascade}.
Although in sections \ref{app} and \ref{sec_dualitycascade} we have focused on the partition function at $N=1$, it would be interesting to extend the analysis for $N\ge 2$ and see how the structures we have found generalize.
\item It was observed in \cite{MNN} that for certain values of ranks the partition functions are divergent.
In the result in appendix \ref{k2} this is reconfirmed by setting the FI parameters $Z_1$ and $Z_3$ to zero.
We would like to understand the physical meaning of the divergences.
In section \ref{sec_dualitycascade} we have also observed that those points where the partition functions diverge in the limit $Z_i\rightarrow 0$ are related, under the Hanany-Witten transitions involving the exchanges of the 5-branes of the same type, to the points where the partition functions are finite or diverge weakly.
On the other hand, it is known that such Hanany-Witten transitions can transform a good theory \cite{Gaiotto:2008ak} into an ugly/bad theory and that the partition function of a bad theory regularized by adding mass/FI parameters diverges when the regularization is turned off \cite{Y,NY1}.
Putting these together we suspect that the divergences we have observed may also signal that the super Chern-Simons theory is bad for the corresponding ranks \cite{Nosaka:2018eip}.
\item If we apply the $D_5$ Weyl group to the inequalities \eqref{funddomain}, we find constraints also for the FI parameters \cite{FMMN}.
From the computational perspective a bound on the FI parameters may arise as follows, although the bounds thus obtained seem different from those predicted from the $D_5$ Weyl group.
In the matrix model \eqref{Zprime1}, the FI terms on the nodes with non-vanishing Chern-Simons levels can be absorbed by shifting the integration variables.
When $Z_1,Z_2\in i\mathbb{R}$, such shift does not change the integration contour.
However, when $|\text{Re}[Z_1]|$ or $|\text{Re}[Z_3]|$ is larger than $k/2$, the integration contour crosses the pole of the one-loop determinant of bifundamental matters, and hence the matrix model for the new contour may not be the same as the original one.
This implies that an analytic expression obtained by continuing from $Z_1,Z_2\in i\mathbb{R}$ would be valid only when $|\text{Re}[Z_1]|$ and $|\text{Re}[Z_3]|$ are smaller than the critical value.
However, the bounds thus obtained seem different from those predicted from the $D_5$ Weyl group.
It is interesting to see the relation between the two bounds and understand their physical interpretation.
\item In this paper we have checked that the grand partition function \eqref{XiHinv} satisfies the bilinear relations \eqref{bilinearwithkappa2} to the second-lowest order in $\kappa$ (namely $N=1$) for various $k$ and ${\bm M}$ by using the exact values of the partition functions.
It would be more desirable to derive the bilinear relations analytically\footnote{
A proof for a part of the 40 bilinear relations of $q\text{P}_\text{VI}$ \cite{JNS} at ${\cal O}(\kappa^1)$ with arbitrary values of $k$, $L_1$ and $L_2$ but $L=0$ was performed in \cite{BGKNT}.
Although the proof was done with the different coefficients $F_{k,{\bm M}}$ and $\Omega_{k,{\bm M}}$ from our current proposal in section \ref{sec_comparewithBGKNT}, the technique itself may be applicable.
}
in $k,{\bm M}$ for all order in $\kappa$ as new identities among matrix models with different ranks.
As far as the authors recognize, such proof was completed only for the matrix model for $q\text{P}_{{\text{III}}_3}$ under the differential limit (which is called the dual 4d limit in \cite{Bonelli:2016idi,BGT3}) \cite{Zamolodchikov:1994uw,Tracy:1995proof}.
An easier problem is to derive the bilinear relations analytically for $k,{\bm M}$ but only at $\kappa=0$ (namely $N=0$).
At this order, the bilinear relations reduces to those for $Z_{k,{\bm M}}(0)$ \eqref{biliearZtilde}.
Since $Z_{k,{\bm M}}(0)$ is a matrix model by itself, this simpler problem may also provide some hints to find identities for the matrix models at $N\ge 0$ which are helpful to prove the bilinear relations.
Moreover, since $Z_{k,{\bm M}}(0)$ is written as an $L\times L$ determinant \eqref{Z0kM0formula}, the bilinear relations \eqref{biliearZtilde} are reminiscent of the Pl\"ucker relations in integrable systems \cite{Satojp,Satoeng,Kazakov:1997ya,KNY0012063}.
Our bilinear relations might be understood based on the Pl\"ucker relations or its generalization.
Also, since ${\bm M}$ are the parameters of the $D_5$ curve, the bilinear relations \eqref{Z0kM0formula} should be generalized for generic complex numbers of ${\bm M}$ eventually.
The most crucial step is probably the definition of $Z_{k,{\bm M}}(0)$ for $L\notin{\mathbb N}$.
\item
In the ABJM theory one can also consider the expectation values of the half-BPS Wilson loops \cite{DT,MPTS} in the grand canonical ensemble, which also enjoy non-trivial integrable structures.
In \cite{MM,Giambelli,JT} it was found that one-point functions of the Wilson loops \cite{Klemm:2012ii,HHMO} enjoy various Pl\"ucker relations in the modified Kadomtsev-Petviashvili integrable hierarchy.
This is even generalized to the two-dimensional Toda lattice integrable hierarchy \cite{2DTL} for two-point functions \cite{KMtwo}.
Since the correspondence is shown with the unique relative rank $M$ kept fixed, this can be regarded as the open string side of the integrable structure.
Comparatively, the integrable structure discussed in this paper relates the grand partition function with different relative ranks ${\bm M}$ and should be regarded as the closed string side.
It would be interesting to relate the two sides of the integrable structure by the open/closed duality \cite{HaOk,closed}.
\item We may assume that similar non-linear $q$-difference equations arise for the grand partition function of more general three-dimensional supersymmetric gauge theories with an unbounded rank $N$.
If this is the case, we might be able to guess the equations for each model by calculating the partition functions at $N=0$ as in section \ref{bilinear}, which are typically more tractable.
We may also gain non-trivial hints from the web of dualities, as was the case in our current setup.
It is interesting to seek $q$-difference equations associated with other theories of M2-branes such as the super Chern-Simons theories with affine $D$-type quivers \cite{Gulotta:2011vp,Assel:2015hsa,Moriyama:2015jsa} under this strategy \cite{BFKNT_wip}.
\item Our result may shed a new light on the study of the five-dimensional gauge theories in the following sense.
In \cite{JNS} it was found that the Nekrasov-Okounkov partition function of the five-dimensional ${\cal N}=1$ $\text{SU}(2)$ super Yang-Mills theory with $N_\text{f}=4$ fundamental matters also satisfy a set of eight bilinear $q$-difference equations of $q\text{P}_{\text{VI}}$.
Indeed, if we relate the parameters between the five-dimensional theory and the super Chern-Simons theory by identifying the classical limit of the curve ${\widehat H}_{k,{\bm M}}$ \eqref{H} with the Seiberg-Witten curve of the five-dimensional theory, we find that the five-dimensional parameters in the bilinear relations in \cite{JNS} are shifted in the same way as those in the corresponding 8 relations in tables \ref{bilinear1} and \ref{bilinear2}.
On the contrary, the coefficients of the three bilinears in each equation in \cite{JNS} do not coincide with those in tables \ref{bilinear1} and \ref{bilinear2} (see also \eqref{S(a)compact}).
This suggests that the Nekrasov-Okounkov partition function $Z_{\text{NO}}$ in \cite{JNS} and the grand partition function are normalized differently as $Z_{\text{NO}}=F_{k,{\bm M}}\Xi_{k,{\bm M}}(\kappa)$, with an overall factor $F_{k,{\bm M}}$.
Note that a similar attempt to find the overall factor relating $Z_{\text{NO}}$ and the Fredholm determinant was made for $L=0$ in \cite{BGKNT}, as mentioned in section \ref{sec_comparewithBGKNT}.
In \cite{BGKNT}, however, the guiding principle for determining $F_{k,{\bm M}}$ from the three-dimensional gauge theory was missing and it was unclear how to generalize the result for $L\neq 0$.
Hence we expect that our new result where the Fredholm determinant is multiplied with the partition function at $N=0$ is a more plausible starting point.
Note that these requirements alone do not fix $F_{k,{\bm M}}$ uniquely due to, for example, the ambiguity of multiplying $F_{k,{\bm M}}$ with an arbitrary periodic function of ${\bm M}$ \cite{BS3}.
Nevertheless, the determination of $F_{k,{\bm M}}$ is interesting because, once we identify it, we may be able to use the Fredholm determinant to evaluate the Nekrasov-Okounkov partition function in the parameter regime where the standard analysis in the five-dimensional theory is difficult \cite{Bonelli:2016idi,Bonelli:2017ptp}.
Namely, while in the five-dimensional theory $Z_{\text{NO}}$ is calculated perturbatively in the instanton counting parameter $t$, in the super Chern-Simons theory (where $t$ is identified with one of the five parameters ${\bm M}$) the grand partition function can be calculated non-perturbatively in ${\bm M}$ at each order of $\kappa$.
\item Although our proposal that the grand partition function $\Xi_{k,{\bm M}}(\kappa)$ serves as the tau function for Painlev\'e equations, $\tau^{q\text{P}_{\text{VI}}}=\Xi_{k,{\bm M}}(\kappa)$, provides a large class of solutions to $q\text{P}_{\text{VI}}$, this does not cover all solutions.
Indeed, the general solution should contain two arbitrary parameters besides the parameters of the Painlev\'e equation $(t,\theta_0,\theta_t,\theta_1,\theta_\infty)$, which correspond to the two initial conditions for the second-order differential equation in the $q\rightarrow 1$ limit.
In the solutions given by the four/five-dimensional Nekrasov-Okounkov partition function (i.e.~the discrete Fourier transform of the Nekrasov partition function), these two parameters are incorporated by the Coulomb modulus $\sigma$ and the momentum $s$ of the Fourier transformation.
In our proposal, while $\kappa$ corresponds to $\sigma$, there is no parameter corresponding to $s$.
Namely, for our three-dimensional super Chern-Simons theory, as suggested by the integral representation of the Airy function, to relate the result of the matrix model to the free energy of topological strings, we need to prepare infinite copies of the grand potentials with the chemical potential $\mu=\log\kappa$ shifted by $2\pi i$ and sum over the copies \cite{HMO2}.
This summation is identified with the summation for the Fourier transform in the Nekrasov-Okounkov partition function.
However, due to the single-valuedness of $\Xi_{k,{\bm M}}(\kappa)$ in $\kappa$ \eqref{XiHinv} this summation does not introduce the parameter $s$, or in other words, freezes it to $s=1$.
It is interesting to incorporate $s$ on the three-dimensional side.
\item Besides the $q$-deformation we can introduce another deformation by quantizing the Painlev\'e equations \cite{Hasegawa,Bershtein:2017swf,Awata:2022idl}.
In the five-dimensional Nekrasov-Okounkov partition function, the two deformation parameters are realized precisely in the $\Omega$ background \cite{Nekrasov:2003rj}.
It is interesting to generalize our three-dimensional grand partition function as well so that it incorporates two parameters.
The analysis in \cite{NY,BS3,BS4,Kim:2019uqw}, where the affine Weyl group actions on non-commutative parameters were constructed, may provide some hints toward this direction.
\item The $q$-Painlev\'e equations for the five-dimensional Nekrasov-Okounkov partition function $Z_{\text{NO}}$ on the self-dual $\Omega$ background $Z_{\text{NO}}^{\text{SD}}$ are believed to be derived by using the Nakajima-Yoshioka blowup equations \cite{NY}.
This is done explicitly for the five-dimensional ${\cal N}=1$ $\text{SU}(2)$ pure Yang-Mills theory where the corresponding $q$-Painlev\'e equation is $q\text{P}_{\text{III}_3}$ \cite{BS4}.
Note that the relations obtained directly from the blowup equations are not $q\text{P}_{\text{III}_3}$ itself, but a couple of different expressions for $Z_{\text{NO}}^{\text{SD}}$ in terms of bilinears of $Z_{\text{NO}}$ on generic non-self-dual $\Omega$ backgrounds.
Then, $q\text{P}_{\text{III}_3}$ is obtained by eliminating these $Z_{\text{NO}}$ from the expressions \cite{BS4}.
In terms of the three-dimensional super Chern-Simons theory, which is the ABJM theory for $q\text{P}_{\text{III}_3}$, the grand partition function $\Xi_{k,{\bm M}}(\kappa)$ corresponds to $Z_{\text{NO}}^{\text{SD}}$, while the counterparts of $Z_{\text{NO}}$ on generic $\Omega$ backgrounds are still unclear.
Interestingly, however, a couple of relations which resemble the blowup equations
are observed even in the ABJM theory (called the Wronskian-like relation) \cite{GHM2}, which express the Fredhom determinant itself in terms of bilinears of those with the chiral projections
${\widehat \Pi}_\pm$, $\text{Det}(1+\kappa{\widehat H}^{-1}{\widehat\Pi}_\pm)$.
It is interesting to study such relations also for the theory studied in this paper with two NS5-branes and two $(1,k)$5-branes, which would be more fundamental than $q\text{P}_{\text{VI}}$.
Also, since such relations would directly correspond to the blowup equations, they would be helpful in generalizing the proof in \cite{BS4} to show explicitly that $Z_{\text{NO}}^{\text{SD}}$ for $N_\text{f}=4$ satisfies $q\text{P}_{\text{VI}}$.
\end{itemize}

\appendix

\section{Computation of partition functions}
\label{computation}

In this appendix we explain how to calculate the expansion coefficients of $\Xi_{k,{\bm M}}(\kappa)$ \eqref{XiHinv} in $\kappa$.
First we notice that from the generalized Fermi gas form \eqref{Z0kM}, $\Xi_{k,{\bm M}}(\kappa)$ is given by (see \cite{BGKNT})
\begin{align}
&\Xi_{k,{\bm M}}(\kappa)
=\sum_{N=0}^\infty \kappa^NZ_{k,{\bm M}}(N)\nonumber \\
&=
e^{i\Theta_{k,{\bm M}}}
Z^{\text{CS}}(L_1)
Z^{\text{CS}}(L_2)
\text{Det}(1+\kappa{\widehat D}_1{\widehat D}_2)
\det\begin{pmatrix}
[\langle\!\langle t_{L,r}|{\widehat d}_1(1+\kappa{\widehat D}_2{\widehat D}_1)^{-1}{\widehat d}_2|{-t_{L,s}}\rangle\!\rangle]_{(r,s)}
\end{pmatrix}.
\label{openclosed}
\end{align}
This expression implies that we can obtain the $N$-th order expansion coefficient $Z_{k,{\bm M}}(N)$ of $\Xi_{k,{\bm M}}(\kappa)$ by computing
\begin{align}
\Tr({\widehat D}_1{\widehat D}_2)^n,\quad
\langle\!\langle t_{L,r}|{\widehat d}_1({\widehat D}_2{\widehat D}_1)^n{\widehat d}_2|{-t_{L,s}}\rangle\!\rangle,
\label{trandmatelem}
\end{align}
with $n\le N$.

In the following we display two different methods to calculate the exact values of $\Tr({\widehat D}_1{\widehat D}_2)^n$ and $\langle\!\langle t_{L,r}|{\widehat d}_1({\widehat D}_2{\widehat D}_1)^n{\widehat d}_2|{-t_{L,s}}\rangle\!\rangle$ for $k,L_1,L_2,L\in\mathbb{Z}_{\ge 0}$ and $L_1,L_2\le k$.
The first method (method I) is the one developed in \cite{PY} based on the technique \cite{Tracy:1995ax}, and the second method (method II) is the integration trick used in \cite{Okuyama:2011su} to calculate the partition function of the ABJM theory at $N=2$.\footnote{
The two methods (I and II) were also used respectively in \cite{MN3} to calculate the partition function with $L_1=L_2=L=\zeta_1=\zeta_2=0$ and in \cite{Nosaka:2020tyv} to calculate the partition function of the mass-deformed ABJM theory \cite{HLLLP2,Gomis:2008vc} for higher ranks.
}
Method I allows us to organize the calculations of $\Tr({\widehat D}_1{\widehat D}_2)^n$ and $\langle\!\langle t_{L,r}|{\widehat d}_1({\widehat D}_2{\widehat D}_1)^n{\widehat d}_2|{-t_{L,s}}\rangle\!\rangle$ in a systematic recursive process, although the method requires us to fix $\zeta_1$ such that\footnote{
One can in principle generalize method I to $2i\zeta_1\in \mathbb{Q}$ though the calculation becomes cumbersome.
}
$2i\zeta_1\in \mathbb{Z}$
(namely, we obtain $Z_{k,{\bm M}}(N)$ as a function of $\zeta_2$ but only at discrete points of $\zeta_1$).
Method II allows us to calculate $Z_{k,{\bm M}}(N)$ as a function of $\zeta_1,\zeta_2$, while it is more difficult to program the recursive process for higher ranks.

Note that in method I we perform the calculations by assuming $\zeta_2\in \mathbb{R}$, and define $Z_{k,{\bm M}}(N)$ for $\zeta_2\in \mathbb{C}$ as the continuation of the analytic expressions obtained for $\zeta_2\in \mathbb{R}$.
In method II we assume $\zeta_1,\zeta_2\in \mathbb{R}$ to obtain the analytic expressions for $Z_{k,{\bm M}}(N)$ and define $Z_{k,{\bm M}}(N)$ for generic $\zeta_1,\zeta_2\in \mathbb{C}$ by the continuation.
We observe that the exact values of $Z_{k,{\bm M}}(N)$ calculated in this prescription of analytic continuation\footnote{
The same prescription was adopted in \cite{Nosaka:2020tyv} for the mass parameters of the mass-deformed ABJM theory to obtain the generalized $q$-Toda bilinear relation.
}
are indeed consistent with each other, compatible with the invariance of $\frac{Z_{k,{\bm M}}(N)}{Z_{k,{\bm M}}(0)}$ under the Weyl group and satisfy the 40 bilinear relations \eqref{bilinearwithkappa2}.

In the rest of this appendix after elaborating the two quantities \eqref{trandmatelem}, we turn to the evaluations with the two methods in appendices \ref{method1} and \ref{sec_Okuyama}.

\subsection{$\Tr({\widehat D}_1{\widehat D}_2)^n$ and $\langle\!\langle t_{L,r}|{\widehat d}_1({\widehat D}_2{\widehat D}_1)^n{\widehat d}_2|{-t_{L,s}}\rangle\!\rangle$}
\label{tracesandmatrixelements}

To explain the two methods, let us rewrite the two quantities in \eqref{trandmatelem}, $\Tr({\widehat D}_1{\widehat D}_2)^n$ and $\langle\!\langle t_{L,r}|{\widehat d}_1({\widehat D}_2{\widehat D}_1)^n{\widehat d}_2|{-t_{L,s}}\rangle\!\rangle$, into recursive expressions.
We first consider the common ingredient $\widehat D_1\widehat D_2$.
After applying similarity transformations, it is rewritten as
\begin{align}
{\widehat D}_1{\widehat D}_2\sim {\widehat \rho}_1{\widehat \rho}_2,
\end{align}
where $\widehat\rho_1$ and $\widehat\rho_2$ are given by
\begin{align}
{\widehat \rho}_1=\sqrt{A({\widehat x})}\frac{1}{2\cosh\frac{{\widehat p}}{2}}\sqrt{B({\widehat x})},\quad
{\widehat \rho}_2=\sqrt{B({\widehat x})}\frac{1}{2\cosh\frac{{\widehat p}}{2}}\sqrt{A({\widehat x})},
\label{TBAstructure}
\end{align}
with
\begin{align}
&A(x)=e^{(-\frac{i\zeta_1}{k}-\frac{L}{k})x}\frac{1}{\prod_{r=1}^{k-L_1}2\cosh\frac{x-t_{k-L_1,r}}{2k}}\frac{1}{\prod_{r=1}^{k-L_2}2\cosh\frac{x-t_{k-L_2,r}+2\pi \zeta_2}{2k}},\nonumber \\
&B(x)=e^{(\frac{i\zeta_1}{k}+\frac{L}{k})x}\frac{1}{\prod_{r=1}^{L_1}2\cosh\frac{x-t_{L_1,r}}{2k}}\frac{1}{\prod_{r=1}^{L_2}2\cosh\frac{x-t_{L_2,r}+2\pi\zeta_2}{2k}}.
\label{AB}
\end{align}
The structure \eqref{TBAstructure} allows us to organize the calculation of $\Tr({\widehat D}_1{\widehat D}_2)^n$ in the following way.

First we notice ${\widehat \rho}_1,{\widehat\rho}_2$ satisfy
\begin{align}
e^{\frac{{\widehat x}}{k}}{\widehat\rho}_1+{\widehat\rho}_1e^{\frac{{\widehat x}}{k}}=\sqrt{A({\widehat x})}e^{\frac{{\widehat x}}{2k}}|0\rangle\!\rangle\langle\!\langle 0|e^{\frac{{\widehat x}}{2k}}\sqrt{B({\widehat x})},\quad
e^{\frac{{\widehat x}}{k}}{\widehat\rho}_2+{\widehat\rho}_2e^{\frac{{\widehat x}}{k}}=\sqrt{B({\widehat x})}e^{\frac{{\widehat x}}{2k}}|0\rangle\!\rangle\langle\!\langle 0|e^{\frac{{\widehat x}}{2k}}\sqrt{A({\widehat x})}.
\end{align}
Combining these two relations we obtain
\begin{align}
e^{\frac{{\widehat x}}{k}}{\widehat\rho}_1{\widehat\rho}_2
-{\widehat\rho}_1{\widehat\rho}_2e^{\frac{{\widehat x}}{k}}
=
\sqrt{A({\widehat x})}e^{\frac{{\widehat x}}{2k}}|0\rangle\!\rangle\langle\!\langle 0|e^{\frac{{\widehat x}}{2k}}\sqrt{B({\widehat x})}{\widehat\rho}_2
-{\widehat\rho}_1\sqrt{B({\widehat x})}e^{\frac{{\widehat x}}{2k}}|0\rangle\!\rangle\langle\!\langle 0|e^{\frac{{\widehat x}}{2k}}\sqrt{A({\widehat x})},
\end{align}
which can further be generalized to the relation for $({\widehat\rho}_1{\widehat\rho}_2)^n$ as
\begin{align}
&e^{\frac{{\widehat x}}{k}}({\widehat\rho}_1{\widehat\rho}_2)^n
-({\widehat\rho}_1{\widehat\rho}_2)^ne^{\frac{{\widehat x}}{k}}\nonumber \\
&=
\sum_{\ell=0}^{n-1}
({\widehat\rho}_1{\widehat\rho}_2)^\ell
\Bigl(
\sqrt{A({\widehat x})}e^{\frac{{\widehat x}}{2k}}|0\rangle\!\rangle\langle\!\langle 0|e^{\frac{{\widehat x}}{2k}}\sqrt{B({\widehat x})}{\widehat\rho}_2
-{\widehat\rho}_1\sqrt{B({\widehat x})}e^{\frac{{\widehat x}}{2k}}|0\rangle\!\rangle\langle\!\langle 0|e^{\frac{{\widehat x}}{2k}}\sqrt{A({\widehat x})}
\Bigr)
({\widehat\rho}_1{\widehat\rho}_2)^{n-1-\ell}.
\end{align}
Taking the matrix element of both sides, we obtain
\begin{align}
\langle x|({\widehat\rho}_1{\widehat\rho}_2)^n|y\rangle=\frac{2e^{\frac{x+y}{2k}}\sqrt{A(x)A(y)}}{e^{\frac{x}{k}}-e^{\frac{y}{k}}}\sum_{\ell=0}^{n-1}\phi_{1,\ell}(x)\phi_{2,n-1-\ell}(y),
\end{align}
with
\begin{align}
&\phi_{1,\ell}(x)=\frac{1}{e^{\frac{x}{2k}}\sqrt{A(x)}}\langle x|({\widehat\rho}_1{\widehat\rho}_2)^\ell \sqrt{A({\widehat x})}e^{\frac{{\widehat x}}{2k}}|0\rangle\!\rangle
=\frac{1}{e^{\frac{x}{2k}}\sqrt{A(x)}}\langle\!\langle 0|e^{\frac{{\widehat x}}{2k}}\sqrt{A({\widehat x})}({\widehat\rho}_1{\widehat\rho}_2)^\ell|x\rangle,\nonumber \\
&\phi_{2,\ell}(x)=\frac{1}{e^{\frac{x}{2k}}\sqrt{A(x)}}\langle x|({\widehat\rho}_1{\widehat\rho}_2)^\ell {\widehat\rho}_1 e^{\frac{{\widehat x}}{2k}}\sqrt{B({\widehat x})}|0\rangle\!\rangle
=\frac{1}{e^{\frac{x}{2k}}\sqrt{A(x)}}\langle\!\langle 0|
e^{\frac{{\widehat x}}{2k}}\sqrt{B({\widehat x})}
{\widehat\rho}_2
({\widehat\rho}_1{\widehat\rho}_2)^\ell
|x\rangle.
\label{definephiaell}
\end{align}
Hence $\Tr({\widehat D}_1{\widehat D}_2)^n=\Tr({\widehat \rho}_1{\widehat \rho}_2)^n$ can be written as
\begin{align}
\Tr({\widehat D}_1{\widehat D}_2)^n
&=\int_{-\infty}^\infty \frac{dx}{2\pi}\lim_{y\rightarrow x}\langle x|({\widehat\rho}_1{\widehat\rho}_2)^n|y\rangle\nonumber \\
&=
k\int\frac{dx}{2\pi}A(x)\sum_{\ell=0}^{n-1}
\Bigl(
\frac{d\phi_{1,\ell}(x)}{dx}\phi_{2,n-1-\ell}(x)
-\phi_{1,\ell}(x)\frac{d\phi_{2,n-1-\ell}(x)}{dx}
\Bigr),
\label{TWsystem1}
\end{align}
which we can calculate once we know the state vectors $\phi_{a,\ell}(x)$ with $a=1,2$ and $\ell=0,1,\cdots,n-1$.
Also, these vectors can be generated by the recursion relation
\begin{align}
&\phi_{a,\ell+1}(x)=\int_{-\infty}^\infty\frac{dy}{2\pi}\frac{1}{e^{\frac{x}{2k}}\sqrt{A(x)}}\langle x|{\widehat\rho}_1|y\rangle e^{\frac{y}{2k}}\sqrt{B(y)}{\widetilde \phi}_{a,\ell}(y)
=\frac{e^{-\frac{x}{2k}}}{k}\int\frac{dy}{2\pi}\frac{e^{\frac{y}{2k}}}{2\cosh\frac{x-y}{2k}}B(y){\widetilde \phi}_{a,\ell}(y),\nonumber \\
&{\widetilde \phi}_{a,\ell}(x)=\int_{-\infty}^\infty \frac{dy}{2\pi}\frac{1}{e^{\frac{x}{2k}}\sqrt{B(x)}}\langle x|{\widehat\rho}_2|y\rangle e^{\frac{y}{2k}}\sqrt{A(y)}\phi_{a,\ell}(y)
=\frac{e^{-\frac{x}{2k}}}{k}\int\frac{dy}{2\pi}\frac{e^{\frac{y}{2k}}}{2\cosh\frac{x-y}{2k}}A(y)\phi_{a,\ell}(y),
\label{TWsystem2}
\end{align}
together with the initial data
\begin{align}
&\phi_{1,0}(x)=\langle x|0\rangle\!\rangle=\frac{1}{\sqrt{k}},\nonumber \\
&\phi_{2,0}(x)=\frac{1}{\sqrt{A(x)}e^{\frac{x}{2k}}}\langle x|{\widehat\rho}_1e^{\frac{{\widehat x}}{2k}}\sqrt{B({\widehat x})}|0\rangle\!\rangle
=\frac{e^{-\frac{x}{2k}}}{k\sqrt{k}}\int_{-\infty}^\infty\frac{dy}{2\pi}\frac{e^{\frac{y}{2k}}}{2\cosh\frac{x-y}{2k}}B(y).
\label{TWsystem3}
\end{align}

In the same way, we can organize the calculation of $\langle\!\langle t_{L,r}|{\widehat d}_1({\widehat D}_2{\widehat D}_1)^n{\widehat d}_2|{-t_{L,s}}\rangle\!\rangle$ as
\begin{align}
&\langle\!\langle t_{L,r}|{\widehat d}_1({\widehat D}_2{\widehat D}_1)^n{\widehat d}_2|{-t_{L,s}}\rangle\!\rangle=\int_{-\infty}^\infty\frac{dx}{2\pi}\langle\!\langle t_{L,r}|x\rangle e^{(-\frac{L}{2k}+\frac{1}{2k})x}B(x)\lambda_{s,n}(x),\nonumber\\
&\lambda_{s,n+1}(x)=\int_{-\infty}^\infty\frac{dy}{2\pi k}\frac{e^{\frac{y}{k}}}{e^{\frac{x}{k}}+e^{\frac{y}{k}}}A(y){\widetilde \lambda}_{s,n}(y),\quad
{\widetilde \lambda}_{s,n}(x)=\int_{-\infty}^\infty\frac{dy}{2\pi k}\frac{e^{\frac{y}{k}}}{e^{\frac{x}{k}}+e^{\frac{y}{k}}}B(y)\lambda_{s,n}(y),\nonumber\\
&\lambda_{s,0}(x)=e^{(-\frac{L}{2k}-\frac{1}{2k})x}\langle x|{-t_{L,s}}\rangle\!\rangle.
\label{TWsystemlambda}
\end{align}
Hence we can calculate $\langle\!\langle t_{L,r}|{\widehat d}_1({\widehat D}_2{\widehat D}_1)^n{\widehat d}_2|{-t_{L,s}}\rangle\!\rangle$ from $\lambda_{s,n}(x)$, which, in turn, can be calculated recursively in $n$ starting from the initial data $\lambda_{s,0}(x)$.

\subsection{Method I}
\label{method1}

In this subsection, we explain how the two quantities \eqref{trandmatelem} are further computed when $2i\zeta_1\in\mathbb{Z}$ using method I, and clarify when the computations are valid.
Using these results, we can then check the bilinear relations \eqref{bilinearwithkappa2} in appendix \ref{checkedlist}.

\subsubsection{$\Tr({\widehat D}_1{\widehat D}_2)^n$}
\label{sec_PY1}
 
To proceed further, let us assume $2i\zeta_1\in\mathbb{Z}$ and define new variables $u=e^{\frac{x}{2k}},v=e^{\frac{y}{2k}}$ from $x,y$ to write the recursive system \eqref{TWsystem1},\eqref{TWsystem2},\eqref{TWsystem3} as
\begin{align}
&\Tr({\widehat D}_1{\widehat D}_2)^n=\frac{k}{2\pi}\int_0^\infty duA(u)\sum_{\ell=0}^{n-1}\Bigl(
\frac{d\phi_{1,\ell}(u)}{du}\phi_{2,n-1-\ell}(u)
-\phi_{1,\ell}(u)\frac{d\phi_{2,n-1-\ell}(u)}{du}\Bigr),\nonumber \\
&{\widetilde \phi}_{a,\ell}(u)=\frac{1}{\pi}\int_0^\infty dv\frac{v}{u^2+v^2}A(v)\phi_{a,\ell}(v),\quad
\phi_{a,\ell+1}(u)=\frac{1}{\pi}\int_0^\infty dv\frac{v}{u^2+v^2}B(v){\widetilde \phi}_{a,\ell}(v),\nonumber \\
&\phi_{1,0}(u)=\frac{1}{\sqrt{k}},\quad
\phi_{2,0}(u)=\frac{1}{\pi\sqrt{k}}\int_0^\infty dv\frac{v}{u^2+v^2}B(v),
\label{TWsysteminu}
\end{align}
where $A(u)$ and $B(u)$ are given by
\begin{align}
&A(u)=e^{-\frac{\pi (k-L_2)\zeta_2}{k}}\frac{u^{2k-\Lambda}}{
\prod_{r=1}^{k-L_1}(u^2+e^{\frac{1}{k}t_{k-L_1,r}})
\prod_{r=1}^{k-L_2}(u^2+e^{\frac{1}{k}t_{k-L_2,r}-\frac{2\pi \zeta_2}{k}})
},\nonumber \\
&B(u)=e^{-\frac{\pi L_2 \zeta_2}{k}}\frac{u^\Lambda}{
\prod_{r=1}^{L_1}(u^2+e^{\frac{1}{k}t_{L_1,r}})
\prod_{r=1}^{L_2}(u^2+e^{\frac{1}{k}t_{L_2,r}-\frac{2\pi \zeta_2}{k}})
},
\label{AuBu}
\end{align}
with $\Lambda=L_1+L_2+2L+2i\zeta_1$ ($\in\mathbb{Z}$).
By using the formula \cite{PY}
\begin{align}
\int_0^\infty duf(u) (\log u)^j=-\frac{(2\pi i)^{j+1}}{j+1}\sum_{w\in\mathbb{C}\smallsetminus \mathbb{R}_{\ge 0}} \text{Res}\Bigl[f(u)B_{j+1}\Bigl(\frac{\log^{(+)}u}{2\pi i}\Bigr),u\rightarrow w\Bigr],
\end{align}
where $f(u)$ is any meromorphic function of $u$ without branch cuts, $\log^{(+)}$ is the logarithm with the branch cut on $\mathbb{R}_{\ge 0}$ and $\{B_j(u)\}_j$ is any set of polynomials satisfying $B_{j+1}(u+1)-B_{j+1}(u)=(j+1)u^j$, we can further write the integrations in \eqref{TWsysteminu} as residue sums
\begin{align}
&\Tr({\widehat D}_1{\widehat D}_2)^n=\frac{k}{2\pi}\sum_{j\ge 0} \Bigl(-\frac{(2\pi i)^{j+1}}{j+1}\sum_{w\in\mathbb{C}\smallsetminus\mathbb{R}_{\ge 0}}\text{Res}\Bigl[A(u)\Psi_{n,j}(u)B_{j+1}\Bigl(\frac{\log^{(+)}u}{2\pi i}\Bigr),u\rightarrow w\Bigr]\Bigr),\nonumber\\
&{\widetilde \phi}_{a,\ell}(u)=\frac{1}{\pi}\sum_{j\ge 0}\Bigl(-\frac{(2\pi i)^{j+1}}{j+1}\sum_{w\in\mathbb{C}\smallsetminus\mathbb{R}_{\ge 0}}\text{Res}\Bigl[\frac{v}{u^2+v^2}A(v)\phi_{a,\ell,j}(v)B_{j+1}\Bigl(\frac{\log^{(+)}v}{2\pi i}\Bigr),v\rightarrow w\Bigr]\Bigr),\nonumber\\
&\phi_{a,\ell+1}(u)=\frac{1}{\pi}\sum_{j\ge 0}\Bigl(-\frac{(2\pi i)^{j+1}}{j+1}\sum_{w\in\mathbb{C}\smallsetminus\mathbb{R}_{\ge 0}}\text{Res}\Bigl[\frac{v}{u^2+v^2}B(v){\widetilde \phi}_{a,\ell,j}(v)B_{j+1}\Bigl(\frac{\log^{(+)}v}{2\pi i}\Bigr),v\rightarrow w\Bigr]\Bigr),\nonumber \\
&\phi_{1,0}(u)=\frac{1}{\sqrt{k}},\quad
\phi_{2,0}(u)=\frac{1}{\pi\sqrt{k}}\Bigl(-2\pi i\sum_{w\in\mathbb{C}\smallsetminus\mathbb{R}_{\ge 0}}\text{Res}\Bigl[\frac{v}{u^2+v^2}B(v)B_1\Bigl(\frac{\log^{(+)}v}{2\pi i}\Bigr),v\rightarrow w\Bigr]\Bigr),
\label{PYresidues}
\end{align}
where $\Psi_{n,j}(u)$, $\phi_{a,\ell,j}(u)$, ${\widetilde \phi}_{a,\ell,j}(u)$ are defined by
\begin{align}
&\sum_{\ell=0}^{n-1}\Bigl(\frac{d\phi_{1,\ell}(u)}{du}\phi_{2,n-1-\ell}(u)
-\phi_{1,\ell}(u)\frac{d\phi_{2,n-1-\ell}(u)}{du}\Bigr)=
\sum_{j\ge 0}\Psi_{n,j}(u)(\log u)^j,\nonumber \\
&\phi_{a,\ell}(u)=\sum_{j\ge 0}\phi_{a,\ell,j}(u)(\log u)^j,\quad
{\widetilde \phi}_{a,\ell}(u)=\sum_{j\ge 0}{\widetilde \phi}_{a,\ell,j}(u)(\log u)^j.
\end{align}
A sufficient set of the poles for the residues to be picked in the calculation of $\phi_{a,\ell}(u)$ and ${\widetilde \phi}_{a,\ell}(u)$ is the union of $\{iu,-iu\}$ and
\begin{align}
&\{e^{\frac{1}{2k}t_{L_1,r}},
ie^{\frac{1}{2k}t_{L_1,r}},
-e^{\frac{1}{2k}t_{L_1,r}},
-ie^{\frac{1}{2k}t_{L_1,r}}\}_{r=1}^{L_1}\nonumber\\
&\quad \cup \{e^{\frac{1}{2k}t_{k-L_1,r}},
ie^{\frac{1}{2k}t_{k-L_1,r}},
-e^{\frac{1}{2k}t_{k-L_1,r}},
-ie^{\frac{1}{2k}t_{k-L_1,r}}\}_{r=1}^{k-L_1}\nonumber\\
&\quad \cup \{e^{\frac{1}{2k}t_{L_2,r}-\frac{\pi\zeta_2}{k}},
ie^{\frac{1}{2k}t_{L_2,r}-\frac{\pi\zeta_2}{k}},
-e^{\frac{1}{2k}t_{L_2,r}-\frac{\pi\zeta_2}{k}},
-ie^{\frac{1}{2k}t_{L_2,r}-\frac{\pi\zeta_2}{k}}\}_{r=1}^{L_2}\nonumber\\
&\quad \cup \{e^{\frac{1}{2k}t_{k-L_2,r}-\frac{\pi\zeta_2}{k}},
ie^{\frac{1}{2k}t_{k-L_2,r}-\frac{\pi\zeta_2}{k}},
-e^{\frac{1}{2k}t_{k-L_2,r}-\frac{\pi\zeta_2}{k}},
-ie^{\frac{1}{2k}t_{k-L_2,r}-\frac{\pi\zeta_2}{k}}\}_{r=1}^{k-L_2},
\label{sufficientpoles221221}
\end{align}
while a sufficient set of the poles for the calculation of $\Tr({\widehat D}_1{\widehat D}_2)$ is \eqref{sufficientpoles221221} itself.

\subsubsection{$\langle\!\langle t_{L,r}|{\widehat d}_1({\widehat D}_2{\widehat D}_1)^n{\widehat d}_2|{-t_{L,s}}\rangle\!\rangle$}
\label{sec_PY2}

When $2 i\zeta_1\in\mathbb{Z}$ we can compute $\langle\!\langle t_{L,r}|{\widehat d}_1({\widehat D}_2{\widehat D}_1)^n{\widehat d}_2|{-t_{L,s}}\rangle\!\rangle$ by the same technique as in \eqref{PYresidues}
\begin{align}
&\langle\!\langle t_{L,r}|{\widehat d}_1({\widehat D}_2{\widehat D}_1)^n{\widehat d}_2|{-t_{L,s}}\rangle\!\rangle\nonumber \\
&\quad =\frac{\sqrt{k}}{\pi}\sum_{j\ge 0}\Bigl(-\frac{(2\pi i)^{j+1}}{j+1}\sum_{w\in\mathbb{C}\smallsetminus\mathbb{R}_{\ge 0}}\text{Res}\Bigl[u^{1-2r}B(u)\lambda_{s,n,j}(u)B_{j+1}\Bigl(\frac{\log^{(+)}u}{2\pi i}\Bigr),u\rightarrow w\Bigr]\Bigr),\nonumber \\
&{\widetilde \lambda}_{s,n}(u)=\frac{1}{\pi}\sum_{j\ge 0}\Bigl(-\frac{(2\pi i)^{j+1}}{j+1}\sum_{w\in\mathbb{C}\smallsetminus\mathbb{R}_{\ge 0}}\text{Res}\Bigl[\frac{v}{u^2+v^2}B(v)\lambda_{s,n,j}(v)B_{j+1}\Bigl(\frac{\log^{(+)}v}{2\pi i}\Bigr),v\rightarrow w\Bigr]\Bigr),\nonumber \\
&\lambda_{s,n+1}(u)=\frac{1}{\pi}\sum_{j\ge 0}\Bigl(-\frac{(2\pi i)^{j+1}}{j+1}\sum_{w\in\mathbb{C}\smallsetminus\mathbb{R}_{\ge 0}}\text{Res}\Bigl[\frac{v}{u^2+v^2}A(v){\widetilde \lambda}_{s,n,j}(v)B_{j+1}\Bigl(\frac{\log^{(+)}v}{2\pi i}\Bigr),v\rightarrow w\Bigr]\Bigr),\nonumber \\
&\lambda_{s,0}(u)=\frac{1}{\sqrt{k}}u^{-2s},
\end{align}
where $\lambda_{s,n,j}(u)$ and ${\widetilde \lambda}_{s,n,j}(u)$ are defined by
\begin{align}
\lambda_{s,n}(u)=\sum_{j\ge 0}\lambda_{s,n,j}(u)(\log u)^j,\quad
{\widetilde \lambda}_{s,n}(u)=\sum_{j\ge 0}{\widetilde \lambda}_{s,n,j}(u)(\log u)^j.
\end{align}
A sufficient set of the poles to be picked in the calculation of $\lambda_{s,n}(u)$ and ${\widetilde \lambda}_{s,n}(u)$ is the union of $\{iu,-iu\}$ and \eqref{sufficientpoles221221}, while a sufficient set of the poles for the calculation of $\langle\!\langle t_{L,r}|{\widehat d}_1({\widehat D}_2{\widehat D}_1)^n{\widehat d}_2|{-t_{L,s}}\rangle\!\rangle$ is \eqref{sufficientpoles221221}.

\subsubsection{Domain of validity}
\label{app_validity}
Note that in order to ensure that method I gives the correct answers, the parameters $k,L_1,L_2,L,\zeta_1$ have to satisfy the following conditions.
\begin{itemize}
\item The derivation of \eqref{openclosed} was based on the assumptions
\begin{align}
k,L_1,L_2,L\in\mathbb{Z},\quad 0\le L\le L_1,L_2\le k.
\label{fundamentalcondition}
\end{align}
\item In method I we have assumed $2i\zeta_1\in\mathbb{Z}$.
\item The calculations of the integrations by residue sums in method I are trustable only when the original integrations are convergent at $x\rightarrow\pm\infty$ corresponding to $u\rightarrow 0,\infty$.
\end{itemize}
Here let us look at the conditions required for the validity of the calculations of $Z_{k,{\bm M}}(0)$ and $Z_{k,{\bm M}}(1)$, which involve the calculations of the quantities
\begin{align}
\phi_{2,0}(u),\quad
\Tr{\widehat D}_1{\widehat D}_2,\quad
\langle\!\langle t_{L,r}|{\widehat d}_1{\widehat d}_2|{-t_{L,s}}\rangle\!\rangle,\quad
{\widetilde \lambda}_{0,s}(u),\quad
\lambda_{1,s}(u),\quad
\langle\!\langle t_{L,r}|{\widehat d}_1{\widehat D}_2{\widehat D}_1{\widehat d}_2|{-t_{L,s}}\rangle\!\rangle,
\end{align}
with the condition \eqref{fundamentalcondition} assumed.
First let us look at the integration for $\phi_{2,0}(u)$ \eqref{TWsysteminu}, where the integrand behaves at $v\rightarrow 0,\infty$ as
\begin{align}
\frac{1}{\pi\sqrt{k}}\frac{v}{u^2+v^2}B(v)\sim\begin{cases}
v^{L_1+L_2+2L+2i\zeta_1},&\quad (v\rightarrow 0),\\
v^{-1-L_1-L_2+2L+2i\zeta_1},&\quad (v\rightarrow \infty).
\end{cases}
\end{align}
To ensure the convergence of the integration $\int_0^\infty dv$, the integrand must behave as $\sim v^p$ with $p>-1$ at $v\rightarrow 0$ and as $\sim v^q$ with $q<-1$ at $v\rightarrow\infty$.
Therefore, we obtain the condition
\begin{align}
L_1+L_2+2L+2i\zeta_1>-2,\quad
L_1+L_2-2L-2i\zeta_1>0,
\label{phi20condition}
\end{align}
for the validity of $\phi_{2,0}(u)$.
Next let us look at the integration for $\Tr{\widehat D}_1{\widehat D}_2$ \eqref{TWsysteminu}.
Once we assume \eqref{phi20condition}, from the residue formula \eqref{PYresidues} we obtain the asymptotics of $\phi_{2,0}(u)$ as
\begin{align}
\phi_{2,0}(u)\sim\begin{cases}
u^{\text{min}(L_1+L_2+2L+2i\zeta_1,0)},&\quad (u\rightarrow 0),\\
u^{\text{max}(-L_1-L_2+2L+2i\zeta_1,-2)},&\quad (u\rightarrow \infty).
\end{cases}
\end{align}
Substituting this to \eqref{TWsysteminu} and examining the asymptotics of the integrand, we end up with the conditions
\begin{align}
L_1+L_2+2L+2i\zeta_1<2k,\quad
L_1+L_2-2L-2i\zeta_1<2k+2,
\label{TrD1D2condition}
\end{align}
for the validity of $\Tr{\widehat D}_1{\widehat D}_2$.

When $L>0$, there are additional constraints from the calculations of $\langle\!\langle t_{L,r}|{\widehat d}_1{\widehat d}_2|{-t_{L,s}}\rangle\!\rangle$, ${\widetilde \lambda}_{0,s}(u)$, $\lambda_{1,s}(u)$ and $\langle\!\langle t_{L,r}|{\widehat d}_1{\widehat D}_2{\widehat D}_1{\widehat d}_2|{-t_{L,s}}\rangle\!\rangle$.
From the convergence of the integration for $\langle\!\langle t_{L,r}|{\widehat d}_1{\widehat d}_2|{-t_{L,s}}\rangle\!\rangle$ given by the integration
\begin{align}
\langle\!\langle t_{L,r}|{\widehat d}_1{\widehat d}_2|{-t_{L,s}}\rangle\!\rangle=\int_{-\infty}^\infty\frac{dx}{2\pi k}\frac{e^{\frac{1}{k}(L+1-r-s+i\zeta_1)x}}{
\prod_{r=1}^{L_1}2\cosh\frac{x-t_{L_1,r}}{2k}
\prod_{r=1}^{L_2}2\cosh\frac{x+2\pi\zeta_2-t_{L_2,r}}{2k}},
\end{align}
the condition
\begin{align}
L+1-r-s+i\zeta_1-\frac{L_1+L_2}{2}<0,\quad
-L-1+r+s-i\zeta_1-\frac{L_1+L_2}{2}<0,\quad (1\le r,s\le L),
\label{Z(0)condition}
\end{align}
is required for the convergence at $x\rightarrow\pm\infty$.
Since the first condition (from $x\rightarrow\infty$) and the second condition (from $x\rightarrow-\infty$) are the most stringent respectively at $r=s=1$ and $r=s=L$, it is sufficient to require the condition \eqref{Z(0)condition} at these values of $r,s$.
Hence we can reduce \eqref{Z(0)condition} as
\begin{align}
L_1+L_2-2L-2i\zeta_1>-2,\quad
L_1+L_2-2L+2i\zeta_1>-2.
\label{Z(0)conditionreduced}
\end{align}
Once this condition (or more generally \eqref{Z(0)conditiongeneral}) is satisfied, $\langle\!\langle t_{L,r}|{\widehat d}_1{\widehat d}_2|{-t_{L,s}}\rangle\!\rangle$ can be calculated by the formula \eqref{Inalphabeta}.
From the convergence of the integration for ${\widetilde \lambda}_{0,s}(u)$ \eqref{TWsystemlambda} we obtain the conditions
\begin{align}
1-2s+L_1+L_2+2L+2i\zeta_1>-1,\quad
-1-2s-L_1-L_2+2L+2i\zeta_1<-1,
\label{lambdatilde0condition}
\end{align}
where the first condition is for the convergence at $v\rightarrow 0$ and the second condition is for the convergence at $v\rightarrow\infty$.
Again,
the condition from $v\rightarrow 0$ is the most stringent for $s=L$ and the condition from $v\rightarrow \infty$ is the most stringent for $s=1$.
Hence it is sufficient to require the conditions \eqref{lambdatilde0condition} only at these two values of $s$, as
\begin{align}
L_1+L_2+2i\zeta_1>-2,\quad
L_1+L_2-2L-2i\zeta_1>-2.
\label{lambdatiode0conditionreduced}
\end{align}
Convergence conditions for $\lambda_{1,s}(u)$ and $\langle\!\langle t_{L,r}|{\widehat d}_1{\widehat D}_2{\widehat D}_1{\widehat d}_2|{-t_{L,s}}\rangle\!\rangle$ can be studied similarly.
Although the calculations are more involved than those demonstrated so far, they can be slightly simplified by noticing that it is sufficient to require the convergence around $u=0$ only for $r=s=L$ and the convergence around $u\rightarrow\infty$ only for $r=s=1$.
Here we only display the final results
\begin{align}
\lambda_{1,s}(u)&:\quad L_1+L_2+4L+2i\zeta_1<2k+2,\quad L_1+L_2-2L-2i\zeta_1<2k+4,\nonumber \\
\langle\!\langle t_{L,r}|{\widehat d}_1{\widehat D}_2{\widehat D}_1{\widehat d}_2|{-t_{L,s}}\rangle\!\rangle&:\quad L_1+L_2+2i\zeta_1>-2,\quad L_1+L_2-2L-2i\zeta_1>-2.
\label{lambda1and<<dDDd>>condition}
\end{align}

To summarize, the conditions required for the method I to be trustable for $Z_{k,{\bm M}}(0)$ and $Z_{k,{\bm M}}(1)$ are
\begin{align}
\eqref{phi20condition}
\text{ AND }
\eqref{TrD1D2condition}
\text{ AND }\big[L=0\text{ OR }[\eqref{Z(0)conditionreduced}\text{ AND }\eqref{lambdatiode0conditionreduced}\text{ AND }\eqref{lambda1and<<dDDd>>condition}]\big],
\label{allconditionsforZ(1)}
\end{align}
which further reduces under the condition \eqref{fundamentalcondition}.
Finally, we obtain the conditions
\begin{align}
&k,L_1,L_2,L,2i\zeta_1\in\mathbb{Z},\quad 0\le L\le L_1,L_2\le k,\nonumber \\
&2i\zeta_1>-\text{min}(2k-L_1-L_2+2L,L_1+L_2-2L)-2,\nonumber \\
&2i\zeta_1<\text{min}(L_1+L_2-2L,2k-L_1-L_2+2-4L,2k-L_1-L_2).
\label{validitycondition}
\end{align}

\subsection{Method II}
\label{sec_Okuyama}

We can also calculate the integrations in appendix \ref{tracesandmatrixelements} such as $\Tr({\widehat D}_1{\widehat D}_2)^n$, $\phi_{a,\ell}(x)$, ${\widetilde \phi}_{a,\ell}(x)$ and $\langle\!\langle t_{L,r}|{\widehat d}_1({\widehat D}_2{\widehat D}_1)^n{\widehat d}_2|{-t_{L,s}}\rangle\!\rangle$, $\lambda_{s,n}(x)$, ${\widetilde \lambda}_{s,n}(x)$ without fixing $\zeta_1$, as in \cite{Okuyama:2011su,Nosaka:2020tyv}.

This method is a repeated use of the formula (which is a generalization of the integration formula \eqref{Inalphabeta})
\begin{align}
I_{k,n}(\alpha,\{\beta_a+i\gamma_a\})&=\strokedint_{-\infty}^\infty\frac{dx}{2\pi k}\frac{e^{\frac{\alpha x}{k}}}{\prod_{a=1}^{n}2\cosh\frac{x-\beta_a-i\gamma_a}{2k}}\nonumber \\
&=\frac{1}{e^{-\pi i\alpha}-(-1)^n e^{\pi i\alpha}}\Biggl(
\sum_{\substack{a\\ (\gamma_a>-\pi k)}}\frac{e^{\frac{\alpha(\beta_a+i\gamma_a)}{k}}}{\prod_{b(\neq a)}2\cosh\frac{\beta_a-\beta_b+i(\gamma_a-\gamma_b+\pi k)}{2k}}\nonumber \\
&\quad +
\frac{e^{-\pi i\alpha}+(-1)^{n}e^{\pi i\alpha}}{2}
\sum_{\substack{a\\ (\gamma_a=-\pi k)}}\frac{e^{\frac{\alpha\beta_a}{k}}}{\prod_{b(\neq a)}2\cosh\frac{\beta_a-\beta_b-i\gamma_b}{2k}}
\Biggr),
\label{Inalphabetagamma}
\end{align}
to the integrations in \eqref{TWsystem1}, \eqref{TWsystem2}, \eqref{TWsystem3}.
Here $\beta_a,\gamma_a$ are real numbers with $-\pi k\le \gamma_a<\pi k$ and the symbol $\strokedintnotdisplay$ stands for the integration with the principal-value prescription with poles on the real axis.
Note that when some of $\beta_a+i\gamma_a$ coincide, we have to resolve the degeneracy by shifting the degenerate values of $\beta_a$ before applying the formula \eqref{Inalphabetagamma}, and take the unresolved limit at the end of the calculation.
Similarly, when $e^{-\pi i\alpha}-(-1)^n e^{\pi i\alpha}=0$, we should first shift $\alpha$, apply the formula \eqref{Inalphabetagamma} and take the limit back to its original value afterwards.

Let us explain with examples how method II works by calculating $\phi_{2,0}$, $\Tr{\widehat D}_1{\widehat D}_2$ and so on for the partition functions at $(k,L_1,L_2,L)=(1,0,0,0)$, $(1,1,0,0)$ and $(1,1,1,1)$.

\subsubsection{$(k,L_1,L_2,L)=(1,0,0,0)$}
First let us consider the case $(k,L_1,L_2,L)=(1,0,0,0)$.
From \eqref{TWsystem1} and \eqref{TWsystem3} we have
\begin{align}
\Tr{\widehat D}_1{\widehat D}_2=-\int_{-\infty}^\infty\frac{dx}{2\pi} A(x)\frac{d\phi_{2,0}(x)}{dx},
\label{method2example1}
\end{align}
with
\begin{align}
&\phi_{2,0}(x)=e^{-\frac{x}{2}}\int_{-\infty}^\infty\frac{dy}{2\pi}\frac{e^{\frac{y}{2}}}{2\cosh\frac{x-y}{2}}B(y),\label{phi20method21000} \\
&A(x)=\frac{e^{-i\zeta_1x}}{(2\cosh\frac{x}{2})(2\cosh\frac{x+2\pi \zeta_2}{2})},\quad
B(x)=e^{i\zeta_1x}.\label{ABmethod21000}
\end{align}
For $\phi_{2,0}(x)$ we can apply the formula \eqref{Inalphabetagamma} (or simply \eqref{Inalphabeta} for the current case) immediately to obtain
\begin{align}
\phi_{2,0}(x)=e^{-\frac{x}{2}}\int_{-\infty}^\infty \frac{dy}{2\pi}\frac{e^{(i\zeta_1+\frac{1}{2})y}}{2\cosh\frac{y-x}{2}}
=e^{-\frac{x}{2}}I_{1,1}\bigl({\textstyle i\zeta_1+\frac{1}{2}},\{x\}\bigr)=\frac{ie^{i\zeta_1x}}{2\sinh\pi \zeta_1}.
\end{align}
Substituting this to $\Tr{\widehat D}_1{\widehat D}_2$ \eqref{method2example1}, we obtain
\begin{align}
\Tr{\widehat D}_1{\widehat D}_2=\frac{\zeta_1}{2\sinh\pi \zeta_1}\int_{-\infty}^\infty\frac{dx}{2\pi}\frac{1}{(2\cosh\frac{x}{2})(2\cosh\frac{x+2\pi \zeta_2}{2})}
=
\frac{\zeta_1}{2\sinh\pi \zeta_1}
I_{1,2}(0,\{0,-2\pi \zeta_2\}),
\end{align}
which is again of the form \eqref{Inalphabetagamma}.
Note that here we assume $\zeta_2\in\mathbb{R}$.
As explained above, $I_{1,2}(0,\{0,-2\pi \zeta_2\})$ should be interpreted with the process of introducing $\alpha\neq 0$ and then sending $\alpha\rightarrow 0$ after applying the formula \eqref{Inalphabetagamma}, which gives
\begin{align}
I_{1,2}(0,\{0,-2\pi \zeta_2\})
=\lim_{\alpha\rightarrow 0}I_{1,2}(\alpha,\{0,-2\pi \zeta_2\})
=\lim_{\alpha\rightarrow 0}\frac{1-e^{-2\pi \alpha \zeta_2}}{4\sin\pi\alpha \sinh\pi \zeta_2}
=\frac{\zeta_2}{2\sinh\pi \zeta_2}.
\end{align}
Hence we finally obtain
\begin{align}
\Tr{\widehat D}_1{\widehat D}_2=\frac{\zeta_1\zeta_2}{4\sinh\pi \zeta_1 \sinh\pi \zeta_2}.
\end{align}

\subsubsection{$(k,L_1,L_2,L)=(1,1,0,0)$}

Next let us consider the case $(k,L_1,L_2,L)=(1,1,0,0)$.
As in the previous case $\Tr{\widehat D}_1{\widehat D}_2$ can be calculated through \eqref{method2example1} with \eqref{phi20method21000}, with $A(x)$ and $B(x)$ \eqref{AB} replaced with
\begin{align}
A(x)=\frac{e^{-i\zeta_1x}}{2\cosh\frac{x+2\pi \zeta_2}{2}},\quad
B(x)=\frac{e^{i\zeta_1x}}{2\cosh\frac{x}{2}}.
\label{ABmethod2example2}
\end{align}
Again it is straightforward to apply the formula \eqref{Inalphabetagamma} to $\phi_{2,0}(x)$ \eqref{phi20method21000}, and obtain
\begin{align}
\phi_{2,0}(x)=\int_{-\infty}^\infty \frac{dy}{2\pi}\frac{e^{-\frac{x}{2}+(i\zeta_1+\frac{1}{2})y}}{(2\cosh\frac{y-x}{2})(2\cosh\frac{y}{2})}
=e^{-\frac{x}{2}}I_{1,2}\bigl({\textstyle i\zeta_1+\frac{1}{2}},\{x,0\}\bigr)
=\frac{1}{2\cosh\pi \zeta_1}\frac{e^{i\zeta_1x}-e^{-\frac{x}{2}}}{2\sinh\frac{x}{2}}.
\end{align}
Before substituting this into $\Tr{\widehat D}_1{\widehat D}_2$ \eqref{method2example1}, to reduce the calculations we integrate \eqref{method2example1} by parts and replace the derivative by $\frac{dA(x)}{dx}=(-i\zeta_1+\frac{1}{2\pi}\frac{d}{d\zeta_2})A(x)$, which is valid for $A(x)$ given in \eqref{ABmethod2example2}, as
\begin{align}
\Tr{\widehat D}_1{\widehat D}_2=\frac{1}{2\cosh\pi\zeta_1}\Bigl(-i\zeta_1+\frac{1}{2\pi}\frac{d}{d\zeta_2}\Bigr)\int_{-\infty}^\infty \frac{dx}{2\pi}A(x)\frac{e^{i\zeta_1x}-e^{-\frac{x}{2}}}{2\sinh\frac{x}{2}}.
\label{trD1D21100integral}
\end{align}
Notice that the integrand is finite and smooth at $x=0$ in total, although each term diverges at $x=0$.
This implies that the integration \eqref{trD1D21100integral} is the same as the integration with $x=0$ treated by the principal-value prescription, which can be calculated separately for each term by using $I_{1,2}(\alpha,\{\beta_a\})$ \eqref{Inalphabetagamma} as
\begin{align}
\Tr{\widehat D}_1{\widehat D}_2=\frac{i}{2\cosh\pi\zeta_1}\Bigl(-i\zeta_1+\frac{1}{2\pi}\frac{d}{d\zeta_2}\Bigr)\bigl[I_{1,2}(0,\{-2\pi \zeta_2,-\pi i\})-I_{1,2}({\textstyle -i\zeta_1-\frac{1}{2}},\{-2\pi\zeta_2,-\pi i\})\bigr].
\end{align}
Evaluating $I_{1,2}(0,\{-2\pi\zeta_2,-\pi i\})$ by regularizing $\alpha=0$, we finally obtain
\begin{align}
&\Tr({\widehat D}_1{\widehat D}_2)=\frac{1}{16\pi\cosh\pi\zeta_1\cosh\pi\zeta_2}\nonumber \\
&\quad \times \Bigl[-2+\pi i(2\zeta_1-i\tanh\pi\zeta_2)(2\zeta_2-i\tanh\pi\zeta_1)+\frac{\pi e^{2\pi i\zeta_1\zeta_2}}{\cosh\pi\zeta_1\cosh\pi\zeta_2}\Bigr].
\label{TrD1D21100}
\end{align}

\subsubsection{$(k,L_1,L_2,L)=(1,1,1,1)$}
Lastly, let us look into the calculation of the partition function at $N=1$, $Z_{k,{\bm M}}(1)$, with $(k,L_1,L_2,L)=(1,1,1,1)$.
From the grand partition function \eqref{openclosed}, $Z_{k,{\bm M}}(1)$ is given as
\begin{align}
\frac{Z_{k,{\bm M}}(1)}{Z_{k,{\bm M}}(0)}=\Tr{\widehat D}_1{\widehat D}_2
-\frac{
\langle\!\langle t_{1,1}|{\widehat d}_1{\widehat D}_2{\widehat D}_1{\widehat d}_2|{-t_{1,1}}\rangle\!\rangle
}{
\langle\!\langle t_{1,1}|{\widehat d}_1{\widehat d}_2|{-t_{1,1}}\rangle\!\rangle}.
\label{1111Z(1)/Z(0)}
\end{align}
The denominator $\langle\!\langle t_{1,1}|{\widehat d}_1{\widehat d}_2|{-t_{1,1}}\rangle\!\rangle$ is calculated by the formula $I_{k,n}(\alpha,\{\beta_a\})$ \eqref{Inalphabetagamma} as
\begin{align}
\langle\!\langle t_{1,1}|{\widehat d}_1{\widehat d}_2|{-t_{1,1}}\rangle\!\rangle=I_{1,1}(i\zeta_1,\{0,-2\pi\zeta_2\})=-\frac{i(1-e^{-2\pi i\zeta_1\zeta_2})}{(2\sinh\pi\zeta_1)(2\sinh\pi\zeta_2)}.
\label{1111<<|d1d2|>>}
\end{align}
The calculation of $\Tr{\widehat D}_1{\widehat D}_2$ goes straightforwardly as above, which ends up with
\begin{align}
\Tr{\widehat D}_1{\widehat D}_2=\frac{-\zeta_1\zeta_2+i\zeta_2}{4\sinh\pi\zeta_1\sinh\pi\zeta_2}.
\label{1111TrD1D2}
\end{align}
To calculate $\langle\!\langle t_{1,1}|{\widehat d}_1{\widehat D}_2{\widehat D}_1{\widehat d}_2|{-t_{1,1}}\rangle\!\rangle$, let us insert the identity for a complete set of basis between each pair of operators and find\footnote{
Here we adopt a different calculation for $\langle \!\langle t_{L,r}|{\widehat d}_1{\widehat D}_2{\widehat D}_1{\widehat d}_2|{-t_{L,s}}\rangle\!\rangle$ which seems simpler for $(k,L_1,L_2,L)=(1,1,1,1)$, compared with the recursive calculations of $\lambda_{s,0}(u),{\widetilde \lambda}_{s,0}(u),\lambda_{s,1}(u)$ explained in \eqref{TWsystemlambda}.
}
\begin{align}
\langle\!\langle t_{1,1}|{\widehat d}_1{\widehat D}_2{\widehat D}_1{\widehat d}_2|{-t_{1,1}}\rangle\!\rangle
=
\int_{-\infty}^\infty\frac{dx}{2\pi}
\frac{e^{i\zeta_1x}}{2\cosh\frac{x}{2}}
\frac{dy}{2\pi}
\frac{1}{2\cosh\frac{y+2\pi \zeta_2}{2}}
\langle x|{\widehat D}_2{\widehat D}_1|y\rangle,
\label{<<|dDDd|>>method2}
\end{align}
with
\begin{align}
\langle x|{\widehat D}_2{\widehat D}_1|y\rangle
=
\frac{e^{\frac{x}{2}}}{2\cosh\frac{x+2\pi i\zeta_2}{2}}
\frac{e^{(i\zeta_1+\frac{1}{2})y}}{2\cosh\frac{y}{2}}
\int_{-\infty}^\infty\frac{dz}{2\pi}
\langle x|\frac{1}{2\cosh\frac{{\widehat p}}{2}}|z\rangle e^{(-i\zeta_1-1)z}\langle z|\frac{1}{2\cosh\frac{{\widehat p}}{2}}|y\rangle.
\label{<x|D2D1|y>}
\end{align}
The $z$-integration in $\langle x|{\widehat D}_2{\widehat D}_1|y\rangle$ \eqref{<x|D2D1|y>} can be performed by the formula \eqref{Inalphabetagamma} as
\begin{align}
&\int_{-\infty}^\infty\frac{dz}{2\pi}
\langle x|\frac{1}{2\cosh\frac{{\widehat p}}{2}}|z\rangle e^{(-i\zeta_1-1)z}\langle z|\frac{1}{2\cosh\frac{{\widehat p}}{2}}|y\rangle\nonumber \\
&\quad =I_{1,2}(-i\zeta_1-1,\{x,y\})
=-\frac{i}{2\sinh\pi \zeta_1}\frac{e^{(-i\zeta_1-1)x}-e^{(-i\zeta_1-1)y}}{2\sinh\frac{x-y}{2}}.
\label{zintegrationin<x|D2D1|y>}
\end{align}
Since this is regular at $x=y$, we can calculate $\langle\!\langle t_{1,1}|{\widehat d}_1{\widehat D}_2{\widehat D}_1{\widehat d}_2|{-t_{1,1}}\rangle\!\rangle$ in the same way as in \eqref{trD1D21100integral}.
Namely, we substitute \eqref{zintegrationin<x|D2D1|y>} into \eqref{<<|dDDd|>>method2}, replace the integration with the principal-value integration and then perform the integration for the two contributions from $e^{(-i\zeta_1-1)x}$ and $-e^{(-i\zeta_1-1)y}$ in \eqref{zintegrationin<x|D2D1|y>} separately.
Since the rest of the integrand is antisymmetric under $x\leftrightarrow y$, the two contributions are actually identical, which leads to
\begin{align}
&\langle\!\langle t_{1,1}|{\widehat d}_1{\widehat D}_2{\widehat D}_1{\widehat d}_2|{-t_{1,1}}\rangle\!\rangle\nonumber\\
&=
-\frac{i}{\sinh\pi \zeta_1}
\strokedint_{-\infty}^\infty\frac{dx}{2\pi}
\frac{e^{-\frac{x}{2}}}{(2\cosh\frac{x}{2})(2\cosh\frac{x+2\pi \zeta_2}{2})}
\strokedint_{-\infty}^\infty\frac{dy}{2\pi}
\frac{e^{(i\zeta_1+\frac{1}{2})y}}{(2\cosh\frac{y}{2})(2\cosh\frac{y+2\pi \zeta_2}{2})}
\frac{1}{2\sinh\frac{x-y}{2}}.
\end{align}
Now the integration can be performed step by step with the formula \eqref{Inalphabetagamma}.
We find
\begin{align}
&\langle\!\langle t_{1,1}|{\widehat d}_1{\widehat D}_2{\widehat D}_1{\widehat d}_2|{-t_{1,1}}\rangle\!\rangle
=\frac{1}{4\sinh^2\pi\zeta_1}\nonumber\\
&\times\Bigl[\frac{1}{\sinh\pi\zeta_2}\big(I_{1,3}({\textstyle-\frac{1}{2}},\{0,0,-2\pi \zeta_2\})-e^{-(i\zeta_1+\frac{1}{2})2\pi\zeta_2}I_{1,3}({\textstyle-\frac{1}{2}},\{0,-2\pi\zeta_2,-2\pi\zeta_2\})\big)\nonumber\\
&\qquad+2i\cosh\pi\zeta_1I_{1,4}(i\zeta_1,\{0,0,-2\pi\zeta_2,-2\pi \zeta_2\})\Bigr].
\end{align}
Substituting this together with \eqref{1111<<|d1d2|>>}, \eqref{1111TrD1D2} into \eqref{1111Z(1)/Z(0)}, we finally find
\begin{align}
&\frac{Z_{k,{\bm M}}(1)}{Z_{k,{\bm M}}(0)}
=\frac{\pi^{-1}\sinh\pi\zeta_1\sinh\pi\zeta_2\cos\pi\zeta_1\zeta_2+\cosh\pi\zeta_1\cosh\pi\zeta_2\sin\pi\zeta_1\zeta_2}{4\sinh^2\pi\zeta_1\sinh^2\pi\zeta_2\sin\pi\zeta_1\zeta_2}\nonumber\\
&\quad -\frac{\zeta_1\cosh\pi\zeta_1\cos\pi\zeta_1\zeta_2}{4\sinh^2\pi\zeta_1\sinh\pi\zeta_2\sin\pi\zeta_1\zeta_2}
-\frac{\zeta_2\cosh\pi\zeta_2\cos\pi\zeta_1\zeta_2}{4\sinh\pi\zeta_1\sinh^2\pi\zeta_2\sin\pi\zeta_1\zeta_2}
-\frac{\zeta_1\zeta_2}{4\sinh\pi\zeta_1\sinh\pi\zeta_2}.
\end{align}

\section{Partition functions at $N=0$}\label{lowestpf}

We summarize the lowest term of the grand partition function in this appendix.
Two subsections are devoted to the case of $k=1$ and $k=2$ respectively.
We denote the lowest term by 
\begin{align}
Z^{(0)}_{M_0,M_1,M_3}=e^{-i\Theta_{k,{\bm M}}}Z_{k,{\bm M}}(N=0)=e^{-i\Theta'_{k,{\bm M}}}Z'_{k,{\bm M}}(N=0).
\end{align}
Here, for simplicity in expressions, we remove both the phase coming from partition functions $e^{i\Theta'_{k,{\bm M}}}$ and the phase introduced to simplify the bilinear relations $e^{i\Theta_{k,{\bm M}}}$.
Also, on the left-hand side, we denote $N=0$ only by superscripts $(0)$ and drop indices of the level $k$ and the FI parameters $Z_1,Z_3$ since the level is common in each subsection and the results obviously depend on the FI parameters.

Note that partition functions at $N=0$ vanish identically for lattice points outside the fundamental domain and we restrict ourselves in the fundamental domain.
It is also interesting to note that the results below and those in appendix \ref{higherrank1} are given cleanly by trigonometric functions (instead of mixtures of trigonometric functions and hyperbolic functions) only when these results are written in terms of $Z_1$ and $Z_3$.
See, for example, $Z^{(0)}_{0,0,0}$ for $k=1$ \eqref{Z0000}.
This may relate to the reality condition on the FI parameters we have discussed above appendix \ref{tracesandmatrixelements}.

\subsection{$k=1$}\label{k1}

In this subsection, we list exact values of partition functions for $k=1$.
For $M_0=\pm 1$, they are given by
\begin{align}
Z^{(0)}_{-1,0,0}=1,\quad
Z^{(0)}_{1,0,0}=e^{-2\pi iZ_1Z_3}.
\end{align}
For $M_0=\pm\frac{1}{2}$, they are given by
\begin{align}
&Z^{(0)}_{-\frac{1}{2},-\frac{1}{2},-\frac{1}{2}}=\frac{e^{-2\pi iZ_1Z_3}}{4}\sec(\pi Z_1)\sec(\pi Z_3),\qquad
Z^{(0)}_{-\frac{1}{2},-\frac{1}{2},\frac{1}{2}}=\frac{1}{2}\sec(\pi Z_1),\nonumber\\
&Z^{(0)}_{-\frac{1}{2},\frac{1}{2},-\frac{1}{2}}=\frac{1}{2}\sec(\pi Z_3),\qquad
Z^{(0)}_{-\frac{1}{2},\frac{1}{2},\frac{1}{2}}=1,
\end{align}
and
\begin{align}
&Z^{(0)}_{\frac{1}{2},-\frac{1}{2},-\frac{1}{2}}=\frac{e^{-2\pi iZ_1Z_3}}{4}\sec(\pi Z_1)\sec(\pi Z_3),\qquad
Z^{(0)}_{\frac{1}{2},-\frac{1}{2},\frac{1}{2}}=\frac{e^{-2\pi iZ_1Z_3}}{2}\sec(\pi Z_1),\nonumber\\
&Z^{(0)}_{\frac{1}{2},\frac{1}{2},-\frac{1}{2}}=\frac{e^{-2\pi iZ_1Z_3}}{2}\sec(\pi Z_3),\qquad
Z^{(0)}_{\frac{1}{2},\frac{1}{2},\frac{1}{2}}=1.
\end{align}
Finally, for $M_0=0$, they are given by
\begin{align}
&Z^{(0)}_{0,-1,0}=-\frac{e^{-2\pi iZ_1Z_3}}{4}\csc^2(\pi Z_1),\qquad
Z^{(0)}_{0,0,-1}=-\frac{e^{-2\pi iZ_1Z_3}}{4}\csc^2(\pi Z_3),\nonumber\\
&Z^{(0)}_{0,0,0}=\frac{e^{-\pi iZ_1Z_3}}{2}\csc(\pi Z_1)\csc(\pi Z_3)\sin(\pi Z_1Z_3),\qquad
Z^{(0)}_{0,0,1}=1,\qquad
Z^{(0)}_{0,1,0}=1.
\label{Z0000}
\end{align}

\subsection{$k=2$}\label{k2}

In this subsection, we list exact values of partition functions for $k=2$.
To simplify the result, we introduce the functions
\begin{align}
X^\pm=e^{\mp\frac{\pi i}{4}}e^{\mp\pi iZ_1Z_3}e^{-\frac{\pi i}{2}(Z_1+Z_3)}\big(1\pm ie^{\pi iZ_1}\pm ie^{\pi iZ_3}+e^{\pi i(Z_1+Z_3)}\big)-2.
\end{align}

\noindent\underline{$M_0=-2$}
\begin{align}
Z^{(0)}_{-2,0,0}=1.
\end{align}
\noindent\underline{$M_0=-\frac{3}{2}$}
\begin{align}
&Z^{(0)}_{-\frac{3}{2},-\frac{1}{2},-\frac{1}{2}}
=\frac{e^{-\pi iZ_1Z_2}}{4\sqrt{2}}\sec(\pi Z_1)\sec(\pi Z_3),\quad
Z^{(0)}_{-\frac{3}{2},-\frac{1}{2},\frac{1}{2}}
=\frac{1}{2\sqrt{2}}\sec(\pi Z_1),\nonumber\\
&Z^{(0)}_{-\frac{3}{2},\frac{1}{2},-\frac{1}{2}}
=\frac{1}{2\sqrt{2}}\sec(\pi Z_3),\quad
Z^{(0)}_{-\frac{3}{2},\frac{1}{2},\frac{1}{2}}
=\frac{1}{\sqrt{2}}.
\end{align}
\noindent\underline{$M_0=-1$}
\begin{align}
&Z^{(0)}_{-1,-1,-1}=\frac{e^{-2\pi iZ_1Z_3}}{16}\csc^2(\pi Z_1)\csc^2(\pi Z_3),\quad
Z^{(0)}_{-1,-1,0}=-\frac{e^{-\pi iZ_1Z_3}}{16}\csc^2(\pi Z_1)\sec\frac{\pi Z_3}{2},\nonumber\\
&Z^{(0)}_{-1,-1,1}=-\frac{1}{4}\csc^2(\pi Z_1),\quad
Z^{(0)}_{-1,0,-1}=-\frac{e^{-\pi iZ_1Z_3}}{16}\sec\frac{\pi Z_1}{2}\csc^2(\pi Z_3),\nonumber\\
&Z^{(0)}_{-1,0,0}=\frac{e^{-\frac{\pi i}{2}Z_1Z_3}}{4}\csc(\pi Z_1)\csc(\pi Z_3)\sin\frac{\pi Z_1Z_3}{2}
,\quad
Z^{(0)}_{-1,0,1}=\frac{1}{4}\sec\frac{\pi Z_1}{2},
\nonumber\\
&Z^{(0)}_{-1,1,-1}=-\frac{1}{4}\csc(\pi Z_3)^2,\quad
Z^{(0)}_{-1,1,0}=\frac{1}{4}\sec\frac{\pi Z_3}{2},\quad
Z^{(0)}_{-1,1,1}=1.
\end{align}
\noindent\underline{$M_0=-\frac{1}{2}$}
\begin{align}
&Z^{(0)}_{-\frac{1}{2},-\frac{3}{2},-\frac{1}{2}}=\frac{e^{-2\pi iZ_1Z_3}}{16\sqrt{2}}\sec^3(\pi Z_1)\sec(\pi Z_3),\quad
Z^{(0)}_{-\frac{1}{2},-\frac{3}{2},\frac{1}{2}}=\frac{e^{-\pi iZ_1Z_3}}{8\sqrt{2}}\sec^3(\pi Z_1),\nonumber\\
&Z^{(0)}_{-\frac{1}{2},-\frac{1}{2},-\frac{3}{2}}=\frac{e^{-2\pi iZ_1Z_3}}{16\sqrt{2}}\sec(\pi Z_1)\sec^3(\pi Z_3),\quad
Z^{(0)}_{-\frac{1}{2},-\frac{1}{2},-\frac{1}{2}}=\frac{e^{-2\pi iZ_1Z_3}}{32\sqrt{2}}\sec^2(\pi Z_1)\sec^2(\pi Z_3)X^-,\nonumber\\
&Z^{(0)}_{-\frac{1}{2},-\frac{1}{2},\frac{1}{2}}=\frac{1}{16\sqrt{2}}\sec^2(\pi Z_1)\sec(\pi Z_3)X^+,\quad
Z^{(0)}_{-\frac{1}{2},-\frac{1}{2},\frac{3}{2}}=\frac{1}{2\sqrt{2}}\sec(\pi Z_1),\nonumber\\
&Z^{(0)}_{-\frac{1}{2},\frac{1}{2},-\frac{3}{2}}=\frac{e^{-\pi iZ_1Z_3}}{8\sqrt{2}}\sec^3(\pi Z_3),\quad
Z^{(0)}_{-\frac{1}{2},\frac{1}{2},-\frac{1}{2}}=\frac{1}{16\sqrt{2}}\sec(\pi Z_1)\sec^2(\pi Z_3)X^+,\nonumber\\
&Z^{(0)}_{-\frac{1}{2},\frac{1}{2},\frac{1}{2}}=\frac{e^{-\pi iZ_1Z_3}}{8\sqrt{2}}\sec(\pi Z_1)\sec(\pi Z_3)X^-,\quad
Z^{(0)}_{-\frac{1}{2},\frac{1}{2},\frac{3}{2}}=\frac{1}{\sqrt{2}},\nonumber\\
&Z^{(0)}_{-\frac{1}{2},\frac{3}{2},-\frac{1}{2}}=\frac{1}{2\sqrt{2}}\sec(\pi Z_3),\quad
Z^{(0)}_{-\frac{1}{2},\frac{3}{2},\frac{1}{2}}=\frac{1}{\sqrt{2}}.
\end{align}
\noindent\underline{$M_0=0$}
\begin{align}
&Z^{(0)}_{0,-2,0}=\frac{e^{-2\pi iZ_1Z_3}}{16}\csc^4(\pi Z_1),\quad
Z^{(0)}_{0,-1,-1}=\frac{e^{-2\pi iZ_1Z_3}}{32}\csc^2(\pi Z_1)\csc^2(\pi Z_3),\nonumber\\
&Z^{(0)}_{0,-1,0}=-\frac{e^{-\frac{3\pi i}{2}Z_1Z_2}}{16}\csc\frac{\pi Z_1}{2}\csc^2(\pi Z_1)\csc(\pi Z_3)\sin\frac{\pi Z_1Z_3}{2},\quad
Z^{(0)}_{0,-1,1}=-\frac{e^{-\pi iZ_1Z_3}}{8}\csc^2(\pi Z_1),\nonumber\\
&Z^{(0)}_{0,0,-2}=\frac{e^{-2\pi iZ_1Z_3}}{16}\csc^4(\pi Z_3),\quad
Z^{(0)}_{0,0,-1}=-\frac{e^{-\frac{3\pi i}{2}Z_1Z_3}}{16}\csc(\pi Z_1)\csc\frac{\pi Z_3}{2}\csc^2(\pi Z_3)\sin\frac{\pi Z_1Z_3}{2},\nonumber\\
&Z^{(0)}_{0,0,0}=-\frac{e^{-\pi iZ_1Z_3}}{16}\csc^2(\pi Z_1)\csc^2(\pi Z_3)\nonumber\\
&\qquad\qquad\times\Big(1+\cos(\pi Z_1)+\cos(\pi Z_3)-\cos(\pi Z_1)\cos(\pi Z_3)-2\cos(\pi Z_1Z_3)\Big),\nonumber\\
&Z^{(0)}_{0,0,1}=\frac{e^{-\frac{\pi i}{2}Z_1Z_3}}{4}\csc(\pi Z_1)\csc\frac{\pi Z_3}{2}\sin\frac{\pi Z_1Z_3}{2},\quad
Z^{(0)}_{0,0,2}=1,\quad
Z^{(0)}_{0,1,-1}=-\frac{e^{-\pi iZ_1Z_3}}{8}\csc^2(\pi Z_3),\nonumber\\
&
Z^{(0)}_{0,1,0}=\frac{e^{-\frac{\pi i}{2}Z_1Z_3}}{4}\csc\frac{\pi Z_1}{2}\csc(\pi Z_3)\sin\frac{\pi Z_1Z_3}{2},\quad
Z^{(0)}_{0,1,1}=\frac{1}{2},\quad
Z^{(0)}_{0,2,0}=1.
\end{align}
\noindent\underline{$M_0=\frac{1}{2}$}
\begin{align}
&Z^{(0)}_{\frac{1}{2},-\frac{3}{2},-\frac{1}{2}}
=\frac{e^{-2\pi iZ_1Z_3}}{16\sqrt{2}}\sec^3(\pi Z_1)\sec(\pi Z_3),\quad
Z^{(0)}_{\frac{1}{2},-\frac{3}{2},\frac{1}{2}}
=\frac{e^{-2\pi iZ_1Z_3}}{8\sqrt{2}}\sec^3(\pi Z_1),\nonumber\\
&Z^{(0)}_{\frac{1}{2},-\frac{1}{2},-\frac{3}{2}}
=\frac{e^{-2\pi iZ_1Z_3}}{16\sqrt{2}}\sec(\pi Z_1)\sec^3(\pi Z_3),\quad
Z^{(0)}_{\frac{1}{2},-\frac{1}{2},-\frac{1}{2}}
=\frac{e^{-\pi iZ_1Z_3}}{32\sqrt{2}}\sec^2(\pi Z_1)\sec^2(\pi Z_3)X^+,\nonumber\\
&Z^{(0)}_{\frac{1}{2},-\frac{1}{2},\frac{1}{2}}
=\frac{e^{-2\pi iZ_1Z_3}}{16\sqrt{2}}\sec^2(\pi Z_1)\sec(\pi Z_3)X^-,\quad
Z^{(0)}_{\frac{1}{2},-\frac{1}{2},\frac{3}{2}}
=\frac{e^{-\pi iZ_1Z_3}}{2\sqrt{2}}\sec(\pi Z_1),\nonumber\\
&Z^{(0)}_{\frac{1}{2},\frac{1}{2},-\frac{3}{2}}
=\frac{e^{-2\pi iZ_1Z_3}}{8\sqrt{2}}\sec^3(\pi Z_3),\quad
Z^{(0)}_{\frac{1}{2},\frac{1}{2},-\frac{1}{2}}
=\frac{e^{-2\pi iZ_1Z_3}}{16\sqrt{2}}\sec(\pi Z_1)\sec^2(\pi Z_3)X^-,\nonumber\\
&Z^{(0)}_{\frac{1}{2},\frac{1}{2},\frac{1}{2}}
=\frac{1}{8\sqrt{2}}\sec(\pi Z_1)\sec(\pi Z_3)X^+,\quad
Z^{(0)}_{\frac{1}{2},\frac{1}{2},\frac{3}{2}}
=\frac{1}{\sqrt{2}},\nonumber\\
&Z^{(0)}_{\frac{1}{2},\frac{3}{2},-\frac{1}{2}}
=\frac{e^{-\pi iZ_1Z_3}}{2\sqrt{2}}\sec(\pi Z_3),\quad
Z^{(0)}_{\frac{1}{2},\frac{3}{2},\frac{1}{2}}
=\frac{1}{\sqrt{2}}.
\end{align}
\noindent\underline{$M_0=1$}
\begin{align}
&Z^{(0)}_{1,-1,-1}=\frac{e^{-2\pi iZ_1Z_3}}{16}\csc^2(\pi Z_1)\csc^2(\pi Z_3),\quad
Z^{(0)}_{1,-1,0}=-\frac{e^{-2\pi iZ_1Z_3}}{16}\csc^2(\pi Z_1)\sec\frac{\pi Z_3}{2},\nonumber\\
&Z^{(0)}_{1,-1,1}=-\frac{e^{-2\pi iZ_1Z_3}}{4}\csc^2(\pi Z_1),\quad
Z^{(0)}_{1,0,-1}=-\frac{e^{-2\pi iZ_1Z_3}}{16}\sec\frac{\pi Z_1}{2}\csc^2(\pi Z_3),\nonumber\\
&Z^{(0)}_{1,0,0}=\frac{e^{-\frac{3\pi i}{2}Z_1Z_3}}{4}\csc(\pi Z_1)\csc(\pi Z_3)\sin\frac{\pi Z_1Z_3}{2},\quad
Z^{(0)}_{1,0,1}=\frac{e^{-\pi iZ_1Z_3}}{4}\sec\frac{\pi Z_1}{2},\nonumber\\
&Z^{(0)}_{1,1,-1}=-\frac{e^{-2\pi iZ_1Z_3}}{4}\csc^2(\pi Z_3),\quad
Z^{(0)}_{1,1,0}=\frac{e^{-\pi iZ_1Z_3}}{4}\sec\frac{\pi Z_3}{2},\quad
Z^{(0)}_{1,1,1}=1.
\end{align}
\noindent\underline{$M_0=\frac{3}{2}$}
\begin{align}
&Z^{(0)}_{\frac{3}{2},-\frac{1}{2},-\frac{1}{2}}=\frac{e^{-2\pi iZ_1Z_3}}{4\sqrt{2}}\sec(\pi Z_1)\sec(\pi Z_3),\quad
Z^{(0)}_{\frac{3}{2},-\frac{1}{2},\frac{1}{2}}=\frac{e^{-2\pi iZ_1Z_3}}{2\sqrt{2}}\sec(\pi Z_1),\nonumber\\
&Z^{(0)}_{\frac{3}{2},\frac{1}{2},-\frac{1}{2}}=\frac{e^{-2\pi iZ_1Z_3}}{2\sqrt{2}}\sec(\pi Z_3),\quad
Z^{(0)}_{\frac{3}{2},\frac{1}{2},\frac{1}{2}}=\frac{e^{-\pi iZ_1Z_3}}{\sqrt{2}}.
\end{align}
\noindent\underline{$M_0=2$}
\begin{align}
Z^{(0)}_{2,0,0}=e^{-2\pi iZ_1Z_3}.
\end{align}

\section{Partition functions of higher ranks for $k=1$}\label{higherrank1}

In this appendix, we list exact values of partition functions at $N=1$ for $k=1$.
We denote them by
\begin{align}
Z^{(1)}_{M_0,M_1,M_3}=e^{-i\Theta_{k,{\bm M}}}Z_{k,{\bm M}}(N=1)=e^{-i\Theta'_{k,{\bm M}}}Z'_{k,{\bm M}}(N=1).
\end{align}
As in appendix \ref{lowestpf} for $N=0$, we remove the phases and, on the left-hand side, we denote $N=1$ only by superscripts $(1)$ and drop indices of the level $k$ and the FI parameters $Z_1,Z_3$.
Partition functions at $N=1$ for lattice points in the fundamental domain, whose ratios to those at $N=0$ are invariant under the Weyl group, are fixed by \eqref{XYZ}.
We only list those with the relative ranks $(M_0,M_1,M_3)$ outside the fundamental domain.

\noindent\underline{$M_0=-\frac{3}{2}$}
\begin{align}
&Z^{(1)}_{-\frac{3}{2},-\frac{1}{2},-\frac{1}{2}}=\frac{1}{4}\sec(\pi Z_1)\sec(\pi Z_3),\quad
Z^{(1)}_{-\frac{3}{2},-\frac{1}{2},\frac{1}{2}}=\frac{1}{2}\sec(\pi Z_1),\nonumber\\
&Z^{(1)}_{-\frac{3}{2},\frac{1}{2},-\frac{1}{2}}=\frac{1}{2}\sec(\pi Z_3),\quad
Z^{(1)}_{-\frac{3}{2},\frac{1}{2},\frac{1}{2}}=e^{2\pi iZ_1Z_3}.
\end{align}
\noindent\underline{$M_0=-1$}
\begin{align}
&Z^{(1)}_{-1,-1,-1}=\frac{e^{-2\pi iZ_1Z_3}}{16}\csc^2(\pi Z_1)\csc^2(\pi Z_3),\quad
Z^{(1)}_{-1,-1,1}=-\frac{1}{4}\csc^2(\pi Z_1),\nonumber\\
&Z^{(1)}_{-1,1,-1}=-\frac{1}{4}\csc^2(\pi Z_3),\quad
Z^{(1)}_{-1,1,1}=e^{2\pi iZ_1Z_3},
\end{align}
\begin{align}
&Z^{(1)}_{-1,-1,0}=-\frac{e^{-\pi iZ_1Z_3}}{8}\csc^3(\pi Z_1)\csc(\pi Z_3)\sin(\pi Z_1Z_3),\nonumber\\
&Z^{(1)}_{-1,0,-1}=-\frac{e^{-\pi iZ_1Z_3}}{8}\csc(\pi Z_1)\csc^3(\pi Z_3)\sin(\pi Z_1Z_3),\nonumber\\
&Z^{(1)}_{-1,0,1}=\frac{e^{\pi iZ_1Z_3}}{2}\csc(\pi Z_1)\csc(\pi Z_3)\sin(\pi Z_1Z_3),\nonumber\\
&Z^{(1)}_{-1,1,0}=\frac{e^{\pi iZ_1Z_3}}{2}\csc(\pi Z_1)\csc(\pi Z_3)\sin(\pi Z_1Z_3).
\end{align}
\noindent\underline{$M_0=-\frac{1}{2}$}
\begin{align}
&Z^{(1)}_{-\frac{1}{2},-\frac{3}{2},-\frac{1}{2}}=\frac{e^{-2\pi iZ_1Z_3}}{32}\sec^4(\pi Z_1)\sec(\pi Z_3),\quad
Z^{(1)}_{-\frac{1}{2},-\frac{3}{2},\frac{1}{2}}=\frac{e^{-2\pi iZ_1Z_3}}{16}\sec^4(\pi Z_1),\nonumber\\
&Z^{(1)}_{-\frac{1}{2},-\frac{1}{2},-\frac{3}{2}}=\frac{e^{-2\pi iZ_1Z_3}}{32}\sec(\pi Z_1)\sec^4(\pi Z_3),\quad
Z^{(1)}_{-\frac{1}{2},-\frac{1}{2},\frac{3}{2}}=\frac{1}{4}\sec(\pi Z_1)\sec(\pi Z_3),\nonumber\\
&Z^{(1)}_{-\frac{1}{2},\frac{1}{2},-\frac{3}{2}}=\frac{e^{-2\pi iZ_1Z_3}}{16}\sec^4(\pi Z_3),\quad
Z^{(1)}_{-\frac{1}{2},\frac{1}{2},\frac{3}{2}}=\frac{e^{2\pi iZ_1Z_3}}{2}\sec(\pi Z_3),\nonumber\\
&Z^{(1)}_{-\frac{1}{2},\frac{3}{2},-\frac{1}{2}}=\frac{1}{4}\sec(\pi Z_1)\sec(\pi Z_3),\quad
Z^{(1)}_{-\frac{1}{2},\frac{3}{2},\frac{1}{2}}=\frac{e^{2\pi iZ_1Z_3}}{2}\sec(\pi Z_1).
\end{align}
\noindent\underline{$M_0=0$}
\begin{align}
&Z^{(1)}_{0,-1,-1}=\frac{e^{-3\pi iZ_1Z_3}}{32}\csc^3(\pi Z_1)\csc^3(\pi Z_3)\sin(\pi Z_1Z_3),\nonumber\\
&Z^{(1)}_{0,-1,1}=-\frac{e^{-\pi iZ_1Z_3}}{8}\csc^3(\pi Z_1)\csc(\pi Z_3)\sin(\pi Z_1Z_3),\nonumber\\
&Z^{(1)}_{0,1,-1}=-\frac{e^{-\pi iZ_1Z_3}}{8}\csc(\pi Z_1)\csc^3(\pi Z_3)\sin(\pi Z_1Z_3),\nonumber\\
&Z^{(1)}_{0,1,1}=\frac{e^{\pi iZ_1Z_3}}{2}\csc(\pi Z_1)\csc(\pi Z_3)\sin(\pi Z_1Z_3).
\label{Z1M0_011}
\end{align}
\noindent\underline{$M_0=\frac{1}{2}$}
\begin{align}
&Z^{(1)}_{\frac{1}{2},-\frac{3}{2},-\frac{1}{2}}=\frac{e^{-4\pi iZ_1Z_3}}{32}\sec^4(\pi Z_1)\sec(\pi Z_3),\quad
Z^{(1)}_{\frac{1}{2},-\frac{3}{2},\frac{1}{2}}=\frac{e^{-2\pi iZ_1Z_3}}{16}\sec^4(\pi Z_1),\nonumber\\
&Z^{(1)}_{\frac{1}{2},-\frac{1}{2},-\frac{3}{2}}=\frac{e^{-4\pi iZ_1Z_3}}{32}\sec(\pi Z_1)\sec^4(\pi Z_3),\quad
Z^{(1)}_{\frac{1}{2},-\frac{1}{2},\frac{3}{2}}=\frac{1}{4}\sec(\pi Z_1)\sec(\pi Z_3),\nonumber\\
&Z^{(1)}_{\frac{1}{2},\frac{1}{2},-\frac{3}{2}}=\frac{e^{-2\pi iZ_1Z_3}}{16}\sec^4(\pi Z_3),\quad
Z^{(1)}_{\frac{1}{2},\frac{1}{2},\frac{3}{2}}=\frac{1}{2}\sec(\pi Z_3),\nonumber\\
&Z^{(1)}_{\frac{1}{2},\frac{3}{2},-\frac{1}{2}}=\frac{1}{4}\sec(\pi Z_1)\sec(\pi Z_3),\quad
Z^{(1)}_{\frac{1}{2},\frac{3}{2},\frac{1}{2}}=\frac{1}{2}\sec(\pi Z_1).
\end{align}
\noindent\underline{$M_0=1$}
\begin{align}
&Z^{(1)}_{1,-1,0}=-\frac{e^{-3\pi iZ_1Z_3}}{8}\csc^3(\pi Z_1)\csc(\pi Z_3)\sin(\pi Z_1Z_3),\nonumber\\
&Z^{(1)}_{1,0,-1}=-\frac{e^{-3\pi iZ_1Z_3}}{8}\csc(\pi Z_1)\csc^3(\pi Z_3)\sin(\pi Z_1Z_3),\nonumber\\
&Z^{(1)}_{1,0,1}=\frac{e^{-\pi iZ_1Z_3}}{2}\csc(\pi Z_1)\csc(\pi Z_3)\sin(\pi Z_1Z_3),\nonumber\\
&Z^{(1)}_{1,1,0}=\frac{e^{-\pi iZ_1Z_3}}{2}\csc(\pi Z_1)\csc(\pi Z_3)\sin(\pi Z_1Z_3),
\end{align}
\begin{align}
&Z^{(1)}_{1,-1,-1}=\frac{e^{-4\pi iZ_1Z_3}}{16}\csc^2(\pi Z_1)\csc^2(\pi Z_3),\quad
Z^{(1)}_{1,-1,1}=-\frac{e^{-2\pi iZ_1Z_3}}{4}\csc^2(\pi Z_1),\nonumber\\
&Z^{(1)}_{1,1,-1}=-\frac{e^{-2\pi iZ_1Z_3}}{4}\csc^2(\pi Z_3),\quad
Z^{(1)}_{1,1,1}=1.
\end{align}
\noindent\underline{$M_0=\frac{3}{2}$}
\begin{align}
&Z^{(1)}_{\frac{3}{2},-\frac{1}{2},-\frac{1}{2}}=\frac{e^{-4\pi iZ_1Z_3}}{4}\sec(\pi Z_1)\sec(\pi Z_3),\quad
Z^{(1)}_{\frac{3}{2},-\frac{1}{2},\frac{1}{2}}=\frac{e^{-2\pi iZ_1Z_3}}{2}\sec(\pi Z_1),\nonumber\\
&Z^{(1)}_{\frac{3}{2},\frac{1}{2},-\frac{1}{2}}=\frac{e^{-2\pi iZ_1Z_3}}{2}\sec(\pi Z_3),\quad
Z^{(1)}_{\frac{3}{2},\frac{1}{2},\frac{1}{2}}=e^{-2\pi iZ_1Z_3}.
\end{align}

\section{List of values checked for bilinear relations}
\label{checkedlist}

\begin{figure}[!t]
\begin{center}
\includegraphics[width=16cm]{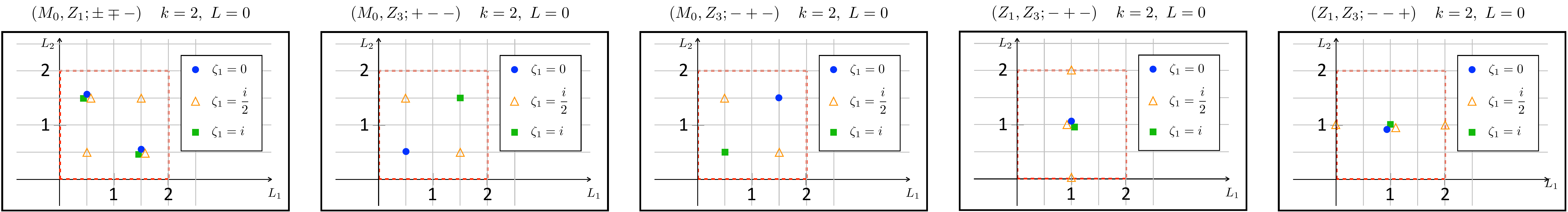}
\caption{
List of points $(k,L_1,L_2,L,\zeta_1)$ where we have checked the bilinear relations in \eqref{checkableatk2} at order $\kappa$ explicitly against the exact values.
Note that all data points are on the grid $2L_1,2L_2\in \mathbb{Z}$, although some of them are drawn with slight displacements for visibility.
The dashed red line is the boundary of the fundamental domain \eqref{funddomain} on each slice.
}
\label{checkedpointsk2}
\end{center}
\end{figure}

\begin{figure}[!t]
\begin{center}
\includegraphics[width=16cm]{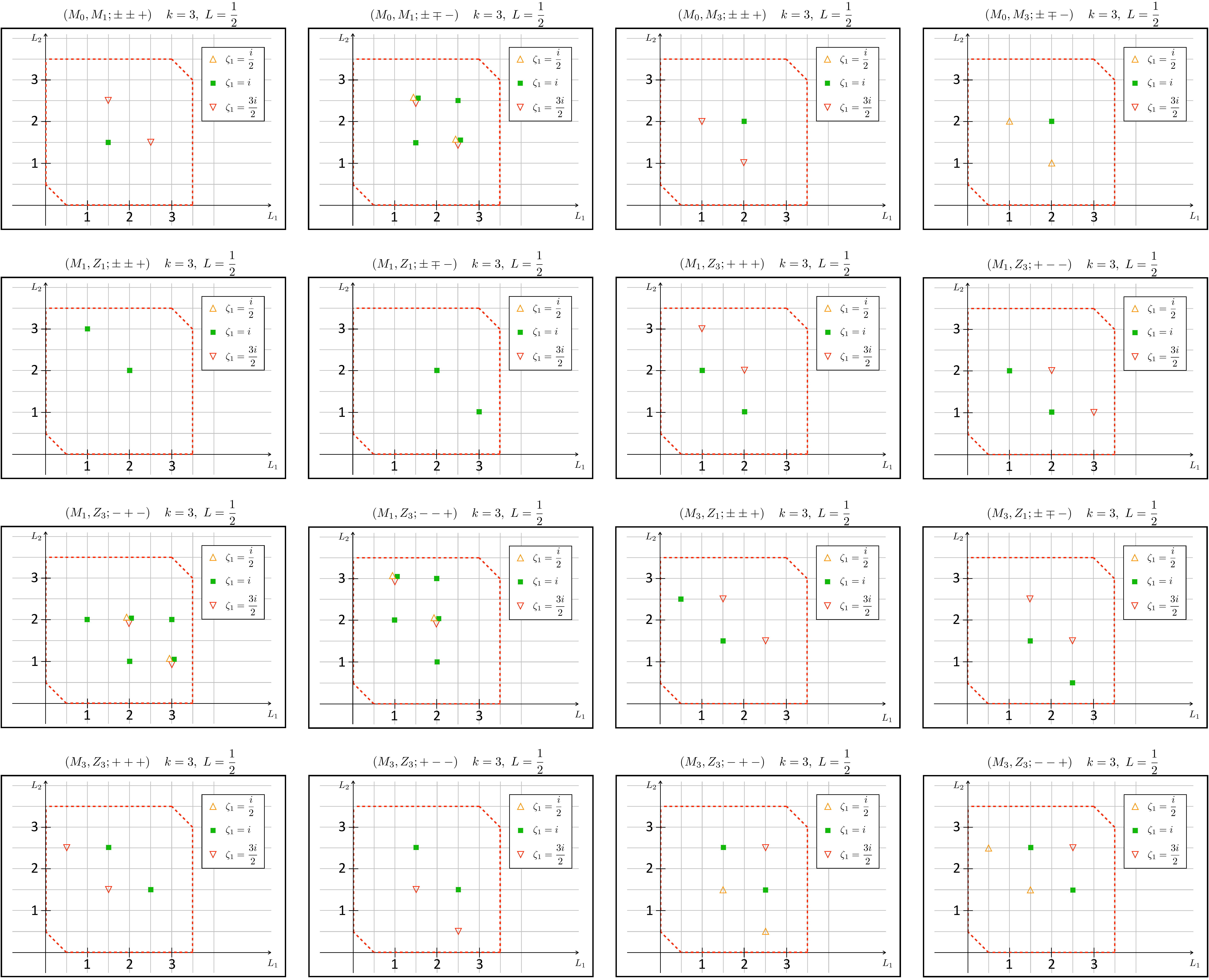}
\caption{
List of points $(k,L_1,L_2,L,\zeta_1)$ where we have checked the bilinear relations in \eqref{checkableatk3} at order $\kappa$ explicitly against the exact values.
}
\label{checkedpointsk3}
\end{center}
\end{figure}

\begin{figure}[!t]
\begin{center}
\includegraphics[width=16cm]{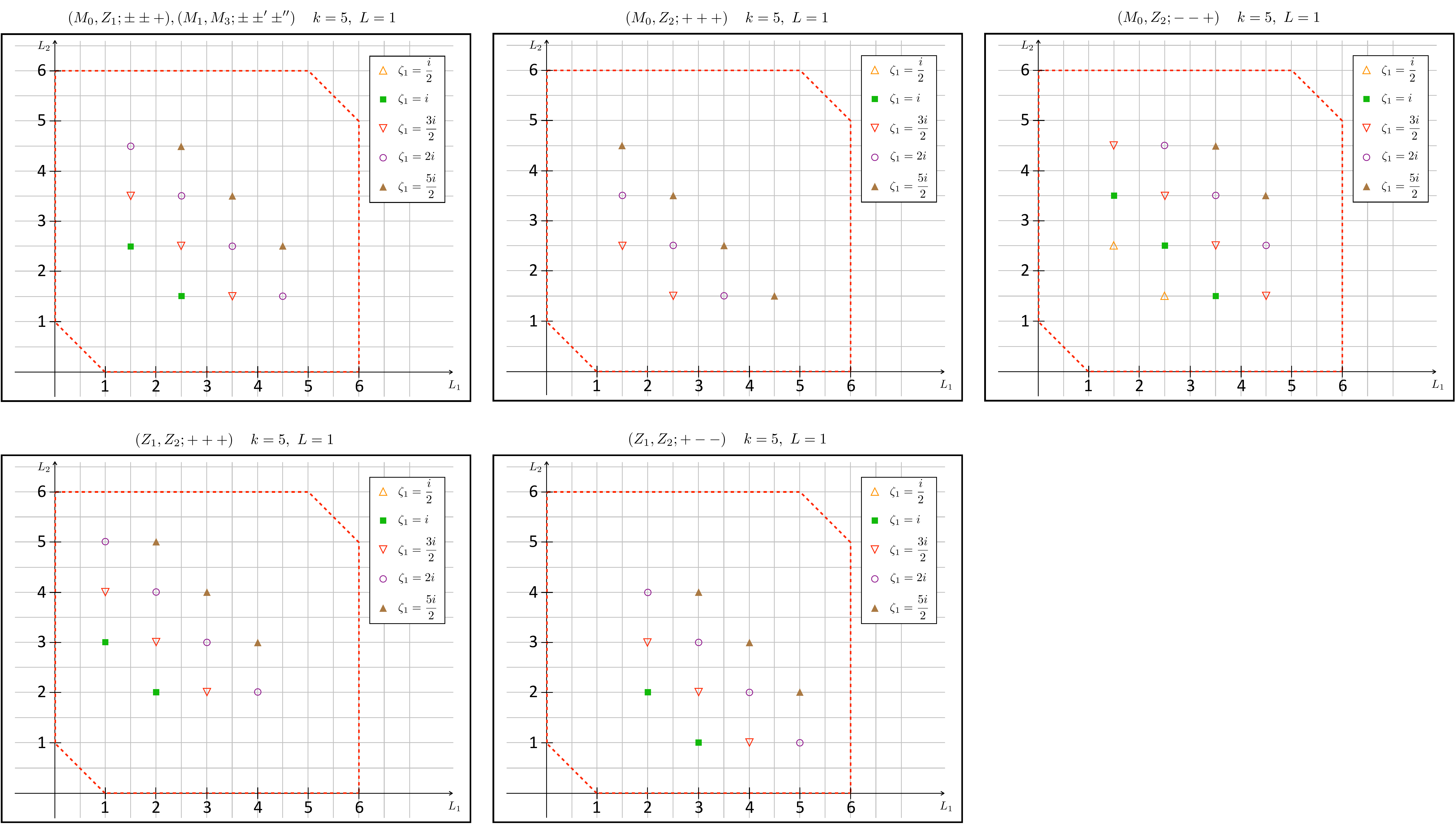}
\caption{
List of points $(k,L_1,L_2,L,\zeta_1)$ where we have checked the bilinear relations in \eqref{checkableatk5} at order $\kappa$ explicitly against the exact values.
}
\label{checkedpointsk5}
\end{center}
\end{figure}

In this appendix we list points $(L_1,L_2,L,\zeta_1)$ where we can check each of the 40 bilinear relations \eqref{bilinearwithkappa2} at the order of $\kappa$, by calculating the exact values with method I explained in section \ref{method1}.
The reason we study with Method I in addition to Method II is the convergence condition discussed in Appendix \ref{app_validity}, where the computations are made in more sound arguments.
For this purpose, we have to choose the level $k$ and the base point $(L_1,L_2,L,\zeta_1)$ so that all of the six shifted points in \eqref{bilinearwithkappa2} satisfy the condition \eqref{validitycondition}.
As one can see from \eqref{validitycondition}, in order for such points to exist $k$ has to be sufficiently large.
In particular, the bilinear relations labelled by
\begin{align}
(a,b;\sigma_1,\sigma_2,\sigma_3)&=(M_0,Z_1;+--),\,
(M_0,Z_1;-+-),\,
(M_0,Z_3;+--),\,
(M_0,Z_3;-+-),\nonumber \\
&\quad\quad (Z_1,Z_3;-+-),\,
(Z_1,Z_3;--+),
\label{checkableatk2}
\end{align}
can be checked for multiple points of $(L_1,L_2,L,\zeta_1)$ only when $k\ge 2$, and those labelled by
\begin{align}
&(M_0,M_1;\sigma_1,\sigma_2,\sigma_3),\,
(M_0,M_3;\sigma_1,\sigma_2,\sigma_3),\,
(M_1,Z_1;\sigma_1,\sigma_2,\sigma_3),\,
(M_1,Z_3;\sigma_1,\sigma_2,\sigma_3),\nonumber \\
&(M_3,Z_1;\sigma_1,\sigma_2,\sigma_3),\,
(M_3,Z_3;\sigma_1,\sigma_2,\sigma_3),
\label{checkableatk3}
\end{align}
can be checked for multiple points of $(L_1,L_2,L,\zeta_1)$ only when $k\ge 3$, while the rest of the bilinear relations which are labelled by
\begin{align}
&(M_0,Z_1;+++),\,
(M_0,Z_1;--+),\,
(M_0,Z_3;+++),\,
(M_0,Z_3;--+),\nonumber \\
&(M_1,M_3;\sigma_1,\sigma_2,\sigma_3),\,
(Z_1,Z_3;+++),\,
(Z_1,Z_3;+--),
\label{checkableatk5}
\end{align}
can be checked only when $k\ge 5$.
In figures \ref{checkedpointsk2}, \ref{checkedpointsk3} and \ref{checkedpointsk5}, we list all points $(L_1,L_2,L,\zeta_1)$ where we can check the bilinear relations in \eqref{checkableatk2} for $k=2$, \eqref{checkableatk3} for $k=3$ and \eqref{checkableatk5} for $k=5$ respectively within the constraint.
We have checked that the bilinear relations \eqref{checkableatk2}-\eqref{checkableatk5} are indeed satisfied at these points.

\section*{Acknowledgements}
We are grateful to Giulio Bonelli, Yasuaki Hikida, Masazumi Honda, Yosuke Imamura, Hiroshi Itoyama, Hiroaki Kanno, Naotaka Kubo, Kimyeong Lee, Kazunobu Maruyoshi, Shun'ya Mizoguchi, Takahiro Nishinaka, Tadakatsu Sakai, Alessandro Tanzini, Yasuhiko Yamada, Shintarou Yanagida, Shuichi Yokoyama for valuable discussions and comments.
The work of S.~M.~is supported by JSPS Grant-in-Aid for Scientific Research (C) \#19K03829 and \#22K03598.
S.~M.~would like to thank Yukawa Institute for Theoretical Physics at Kyoto University for warm hospitality.
Part of the exact values used to check the bilinear relations \eqref{bilinearwithkappa2} was obtained by using the high-performance computing facility provided by Yukawa Institute for Theoretical Physics (Sushiki server).
Preliminary results of this paper were presented in international conferences including ``KEK Theory Workshop 2022'' at Ibaraki, Japan, ``6th International Conference on Holography, String Theory and Spacetime in Da Nang'' at Danang, Vietnam and ``Quantum Field Theories and Representation Theory'' at Osaka, Japan.
We are grateful to the organizers and also thank the participants for various valuable discussions.

\end{document}